\documentclass[fleqn,usenatbib]{mnras}

\usepackage{newtxtext,newtxmath}

\usepackage[T1]{fontenc}
\usepackage{physics}
\usepackage[flushleft]{threeparttable}
\usepackage{siunitx}
\DeclareRobustCommand{\VAN}[3]{#2}
\let\VANthebibliography\thebibliography
\def\thebibliography{\DeclareRobustCommand{\VAN}[3]{##3}\VANthebibliography}

\usepackage{graphicx}	

\title[Cloud and ammonia in Jupiter's atmosphere]{Cloud and ammonia vertical profiles in the equatorial atmosphere of Jupiter determined from visible to near-IR observations made by VLT/MUSE, Cassini/VIMS, IRTF/SpeX and Juno/JIRAM}

\author[P.G.J. Irwin et al.]{Patrick G. J. Irwin$^{1}$\thanks{E-mail: patrick.irwin@physics.ox.ac.uk (PGJI)},
Asier Anguiano-Arteaga$^{2,1}$,
Michelle Colantoni$^{1}$,
Joseph Penn$^{1}$,
\newauthor
Santiago P\'{e}rez-Hoyos$^{2}$,
Davide Grassi$^{3}$,
Alessandro Mura$^{3}$,
Charlotte L. B. Alexander$^{4}$,
\newauthor
Leigh N. Fletcher$^{5}$,
Simon C. A. Toogood$^{5}$
and Michael T. Roman$^{5,6}$
\\
$^{1}$Atmospheric, Oceanic and Planetary Physics, Department of Physics, University of Oxford, Parks Rd, Oxford, OX1 3PU, UK\\
$^{2}$Dpto. F\'isica Aplicada, EIB, Universidad del Pa\'is Vasco UPV/EHU, Bilbao, Spain\\
$^{3}$Istituto di Astrofisica e Planetologia Spaziali, Istituto Nazionale di Astrofisica, Via del Fosso del Cavaliere, snc, I-00133 Roma, Italy\\
$^{4}$Global Systems Institute, Mathematics and Statistics Department, University of Exeter, Exeter, EX4 4QE, UK\\
$^{5}$School of Physics and Astronomy, University of Leicester, University Road, Leicester, LE1 7RH, UK\\
$^{6}$Facultad de Ingeniería y Ciencias, Universidad Adolfo Ibáñez, Av. Diagonal Las Torres 2640, Peñalolén, Santiago, Chile}

\date{Accepted XXX. Received YYY; in original form ZZZ}

\pubyear{2026}

\begin{document}
\label{firstpage}
\pagerange{\pageref{firstpage}--\pageref{lastpage}}
\maketitle

\begin{abstract}
We present a combined cloud/ammonia model for Jupiter's equatorial atmosphere from 0.1 to 10 bar, consistent with observations made at a range of observation geometries from 0.35 to 5.15 $\mu$m by VLT/MUSE, Cassini/VIMS, IRTF/SpeX and Juno/JIRAM. Our cloud model has three components: 1) an optically-thick lower cloud (radius $r$$\sim$$10$ $\mu$m) at 1--2 bar; 2) an optically-thin upper cloud ($r$$\sim$$10$ $\mu$m) at $\sim$0.55 bar; and 3) a layer of blue-absorbing chromophore particles ($r$$\sim$$0.2$ $\mu$m) situated within the main lower cloud.  The ammonia profile is intimately linked with the cloud profile with the lower cloud coinciding with an initial drop in ammonia abundance and the upper cloud coinciding with the ammonia condensation level. The large lower cloud particles are highly scattering at visible wavelengths, allowing sunlight to scatter through the clouds and be Rayleigh-scattered from the deep atmosphere. At 5 $\mu$m, the lower cloud particles are found to be more absorbing, with the belt/zone differences 
mostly accounted for by changes in the single-scattering albedo of these particles and secondarily by changes in the cloud opacity. The spectral properties of these lower cloud particles are possibly consistent with a component of water ice. The upper cloud particles need a distinct absorption band near 3 $\mu$m, possibly consistent with a component of ammonia ice. We note that we do not need a separate upper-level photochemical haze in our model. Instead, we find that the features seen at methane-absorbing wavelengths are caused by variations in the opacity and vertical extent of the upper cloud layer. 
\end{abstract}


\section{Introduction}\label{Section:Introduction}

Recent studies by \citet{hill24} and \citet{irwin25} have shown that visible-wavelength observations of Jupiter, made both by backyard astronomers and professional observatories, can be used to determine maps of `deep' ammonia abundance (i.e., below its condensation level) and cloud-top pressure in Jupiter's atmosphere using simple band-depth models \citep[e.g.,][]{combes79}. These estimates can be compared with those made at longer wavelengths by ground- and space-based telescopes, and also spacecraft observations. Although there are similarities between these different analyses, the visible-wavelength band-depth studies find puzzlingly high cloud-top pressures. In addition, the deep ammonia abundances determined at visible wavelengths are systematically lower than those determined at microwave wavelengths by Juno/MWR \citep{li17,guillot20li} and VLA \citep{moeckel23}, and also  by Juno/JIRAM at 5 $\mu$m \citep{grassi20} by a factor of $\sim$1.5. The deep ammonia abundance is found to peak sharply at the northern edge of the Equatorial Zone (EZ) at $\sim$5$^\circ$N, have a deep minimum in the North Equatorial Belt (NEB) at 6 -- 20$^\circ$N, and then have smaller belt-zone variations at other latitudes. Higher in the atmosphere, above ammonia's expected condensation level, the latitudinal distribution appears to be more symmetric about the equator, with maximal relative humidity seen by Cassini/CIRS and IRTF/TEXES above the zones and minima seen above the belts \citep{irwin04,achterberg06,fletcher16}.

Finding a single aerosol and ammonia abundance model that is simultaneously consistent with observations at all wavelengths is a remarkably difficult problem. However, inspired by our co-analysis of observations of Uranus and Neptune using observations spanning 0.3 -- 2.4 $\mu$m \citep{irwin22}, we here attempt to derive an aerosol-ammonia model that is simultaneously consistent with observations of Jupiter from 0.48 to 5.15 $\mu$m by combining  spectrally-resolved VLT/MUSE observations made from 2018 to 2022 (475 -- 933 nm), Cassini/VIMS observations made during its flyby of Jupiter in 2000 (0.35 -- 5.15 $\mu$m), an IRTF/SpeX observation from 2002 (0.8 -- 2.5 $\mu$m), and Juno/JIRAM observations made in 2016 (2.0 -- 5.0 $\mu$m). Using our NEMESIS radiative transfer and retrieval tool \citep{irwin08} we here determine combined ammonia/aerosol profiles that are consistent with representative spectra in Jupiter's Equatorial Zone (EZ), North Equatorial Belt (NEB) and small North Equatorial Dark Features (NEDFs), also known as `5-micron hotspots'. The first two regions were chosen as archetypal belt and zone regions, whose limb-darkening is well measured in our datasets, while an NEDF was chosen as these regions provide a unique `window' in the EZ, one of which was been measured in situ by the NASA Galileo entry probe in 1995.

The atmosphere of Jupiter, composed mostly of hydrogen and helium, contains several condensible gases, which are expected to condense to form clouds of aqueous-ammonia or water-ice, ammonium hydrosulphide (NH$_4$SH) and ammonia at pressure levels $\sim$10, 2 and 0.7 bar, respectively \citep{west86, atreya99,irwin09}, assuming heavy elements are enriched by $\sim 5\times$ solar.  Since ammonia is known to be abundant in Jupiter's atmosphere and is predicted to condense at the lowest pressures, it is widely expected, and commonly thought, that the main observable clouds we see on Jupiter are ammonia ice clouds. Higher in the atmosphere, photolysis becomes important, which breaks down molecules such as methane, ammonia and phosphine to form complex photochemical products such as ethane and acetylene in the stratosphere (which have been detected) and potential products such as hydrazine and diphosphine in the upper troposphere, which may contribute to the colour-carrying chemicals, or `chromophores', that colour Jupiter's clouds, giving them their characteristic red and ochre hues. 

Jupiter was visited by the Pioneer and Voyager spacecraft in fly-bys during the 1970s and 1980s, but the NASA Galileo spacecraft became the first spacecraft to enter into orbit about the Jovian system in 1995. The Galileo orbiter included the Near Infrared Mapping Spectrometer \citep[NIMS,][]{carlson92,carlson96}, which recorded spectra from 0.7 to 5.2 $\mu$m, covering wavelengths of both reflected sunlight (0.7 -- 3.5 $\mu$m) and thermal emission in the `5-$\mu$m window' (4.7 -- 5.2 $\mu$m), where radiation thermally emitted from the 5 -- 10 bar region, modulated by the absorption of gases and overlying clouds, is detected. The Galileo orbiter also included the Solid State Imaging (SSI) camera, which allowed filter-averaged imaging at various wavelengths from 400 to 1100 nm. Unfortunately, Galileo observations were handicapped by the main communication antenna not deploying fully and thus data could only be transmitted back to Earth with the low gain antenna at greatly reduced data rate \citep{gershman94,statman97}. Hence, very few complete spectra were returned by NIMS. However, one set of four complete spectra (known as the `real-time' spectra) were recorded at a time close to Jupiter orbit insertion, targeting a low cloud-opacity region known as a `5-$\mu$m hotspot' \citep{orton96} at the northern edge of the Equatorial Zone (EZ). \citet{irwin98} found that these spectra were consistent with vertically-thin NH$_4$SH and NH$_3$ clouds, and a detached haze layer near $\sim$0.2 bar. However, further analysis of these spectra \citep{irwin01} found that beneath an upper level haze a single extended vertical distribution of aerosol, with peak opacity (in terms of specific density, i.e., particles/gram) at $\sim$1.4 bar was more consistent with these observations, suggesting that there was generally no need, or evidence, for a distinct, significant ammonia condensation cloud at $\sim$0.7 bar. \citet{irwin01} also found that the main source of the near-IR reflectivity variability, which is seen to be anticorrelated with 5-$\mu$m brightness, is due to cloud opacity changes of this lower cloud. Similarly, \citet{roosserote06} determined the limb-darkening of pairs of 5-$\mu$m NIMS spectral observations of the Equatorial Zone (EZ) and North Equatorial Belt (NEB), from which it was concluded that the vertical location of the 5-$\mu$m absorber was strongly constrained to lie at pressures $p < 2$ bar. More recent analysis of Cassini-VIMS observations by \cite{giles15} suggests that the 5-$\mu$m-absorbing cloud must reside at even lower pressures of $p < 1.2$ bar.

\begin{figure}
	\includegraphics[width=\columnwidth]{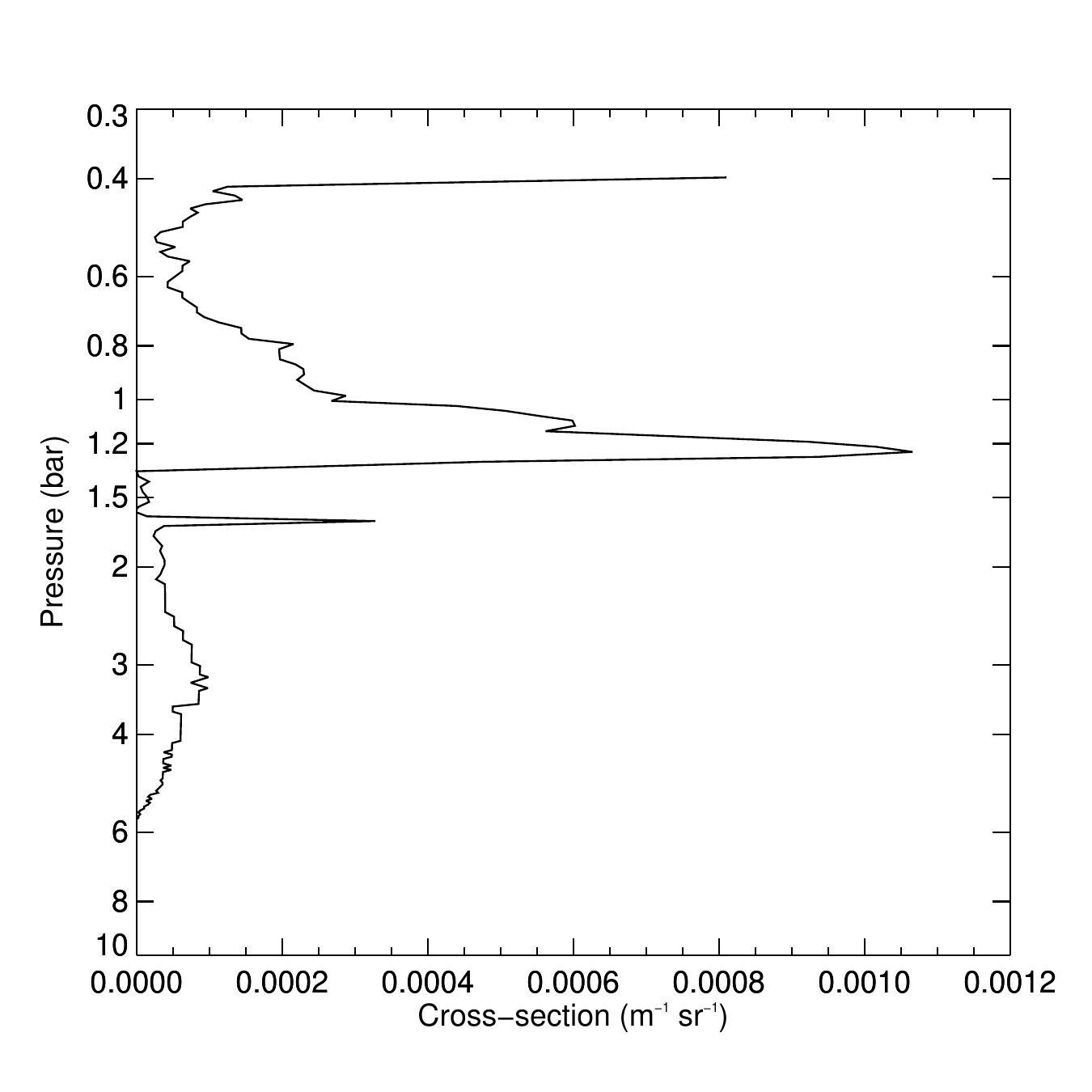}
    \caption{Observed particle back-scatter (178$^\circ$ phase angle) cross-section from the Galileo Probe Nephelometer \citep{ragent98}, recorded during the Galileo probe's descent into Jupiter's atmosphere on December 7th 1995. Here we have used the calibration where the baseline has been adjusted to zero at $p = 1.345$ bar, using pre-launch calibration data extrapolated to cover out-of-range temperatures experienced during the probe descent.}
    \label{fig:nephelometer}
\end{figure}

In addition to the orbiter, the Galileo mission also comprised an entry probe, which entered and descended through Jupiter's atmosphere on December 7th, 1995, at 6.5$^\circ$N (planetocentric), 4.4$^\circ$W (System III), sampling the atmosphere for $\sim$58 minutes from pressure levels of approximately 0.46 to 22 bar. Although intended to fall through a cloudy region of the Equatorial Zone (EZ), the Galileo probe actually entered the atmosphere in a 5-$\mu$m hotspot, where the cloud opacity is anomalously low. 
The probe nephelometer \citep{ragent92,ragent98}, which observed light reflected from a GaAs laser source (904 $\pm$ 5 nm), found evidence for an upper level haze ($p < 0.5$ bar), a main vertically-extended cloud, based at $\sim$1.35 bar,  a vertically-thin layer at $\sim$ 1.7 bar, and then a very broad region of low reflectivity centred at 3 -- 4 bar (Fig. \ref{fig:nephelometer}). The long wavelength of the GaAs laser was chosen to easily distinguish the laser reflection from background reflected sunshine, and also to make the instrument less sensitive to Rayleigh scattering from the air. The Galileo probe also included a Net Flux Radiometer \citep[NFR,][]{sromovsky96,sromovsky98}, which determined that the net heating peaked at $\sim$1.35 bar, consistent with the vertical location of the main aerosol layer seen by the nephelometer. \citet{sromovsky98} suggested that the upper haze layer at 0.4--0.5 bar was an ammonia ice cloud, composed of small particles (r $\sim 0.5 \mu$m), which is consistent with their retrieved ammonia mole fraction profile with a deep mole fraction of $\sim$250 ppm that started to decrease with altitude at 3--4 bar and did not reach the saturated vapour pressure curve until $p < 0.4$ bar. \citet{sromovsky98} concluded that the main aerosol layer was likely composed of moderately-sized NH$_4$SH particles (r $\sim 3 \mu$m). Later, \citet{sromovsky02} found that this model was also consistent with HST/WFPC2 observations over dark and bright regions, with most of the variability accounted for by variation of the opacity of the  $\sim$1.35-bar cloud. For various technical reasons the Galileo Probe Mass Spectrometer \citep{niemann98} was unable to determine a vertical profile of NH$_3$, but only an upper limit (relative to H$_2$) of 2300 ppm, which was later revised down to 664 ppm at $\sim$10 bar by \citet{wong04}. However, measurements of the probe radio signal attenuation \citep{folkner98}, found the ammonia mole fraction steadily increased from 0 ppm at $\sim$1.5 bar to a maximum of $\sim$700 ppm at $\sim$8 bar, and then decreased slightly to $\sim$600 ppm at the maximum measured pressure of 15 bar. Similarly, the Galileo mass spectrometer found the abundance of H$_2$S to increase with depth to a deep abundance (relative to H$_2$) of  89 ppm \citep{wong04}. Further evidence for a steady increase of ammonia mole fraction with depth beneath the ammonia condensation level (perhaps caused by NH$_4$SH condensation or other processes) comes from microwave, and radio-wave observations \citep[e.g.,][]{showmanpater05,depater16,depater19,fletcher21,moeckel23}. 

Although \citet{irwin01} found no evidence for significant opacity at the expected ammonia condensation level of $\sim$0.7 bar, analysis of Galileo/SSI observations \citep{banfield98} reported the presence of a vertically thin component at the base of the tropospheric haze at $\sim$0.75 bar, presumed to be composed of ammonia ice, whose optical depth variations were deduced to be primarily responsible for the features seen in Jupiter's atmosphere at red and longer wavelengths. The Galileo/ISS filters used by \citet{banfield98} were centred at 756, 727 and 889 nm, which cover a wide range of pressure levels, albeit with coarse resolution, but are most sensitive to upper tropospheric particulates, whose properties were found to be consistent with a review of the field made after the Voyager encounters \citep{west86}. Later analysis of Galileo/NIMS observations \citep{baines02} found evidence for a component in Jupiter's clouds with significant absorption near 2 $\mu$m, which was attributed to ammonia ice (referred to as Spectrally-Identifiable Ammonia Clouds - SIACs). However, SIACs were seen to be very uncommon, covering less than 1\% of the planet, and were restricted to small regions of turbulent convective activity such as that to the northwest of Jupiter's Great Red Spot (GRS). 

In addition to the Galileo near-infrared observations, Jupiter was also observed in 1996 by the Short Wave Spectrometer (SWS) of the Infrared Space Observatory (ISO) from 2.4 to 3.2 $\mu$m \citep{brooke98}. The observed reflection spectrum was matched with an optically thick cloud near 1.4 bar ($\tau(0.6\ \mu\textrm{m}) = $  5 to 20) combined with a thin upper cloud ($\tau(0.6\ \mu\textrm{m})=$ 0.5 -- 0.8) at $p\sim 0.55$ bar. The ammonia abundance profile was found to decrease from its deep value of 400 ppm in the 1.4-bar region, then remaining constant at $\sim$ 35 ppm until reaching its condensation level near 0.55 bar, coincident with the upper cloud. Significant additional cloud absorption was found to be necessary in the upper cloud to match the observed spectrum from 2.8 to 3.1 $\mu$m, which \citet{brooke98} attributed to the presence of large ammonia ice crystals ($r \sim 10$ $\mu$m). The nature of this `ubiquitous 3-micron absorber' has been further explored in Cassini/VIMS observations \citep{sromovsky10a, sromovsky10b}, Horizons/LEISA  \citep{sromovsky18}, and more recently in Juno/JIRAM observations \citep{biagiotti25}, who suggest these absorbers may be photochemically-produced `tholins'.

In the mid-infrared, Cassini/CIRS observations also detected the presence of aerosols. \citet{matcheva05} found aerosol absorption was needed to match CIRS observations near 7.2 $\mu$m, with the cloud opacity profile peaking at 1.1 bar in regions of thin cloud, reducing to 0.9 bar in regions of thick cloud. Meanwhile, 
\citet{wong04siac} found that spectral variations seen at longer wavelengths (9.5 -- 10 $\mu$m) could be matched with the presence of opacity at the ammonia condensation level, and that spectrally identifiable ammonia ice absorption features were seen near the equator and also at 23$^\circ$N (planetographic). \citet{fletcher16} also found that aerosol opacity of clouds at pressures less than 1 bar was necessary to match mid-IR CIRS and TEXES observations. Most recently, in an analysis of JWST/MIRI observations,   \citet{harkett24} find that a layer of aerosol based at $p \sim 1$ bar is necessary to match the observations.

Whatever the clouds are made of, and whatever their vertical location and distribution, we know that they are generally not pure ice, since we would then expect brilliant white clouds, not the blue-absorbing clouds that are actually seen. It is clear that photochemistry plays a vital role in the formation and scattering properties of Jupiter's aerosols and there have been many hypotheses put forward to explain what chromophores are composed of, including photochemical products of ammonia or methane, or even allotropes of sulphur or phosphorus \citep{noy81}. In laboratory experiments, \citet{carlson16} found that when irradiated with UV light ($\lambda = 214$ nm) mixtures of ammonia and acetylene (C$_2$H$_2$) reacted to form a blue-absorbing material, whose properties are consistent with those of Jupiter's chromophores. It was proposed that this material caused the reddening of the GRS, with ammonia upwelling from below meeting photochemically-produced acetylene downwelling from above, and photolysing to form the chromophore, giving rise to the so-called `Cr\`eme Br\^ul\'ee' model \citep{baines19}, with the chromophore present in the upper regions of the clouds. \citet{sromovsky17} extended this analysis to other locations on Jupiter by modelling Cassini/VIMS spectra and found that inclusion of different abundances of the `Carlson' chromophore at the top of the clouds was consistent with the measured spectra of many other locations on the planet, suggesting this might be a `universal chromophore'.

To probe more deeply in the atmosphere, the NASA Juno mission includes a Microwave Radiometer \citep[MWR,][]{janssen17}, which can probe from 0.6 to 120 bar \citep[e.g.,][]{fletcher21}. It had been expected that the deep ammonia abundance would be high in the visibly bright `zones', where the air is thought to be upwelling, and low in the dark `belts', where the air is thought to be downwelling. However, by inverting these MWR data, \citet{li17} found that while there were some belt-zone variations, the ammonia distribution was dominated by a very strong maximum at  $\sim$5$^\circ$N and strong depletion at 6 -- 20$^\circ$N. Although the strong ammonia peak at $\sim$5$^\circ$N and strong minimum at 6 -- 20$^\circ$N had previously been noted in the upper troposphere \citep{gierasch86, achterberg06, fletcher16} and has been seen since in other studies  \citep{grassi20,alexander24,irwin25,Alexander2026}, MWR found that these distributions extended very deep into Jupiter's atmosphere, with the latitudinal variation of ammonia abundance inverting at pressures exceeding 10 -- 15 bar \citep{oyafuso20,fletcher21,duer21}. These data were reanalysed by \citet{guillot20li}, who found similar results, while \citet{moeckel23} combined the MWR data with radio-wave observations made with the Very Large Array (VLA), to constrain the deep abundance of ammonia at pressures greater than $\sim$20 bar to be $340.5^{+34.8}_\mathrm{-21.2}$ ppm. \citet{moeckel23} also found that, away from 5$^\circ$N, ammonia was depleted everywhere at pressures less than $\sim$20 bar, which is consistent with the previous determinations, and also with a newly suggested depletion process, where ammonia is trapped in an exotic phase with water ice to form large `mushballs' \citep{guillot20} that precipitate out to deeper levels , stripping the upper atmosphere of gaseous ammonia. Alternatively, the extreme ammonia depletion at  6 -- 20$^\circ$N may be caused by powerful downdrafts.   

Higher in the atmosphere, the clouds have been observed over several decades by Galileo/SSI \citep{simon-miller01}, Cassini/ISS \citep{zhang13}, and HST/WFC3 \citep{perez-hoyos20,asier21,asier23,fry23}. In addition, the upper atmospheric cloud structure was inferred from a Ganymede transit \citep{lopez-puertas18}. These studies tend to place more cloud opacity in the upper atmosphere than is consistent with that determined from visible/near-IR studies. The reason for this is unclear, but at the very shortest wavelengths bands, which are in the UV, the radiative transfer is much more dominated by Rayleigh-scattering, and the effects of the blue-absorbing chromophore become more enhanced, especially if such particles are very small. Work is underway to attempt to reconcile the vis/near-IR determinations with these vis/UV studies and will be addressed in a future study. Observations with HST/WFC3 are also used to determine zonal wind-speeds by cloud-tracking.  HST/WFC3 and JWST/NIRCam observations probe a range different altitudes and so can be used to give estimates of windshear, but to do this reliably we need to know what pressure levels the features are located at. HST/WFC3 wind-speeds are usually assumed to refer to the expected NH$_3$-condensation level of $\sim$0.7 bar. However, an analysis of JWST/MIRI observations of the GRS \citep{harkett24} showed that the main cloud deck needs to be at a pressure level significantly greater than $\sim$0.7 bar to make the difference between Hubble and JWST/NIRCam winds (measured at $\sim$240 mb) around the GRS consistent with the measured temperature field.

The clouds of Jupiter are also mapped regularly by Juno using JunoCam, a public outreach camera that observes in four filters: red, green, blue, and a narrowband methane filter centred at 890 nm. These observations have been processed by the public community to produce ground-breaking new images of Jupiter's clouds from close distances. Juno's orbit is highly elliptical in order to achieve the close passes needed for gravitational mapping, with long periods far away from the planet to keep the spacecraft clear of Jupiter's very strong radiation belts. Hence, imaging observations from JunoCam, and also Juno's infrared spectral mapper JIRAM \citep[e.g.,][]{grassi20,grassi21} are 
limited in spatial coverage due to the constraints of the orbit, with wider coverage of the polar regions, providing a unique new perspective on Jupiter's poles that are very difficult to observe from the Earth due to Jupiter's small obliquity of only 3$^\circ$. However, in many sequences taken close to perijove, small white clouds - almost certainly SIACs - can be seen `poking up' above the main cloud deck in JunoCam images \citep[e.g.,][]{guillot23}. More generally, because Juno's observations of equatorial latitudes are typically limited to narrow swaths, the mission has been supported by ground-based observations of the wider planet close to the perijoves. One such set of supporting observations are those that have been made using the Multi Unit Spectroscopic Explorer (MUSE) instrument at the Very Large Telescope (VLT) \citep{irwin19jupiter, braude20, alexander24,Alexander2026}, and also with the New Mexico State University Acousto-optic Imaging Camera at the Apache Point Observatory \citep{dahl21}. By inverting these visible/near-IR spectra, \citet{braude20}, \citet{dahl21}, \citet{alexander24} and \citet{Alexander2026} find that the main cloud deck on Jupiter is, once again, based near the 1.5-bar level, and find that the Cr\`eme Br\^ul\'ee model is consistent with these observations, if appropriate modifications are made to the imaginary index of refraction spectrum of the `Carlson' chromophore. \citet{braude20}, \citet{alexander24} and \citet{Alexander2026} also present retrievals of the abundance of ammonia at a quoted pressure of $\sim$1 bar along the central meridian, finding enhanced values at the northern edge of the EZ and greatly depleted values in the NEB, with moderately constant values at other latitudes of $\sim 60$ ppm. 

\subsection{Structure of paper}

We here attempt to derive a combined aerosol-ammonia model for Jupiter using our NEMESIS radiative transfer and retrieval tool \citep{irwin08} that is simultaneously consistent with all Jovian observations from 0.48 to 5.15 $\mu$m in the equatorial region of Jupiter's atmosphere. We do not consider the polar regions. We combine spectrally-resolved VLT/MUSE observations of Jupiter made from 2018 -- 2022 (475 -- 933 nm) with Cassini/VIMS observations made during its flyby of Jupiter in 2000 (0.35 -- 5.15 $\mu$m), an IRTF/SpeX observation made in 2002 (0.8 -- 2.5 $\mu$m), and Juno/JIRAM observations made in 2016 (2.0 -- 5.0 $\mu$m). 
In Section \ref{Section: Observations} we describe details of the various datasets analysed and in Section \ref{Section:Modelling} we describe the parameters and models used to simulate the Jovian spectra. In Section \ref{Section:Initial} we describe preliminary analyses of these data, which led us to our preferred vertical cloud and ammonia structure. This model is then tested against the combined MUSE and VIMS datasets covering the EZ, NEB and an NEDF in Section \ref{sec:final_retrieval}. In Section \ref{Section:Discussion} we discuss the implications of this model and in Section \ref{Section:Conclusions} we present our conclusions and outline future avenues for research.

\section{Observations} \label{Section: Observations}
\subsection{VLT/MUSE observations} \label{Sec:MUSEobs}

The Multi Unit Spectroscopic Explorer (MUSE) instrument, at ESO's Very Large Telescope (VLT) in Chile, is an Integral-Field Unit (IFU) Spectrograph. VLT/MUSE records 300 $\times$ 300 pixel images from a field of view of \ang{;;60} $\times$ \ang{;;60} in Wide Field Mode, where each \ang{;;0.2} $\times$ \ang{;;0.2} pixel contains a complete visible/near-infrared spectrum from 475 to 933 nm, with a spectral resolving power $(\lambda/\Delta\lambda)$ of 2000 -- 4000 (i.e., a spectral resolution of $\sim $0.2 nm). Since the MUSE data have three dimensions they are often referred to as `cubes'. MUSE observations of Jupiter have good spatial resolution due to the generally excellent atmospheric `seeing' at the location of VLT, of typically $<$\ang{;;0.8}, but observations can also be made with adaptive optics (AO), which improves the spatial resolution still further.  VLT/MUSE observations of Jupiter have been made in support of the NASA/Juno mission and cover many Juno perijoves between 2016 and 2023. In addition, there are commissioning observations from 2014. Previous MUSE Jupiter observations have been analysed by \citet{irwin18}, \citet{irwin19jupiter}, \citet{braude20} and  \citet{irwin25}. More recent observations have been being analysed by \citet{alexander24,Alexander2026}. 

\begin{figure*}
	\includegraphics[width=\textwidth]{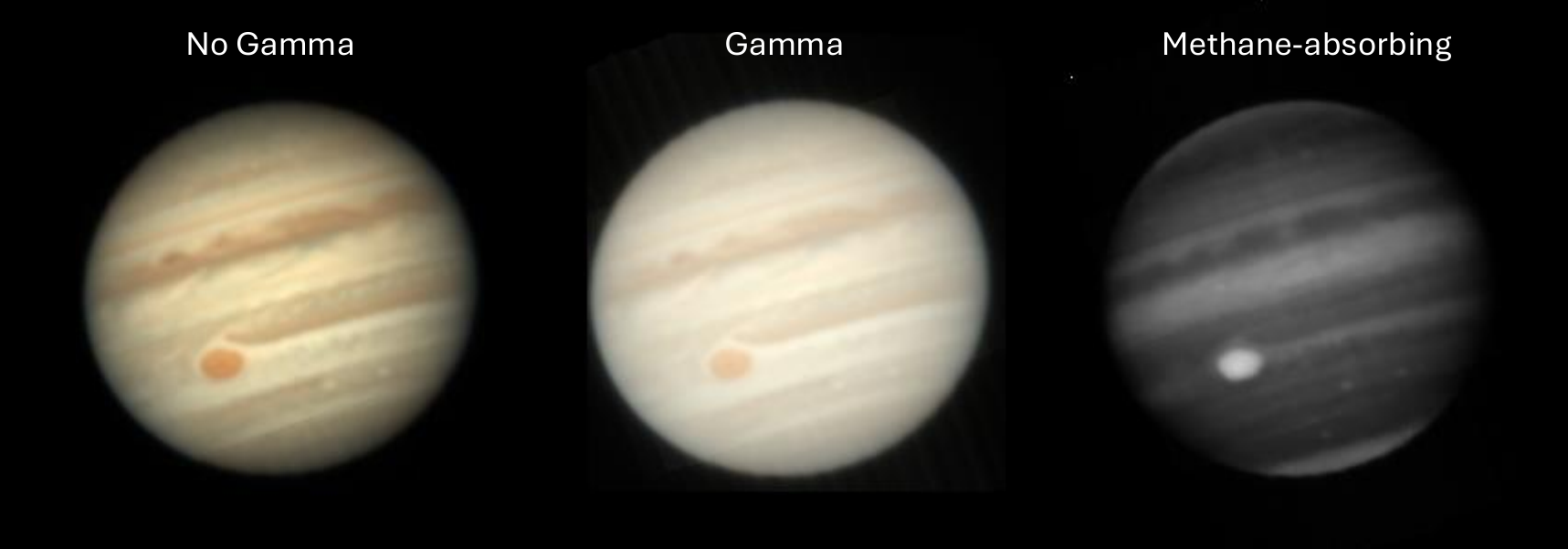}
    \caption{Jupiter as observed by VLT/MUSE on 9th April 2018. The left and middle images show the reconstructed `true-colour' representations following the approach of \citet{irwin25}, where the MUSE spectra at wavelengths less than the MUSE minimum wavelength of 475 nm have been set to the disc‐averaged Jupiter spectrum of \citet{kark94}, scaled to match the MUSE spectra at overlapping wavelengths and the resulting combined spectra converted to true colour. The left hand image shows the appearance when no gamma-correction has been applied, while the middle image shows the gamma-corrected version, which is closer to what the average human observes with their naked eye through a telescope, but has reduced contrast and is less colour-enhanced. The right-hand image shows Jupiter's appearance averaged over the strong methane-absorbing band at 890 nm, revealing the spatial variation of particulates in the upper troposphere near $p < 0.5$ bar.}
    \label{fig:jupiter_muse}
\end{figure*}

In this study, a non-AO MUSE cube observed on 9th April 2018 was used (Fig. \ref{fig:jupiter_muse}). The native MUSE resolution exceeds that of the only methane absorption dataset available to analyse these data over the whole MUSE spectral range \citep{kark10}. Hence, the data were smoothed with a Gaussian line shape with full-width-half-maximum (FWHM) of 0.002 $\mu$m, and sampled every 0.001 $\mu$m. Before smoothing, however, the centre-of-disc spectrum was compared with the high-resolution solar spectrum of \citet{meftah23}, and used to slightly adjust the wavelength calibration of the MUSE data by aligning on the H$\alpha$ line at 656.28 nm. The radiance spectra of the whole cube, appropriately wavelength-shifted, were then divided by the solar spectrum of \citet{meftah23}, smoothed to MUSE resolution, and multiplied by a lower-resolution solar spectrum based on that of \citet{kurucz93}, smoothed to a resolution of 0.02 $\mu$m, sampled every 0.01 $\mu$m. This process was done to remove the solar spectrum Fraunhofer lines and so just leave Jovian atmospheric spectral features. In addition, the radiances were corrected to a standard Jupiter distance of 5.2 AU from the Sun. Finally, the radiance spectra were smoothed to their final resolution and wavelength step. Fig. \ref{fig:jupiter_muse} shows a true colour representation of the MUSE cube, generated in the same way as the MUSE cubes shown by \citet{irwin25}. 

To analyse this MUSE cube, the spectra were averaged into latitude bins of width 4$^\circ$, separated by 2$^\circ$ (to achieve Nyquist sampling), with the Great Red Spot (GRS) masked. In each latitude bin, the observed angular dependence of the reflectances was found to be well fitted by the Minnaert model \citep{minnaert41}:
\begin{equation}\label{eq:minnaert}
(I/F) = (I/F)_\mathrm{0} \mu_\mathrm{0}^{k} \mu^{k-1}.
\end{equation}
where $\mu$ and $\mu_0$ are the cosines of the viewing and solar zenith angles, $(I/F)_0$ are the fitted nadir reflectances, and $k$ are the fitted limb-darkening parameters. Once fitted, the Minnaert parameters were used to generate two synthetic `measurement' spectra for each latitude band, one with the Sun and observer at zenith and one with both the viewing and solar zenith angles set to 42.37$^\circ$ (back-scattering direction). Both these angles are sampled by the 5-point zenith quadrature scheme used in our radiative transfer model described later in Section \ref{sec:radtrans}. A smaller upper zenith angle was used here compared to our previous Uranus `holistic' model \citep{irwin22} in the interests of speed (we have more wavelengths to reproduce here) and also because we combine these spectra with VIMS-IR observations (Section \ref{sec:VIMS}), where the zenith angle variations are less well sampled.

\begin{figure*}
	\includegraphics[width=\textwidth]{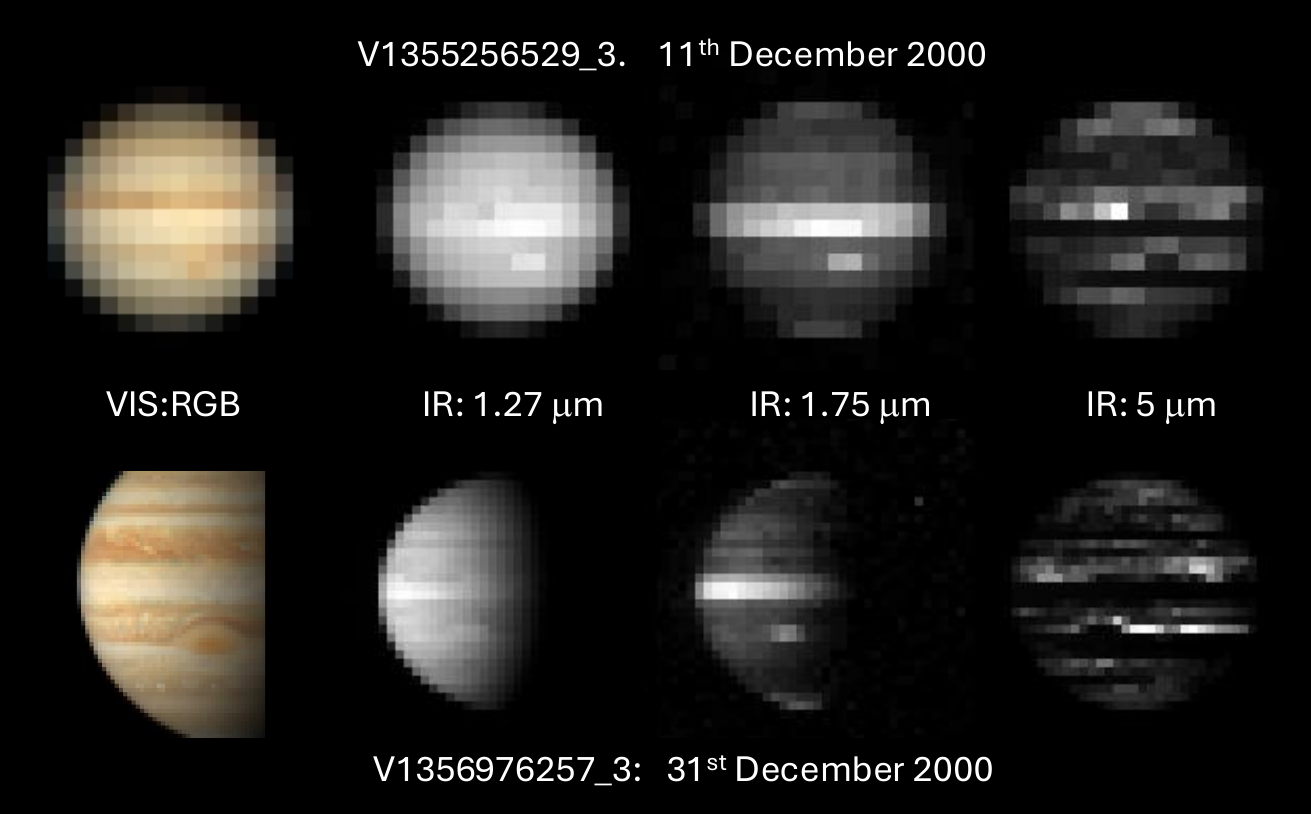}
    \caption{Cassini/VIMS observations of Jupiter on 11th December 2000 (V1355256529\_3) and 31st December 2000 (V1356976257\_3), in both VIS and IR channels. Observations in the VIS channel have been converted to `true colour', as described earlier, and are not gamma-corrected. Observations in the IR channel are shown averaged over three wavelength bands, one centred at 1.27 $\mu$m showing the overall cloud reflectivity, one centred at 1.75 $\mu$m, a band of strong methane absorption, and one averaged over the 5-$\mu$m window showing thermal emission from the deeper atmosphere.}
    \label{fig:jupiter_vims}
\end{figure*}

\subsection{Cassini/VIMS observations}\label{sec:VIMS}

To extend the analysis to longer wavelengths, we used Cassini/VIMS observations of Jupiter acquired in December 2000. These data are also spectral-imaging ``cubes'' and were downloaded from the Planetary Data System. The processing made use of the Integrated Software for Imagers and Spectrometers (ISIS; \citealt{Anderson2004}) to recover the pointing and geometric information, generate geometry backplanes, and access label metadata used in the calibration. The data were calibrated to $I/F$ using a dedicated workflow developed for the Cassini/VIMS Jupiter data set. The VIS and IR channels were calibrated separately. In the VIS channel, the calibration uses the responsivity matrices developed by \cite{Filacchione2006}, thereby naturally incorporating the approximate 10\% upward rescaling discussed by \citet{sromovsky17}. In the IR channel, a custom responsivity-based calibration was adopted; for the IR cubes analysed here, this closely matches the standard ISIS calibration. Pointing-related mismatches between the radiometric cubes and the geometric backplanes were also corrected, and the final products were written as multi-extension FITS (Flexible Image Transport System; \citealt{Pence2010}) files containing the calibrated spectral cubes together with geometry backplanes and wavelength information. The calibrated FITS files are publicly available (see Data Availability section), and the full processing workflow is described in \citet{asier26a}. 
 
After this calibration step, we used the same observations as analysed by \citet{sromovsky10b}. One of these observations, V1355256529\_3, was recorded on approach to Jupiter
on 11th December 2000 at a phase angle of 2.5$^\circ$. In addition, we also considered the V1356976257\_3 observation, obtained on 31st December 2000 at a phase angle of 67.8$^\circ$ and whose spatial resolution is roughly twice as good as the approach observation. Images extracted from these observations are shown in Fig. \ref{fig:jupiter_vims}. The latter observation was also used by \citet{sromovsky17}, who additionally analysed another approach dataset, V1354610545\_3, observed on 4th December 2000.  Cassini/VIMS has two spectral ranges: The VIS mode observes at 96 wavelengths from $\sim$0.35 to 1.05 $\mu$m at a spectral resolution of 0.007368 $\mu$m, while the IR mode observes 256 wavelengths from $\sim$0.88 to 5.15 $\mu$m at a spectral resolution of $\sim$0.0166 $\mu$m. For the two VIMS cubes considered here, V1355256529\_3 has $32 \times 32$ pixels in both VIS and IR, with both channels in nominal mode, whereas V1356976257\_3 has $64\times 64$ pixels in both VIS and IR, but with VIS in high-resolution mode and IR in nominal mode. The effective IFOV is 0.5 mrad $\times$ 0.5 mrad per spatial element for both VIS and IR in nominal mode, and 0.17 mrad $\times$ 0.17 mrad per spatial element for the VIMS-VIS high-resolution mode \citep{brown04}. Both VIS and IR instruments have Gaussian Instrument Line Shapes (ILS). 

The spectral calibration of VIMS-IR varies with wavelength and was also found to vary with time, summarised in the final calibration report of VIMS \citep{clark18} and shown in Fig. \ref{fig:vims_calibration}. The recommended interpolated offsets for our observation dates are considerably offset from the reference 2004 calibration, but \citet{sromovsky10b} report that the reference calibration was accurate to within 1 nm for these Jupiter observations, at odds with \citet{mccord04}, who found that a large shift was required when analysing spectra of the Galilean satellites during the Jupiter flyby. To assess this effect, we ran retrievals over the 0.89 -- 4.0 $\mu$m and 4 -- 5.1 $\mu$m spectral ranges for a range of offsets from the nominal 2004 wavelength scale. Like \citet{sromovsky10b} we find that the optimal offset is very small and thus we used the reference 2004 wavelength calibration in our analyses. We find this calibration compares very favourably with later Juno/JIRAM observations (Section \ref{Section:JIRAM-EZ} and Fig. \ref{fig:jiram_vims_comparison}).

Since the approach observation was observed with a low phase angle, we processed the data as we did with VLT/MUSE, using the Minnaert model. However, in this case the spatial resolution of the VIMS data was not sufficient to extract Minnaert coefficients over a uniformly varying, Nyquist-sampled latitude grid. Instead, since the VIMS frames are aligned with respect to Jupiter's polar axis and Cassini's flyby was close to Jupiter's equatorial plane, we Minnaert-analysed the data from each row of the VIMS-VIS and VIMS-IR cubes. We tabulated the Minnaert coefficients at the mean latitude of each row and then again generated two synthetic `measurement' spectra for each latitude row, one with  both the Sun and observer at zenith, and one with both the solar and viewing zenith angles set to 42.37$^\circ$ (back-scattering direction).

\begin{figure}
	\includegraphics[width=\columnwidth]{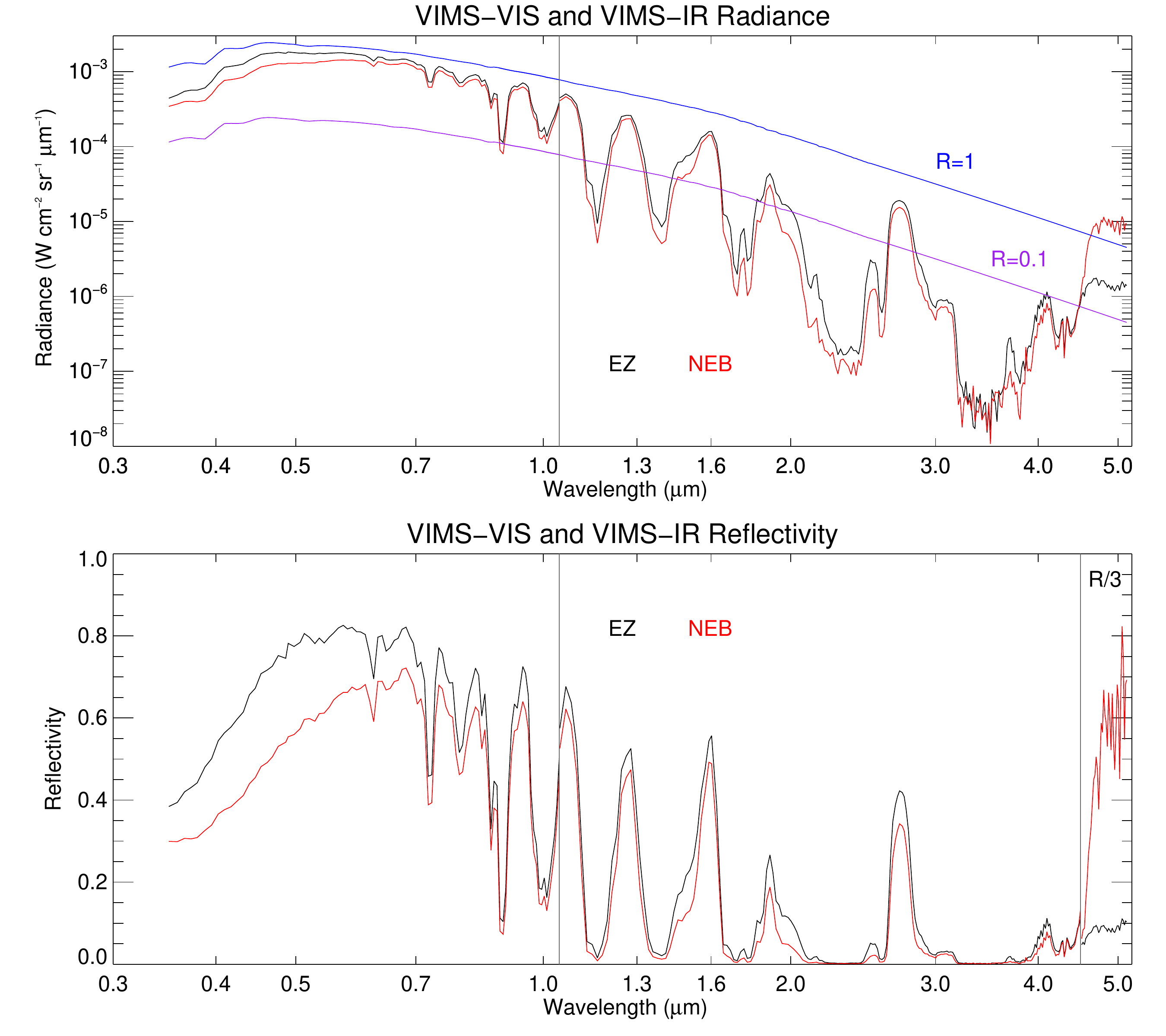}
    \caption{Top panel shows the $0^\circ$ zenith angle radiance spectra extracted from the VIMS-VIS and VIMS-IR V1355256529\_3 approach observation in the Equatorial Zone (EZ) and North Equatorial Belt (NEB). Also shown are the modelled reflected solar spectra from Lambertian surfaces at Jupiter with reflectivities of 1.0 and 0.1, respectively. Bottom panel shows the observed radiance spectra converted to reflectivity. The junction of the VIS and IR data is indicated by the vertical grey lines. In the reflectivity plot, the reflectivities have been divided by three at wavelengths longer than 4.5 $\mu$m, to keep the plotted values less than 1. A log scale has been used for wavelength to more easily interpret the spectra over the combined range.}
    \label{fig:VIMS_spx_comparison}
\end{figure}

Figure \ref{fig:VIMS_spx_comparison} shows the nadir spectra of the Equatorial Zone (EZ) and North Equatorial Belt (NEB) extracted from the approach VIMS-VIS and VIMS-IR cubes using our Minnaert model, showing the very good cross-calibration between the VIMS-VIS and VIMS-IR observations.

\subsection{Juno/JIRAM observations} \label{sec:JIRAMBIN}
NASA's Juno spacecraft, which was launched in 2011 and went into orbit about Jupiter in 2016, has several instruments for characterising the Jovian atmosphere. One of these instruments is the Jovian InfraRed Auroral Mapper (JIRAM), which comprises both an imager, designed for mapping at 3.45 $\mu$m (for auroral observations) and 4.78 $\mu$m (for deep thermal emissions), and a spectrometer (JIRAM-SPE) covering the spectral range 2.0 -- 5.0 $\mu$m in 336 spectral bins \citep{adriani17}. The JIRAM-SPE spectrometer uses a slit that is 256 samples wide and spans 3.52$^\circ$ with an Instantaneous Field of View (IFOV) of 250 $\mu$rad/pixel. As a spectrometer, JIRAM-SPE has a Gaussian ILS with a FWHM that varies from 10.56 nm at 2.1 $\mu$m,  12.18 nm at 3.5 $\mu$m, and 13.58 nm at 4.9 $\mu$m \citep[averaging samples 130, 135 and 140, following][]{adriani17}.  JIRAM is optimised for observing Jupiter's polar regions and usually observes Jupiter at very high solar zenith angle or indeed on the nightside. Hence, it is generally less well suited for studying the reflectivity from Jupiter's clouds at equatorial to mid-latitudes. However, several observations were made of such regions during the early stages of Juno's mission \citep[e.g.,][]{biagiotti25}, which we have analysed here to help verify both the radiometric and wavelength calibration of VIMS-IR, and provide additional observations to constrain our model. The data from JIRAM-SPE have previously been analysed by \citet{grassi20,grassi21}, who note that dayside observations at shorter wavelengths are occasionally sensitive to scattered light, especially at closer Jupiter distances (e.g., over north pole for PJ1), most likely related to out-of-field stray light, rather than in-instrument scattering.

\begin{figure*}
	\includegraphics[width=\textwidth]{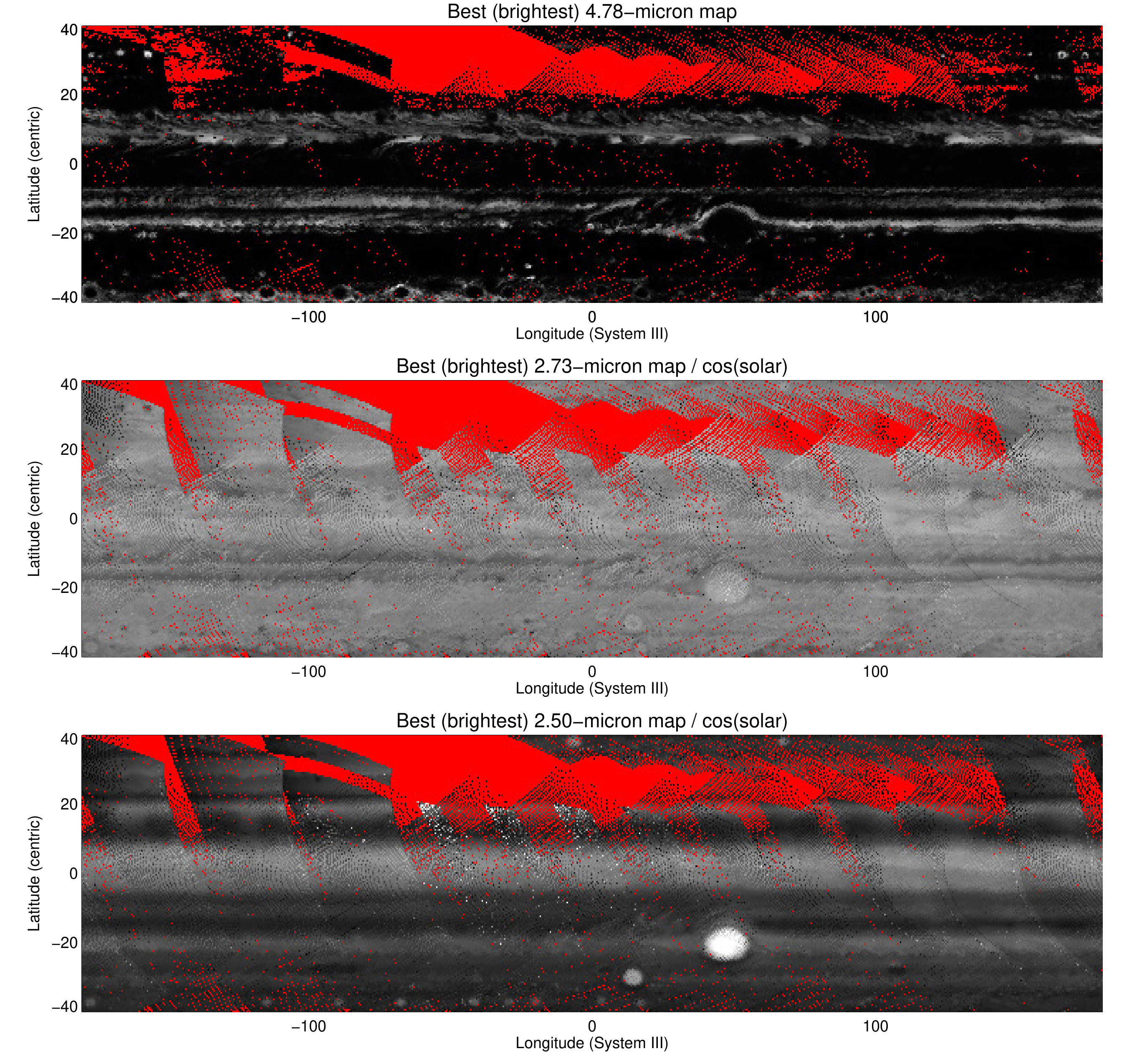}
    \caption{JIRAM-SPE reconstructed cylindrical maps of Jupiter observed on 27th August 2016 (JM0003 planning period) from 40$^\circ$S to 40$^\circ$N (planetocentric), showing the brightest radiances in each bin at 4.78, 2.73, and 2.50 $\mu$m. Longitudes are System III, East is positive. The 4.78-$\mu$m map shows thermal emission from the 5 -- 10 bar level and is not further corrected. However, for the 2.73-$\mu$m map, which is sensitive to reflected sunshine to levels as deep as 5 -- 10 bar, the brightest radiance is divided by the cosine of the solar zenith angle to roughly correct for solar illumination variations. This solar illumination correction is also applied to the 2.50-$\mu$m map, which shows sunlight reflected from the upper level particulates at $p < 0.5$ bar. Red areas in these panels indicate locations that were not sampled by JIRAM-SPE.}
    \label{fig:jiram_images}
\end{figure*}

In Fig. \ref{fig:jiram_images} we have combined the JIRAM-SPE observations from Juno's first orbit (JM0003 planning period) onto a cylindrical map covering all longitudes and planetocentric latitudes from 40$^\circ$S to 40$^\circ$N in bins of width $0.5^\circ \times 0.5^\circ$. The mapped radiances are shown at three wavelengths: 2.50 $\mu$m, 2.73 $\mu$m and 4.78 $\mu$m, the latter wavelength chosen to coincide with the JIRAM-imaging wavelength near 5 $\mu$m. To construct these binned images we noted all observations (and their respective viewing angles) within each latitude/longitude bin. For the 4.78-$\mu$m map we just selected the brightest observation for each bin. Since at this wavelength the radiance is dominated by thermal emission, this approach meant we effectively chose the observation with the lowest viewing zenith angle in each bin and we did not need any other correction to give a clear map. At the shorter wavelengths, though, the radiance is predominantly reflected sunlight and to give clear artefact-free maps we divided the brightest radiance by the cosine of the respective solar zenith angle to roughly correct for solar illumination variations. The 4.78-$\mu$m map shows thermal emission from pressure levels at 5 -- 10 bar, while at 2.73 $\mu$m we see reflected light from levels potentially as deep as this and thus to first order the maps are anti-correlated. At 2.50 $\mu$m, however, gaseous methane absorbs strongly and we can only see light reflected from particulates at $p < 0.5 $ bar (see the calculated two-way transmission to different levels shown later in  Section \ref{sec:probe_level} and  Fig. \ref{fig:jupiter_vims_transmission}) and thus we are only sensitive to the upper level cloud/haze, which can be seen to have a rather different spatial distribution, with enhanced reflection above the EZ and Great Red Spot (GRS). This spatial distribution is very similar to that seen by VLT/MUSE at 890 nm (Fig. \ref{fig:jupiter_muse}), and Cassini/VIMS-IR at 1.75 $\mu$m (Fig. \ref{fig:jupiter_vims}), which are also strong methane absorption bands.

\subsection{NASA IRTF/SpeX observations}
The MUSE, VIMS-VIS and VIMS-IR spectra overlap near 1 $\mu$m, which allows these observations to be potentially combined. However, there are few measurements available that record the entire 0.8 -- 1.2 $\mu$m spectral range simultaneously with a single instrumental setup. A notable exception is the SpeX instrument at NASA's Infrared Telescope Facility in Hawai'i. This instrument, which is a long-slit spectrometer, has made reference observations of all the Giant Planets, with the slit aligned along the central meridian and the fluxes integrated along the slit. The standard reference spectrum for Jupiter is available on the IRTF Spectral Library website\footnote{\url{http://irtfweb.ifa.hawaii.edu/~spex/IRTF_Spectral_Library}}, reported by \citet{rayner09}, and was made in SXD mode (0.8 -- 2.5 $\mu$m) on December 4th 2002. These data have a spectral resolution of 0.002 $\mu$m.

\begin{figure}
	\includegraphics[width=\columnwidth]{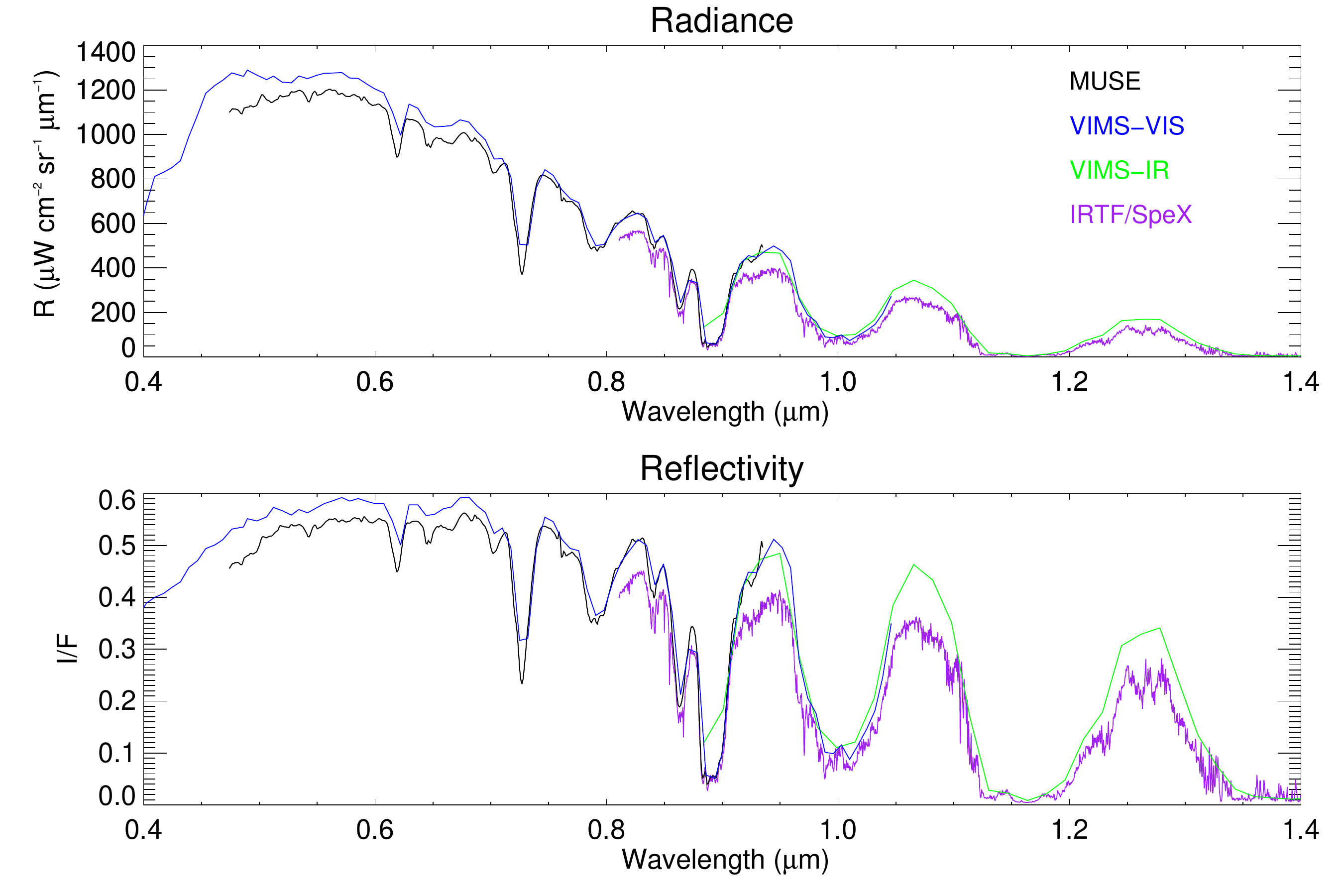}
    \caption{Comparison of MUSE, approach VIMS-VIS, approach VIMS-IR and IRTF/SpeX spectra, averaged along the central meridian of Jupiter. 
    The upper panel compares the radiance spectra (normalised to the same distance from the Sun of 5.2 AU), while the lower panel compares the reflectivities.} 
    \label{fig:Spex_comp}
\end{figure}

To compare this IRTF/SpeX spectrum with our other observations, we integrated the radiance of our MUSE cube and VIMS approach cubes along the central meridian and in Fig. \ref{fig:Spex_comp} we show the four spectra. As can be seen, the MUSE and VIMS-VIS radiances are very slightly lower than the SpeX radiances at overlapping wavelengths, while the VIMS-IR radiances are slightly higher. However, there is not a significant difference in the overlap region. At shorter wavelengths the MUSE and VIMS-VIS spectra diverge slightly, but we must bear in mind that these spectra were all recorded at different times and thus the EZ may not have been equally bright in all spectra. Hence, we conclude that the calibrations we have used for the MUSE and VIMS observations are satisfactorily accurate.

\begin{figure}
	\includegraphics[width=\columnwidth]{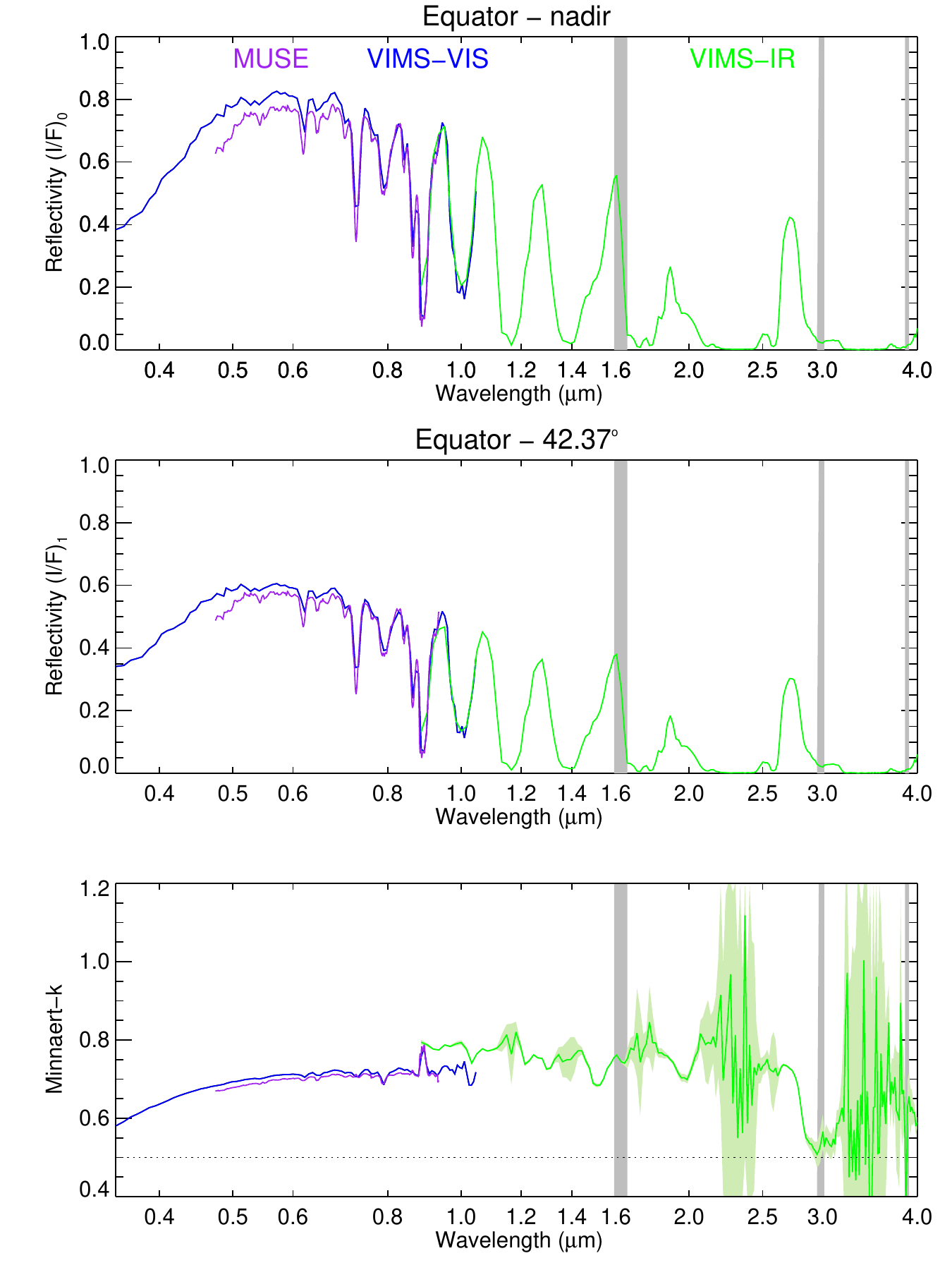}
    \caption{Comparison of Minnaert-reconstructed MUSE, VIMS-VIS and VIMS-IR spectra from 0.3 to 4 $\mu$m in the EZ (0$^\circ$N). The top row shows the reconstructed spectra at nadir, the middle row shows reconstructed spectra at 42.37$^\circ$, and the bottom row shows the fitted Minnaert-$k$ spectra. A log scale has been used for wavelength to more easily compare the spectra. In the Minnaert-$k$ spectra, the estimated noise for VIMS-IR is shaded in light green. For the other instruments, the noise is too small to see. The greyed regions indicate known problems in the VIMS-IR spectra due to order-sorting filter boundaries, etc. The dotted line in the bottom panel at $k=0.5$, indicates the boundary between limb-darkening ($k > 0.5$) and limb-brightening ($k < 0.5$).}
    \label{fig:jupiter_spectra_comparison}
\end{figure}

\subsection{Combined MUSE/VIMS observations}
Figure \ref{fig:jupiter_spectra_comparison} compares the spectra of the EZ extracted from the Minnaert analyses of our MUSE observation, and the VIMS-VIS and VIMS-IR approach observations (the same plot for the NEB is shown in Fig.  \ref{fig:jupiter_spectra_comparison1}). It can be seen that at visible wavelengths the MUSE observations have superior spectral resolution to the VIMS-VIS data, which do not wholly capture the fine details of the ammonia absorption bands seen near 647 and 550 nm. However, all three datasets are reasonably consistent with each other in the spectra reconstructed from the Minnaert coefficients, both at $0^\circ$ zenith angle, $(I/F)_0$, and also at a zenith angle of  $42.37^\circ$, $(I/F)_1$. As noted earlier in Section \ref{Sec:MUSEobs} we chose a lower second zenith angle than we used for our Uranus/Neptune analysis \citep{irwin22} of $61.45^\circ$ since the VIMS limb-darkening is less well sampled than MUSE, and because we have a large number of wavelengths and large number of unknowns, which made our retrieval model computationally expensive. Modelling the second angle of $42.37^\circ$ rather than $61.45^\circ$ required many fewer azimuth components in the Fourier decomposition of the radiance field and was thus much less computationally demanding. The fitted Minnaert limb-darkening spectra, $k(\lambda)$, are compared in the bottom panels of Fig. \ref{fig:jupiter_spectra_comparison}. It can be seen that the observed Jupiter spectra are strongly limb-darkened at all wavelengths with $0.8 < k(\lambda) < 1.2$. Interestingly, both MUSE and VIMS-VIS show enhanced limb-darkening in the strong methane band at 890 nm, which probes what has generally been considered to be a haze at $p < 0.5$ bar \citep[e.g., ][]{west86}. This is surprising as such hazes in other planetary atmospheres usually show limb-brightening, not limb-darkening. Longer wavelengths in the VIMS-IR spectra also probe this level, but here the radiances are much smaller and the uncertainty higher. However, in regions of strong methane absorption in the VIMS-IR spectrum, at wavelengths less than 2 $\mu$m, enhanced limb-darkening is also seen. Evidence for two main components to Jupiter's cloud profile is strong. In the centre of the 890 band we only see the upper component, which has a different spatial distribution to that of the main cloud deck (Figs. \ref{fig:jupiter_muse} and \ref{fig:jupiter_vims}), which only becomes apparent when the two-way transmission to 1 bar drops below $\sim4$\% as we will see later in Section \ref{sec:probe_level}.

Since there was good overall agreement between the MUSE and VIMS-VIS datasets, and the MUSE data have superior spectral resolution,  we decided to use the MUSE data at visible wavelengths, while at wavelengths longer than 0.93 $\mu$m we used VIMS-IR data in our combined spectra.

\section{Model Setup}\label{Section:Modelling}

To analyse the combined VLT/MUSE and Cassini/VIMS datasets we need to construct a reference atmospheric model and also a radiative transfer scheme able to model the observed spectra over a wide range of wavelengths and at a wide range of spectral resolutions.

\subsection{Atmospheric Model}

\begin{figure*}
	\includegraphics[width=\textwidth]{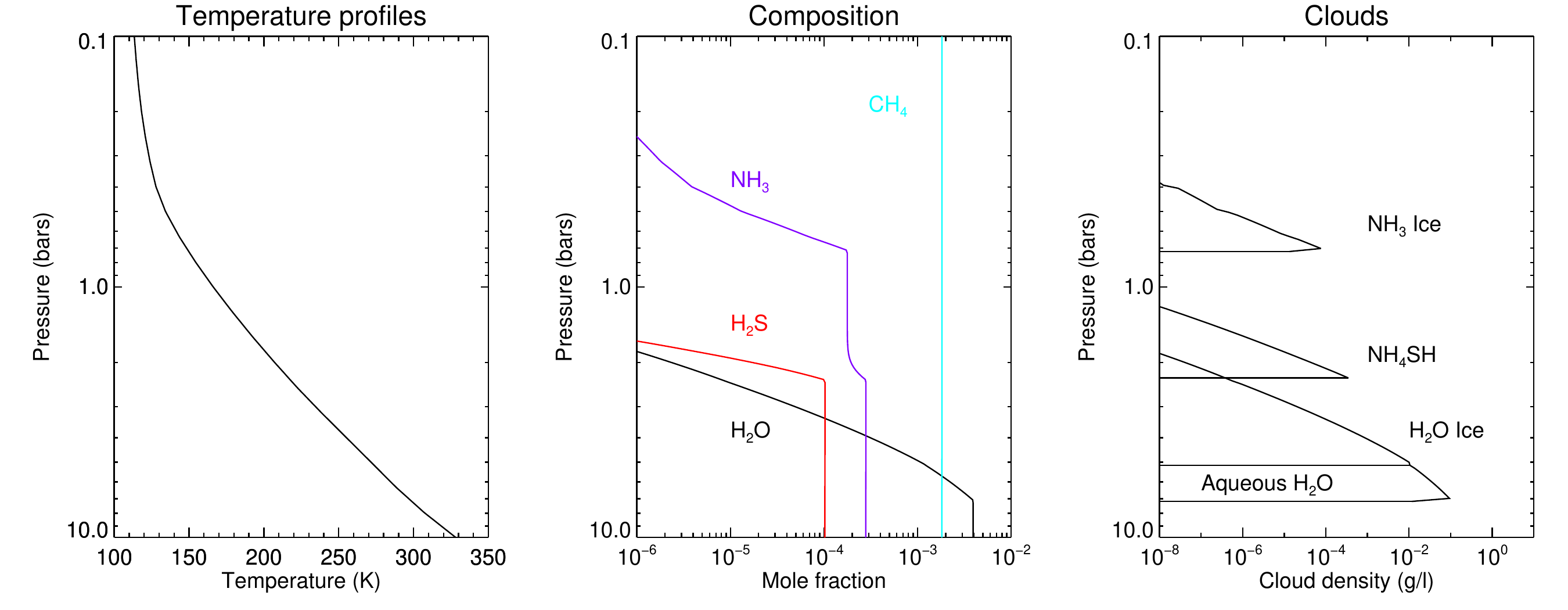}
    \caption{Model Jovian atmospheric profiles. Left panel shows the  variation of temperature with pressure in our reference model atmosphere. Middle panel shows the \textit{a priori}  variation with pressure of the mole fraction of H$_2$O, NH$_3$, H$_2$S and CH$_4$, calculated using an Equilibrium Cloud Condensation Model, while the right panel shows the associated cloud profiles, showing that we expect water to condense at 5 -- 7 bar, NH$_4$SH(s) to condense near 2.3 bar and ammonia ice to condense near 0.7 bar.}
    \label{fig:jupiter_atmosphere_model}
\end{figure*}

Our standard Jovian atmospheric temperature-pressure model is based on that of \citet{irwin09}, which is itself based on a model used by the Cassini/CIRS team for radiative transfer model comparison in the early 2000s. We assumed the following deep abundances: He/H$_2$ = 0.1566, CH$_4$/H$_2$ = $2.11\times 10^{-3}$,  H$_2$O/H$_2$ = $4.57\times 10^{-3}$, NH$_3$/H$_2$ = $3.25\times 10^{-4}$, H$_2$S/H$_2$ = $1.19\times 10^{-4}$. The condensible species were limited by their saturated vapour pressures and the mole fractions of non-condensing gases adjusted to sum to unity, keeping their relative abundances to H$_2$ fixed. Our reference modelled variations of temperature and gaseous abundance are shown on Fig. \ref{fig:jupiter_atmosphere_model}, which also shows the output of an Equilibrium Cloud Condensation Model (ECCM) for these atmospheric conditions, showing that we expect water to condense at 5 -- 7 bar, NH$_4$SH(s) to condense near 2.3 bar and ammonia ice to condense near 0.7 bar. For methane we considered two isotopologues separately C$^{12}$H$_4$ and C$^{12}$H$_3$D, with abundances C$^{12}$H$_4$/H$_2$ = $2.11 \times 10^{-3}$ \citep{niemann98} and C$^{12}$H$_3$D/C$^{12}$H$_4$ = $8 \times 10^{-5}$. In addition, to model features in the 5-$\mu$m window, we included phosphine (PH$_3$) with a reference deep mole fraction of $6 \times 10^{-7}$, decreasing in abundance at pressures $< $ 1 bar with a fractional scale height of 0.3, and also arsine (AsH$_3$) and germane (GeH$_4$), with reference mole fractions of $7 \times 10^{-10}$ and $2 \times 10^{-10}$, respectively.

Finally, although H$_2$S is expected in Jupiter's atmosphere and may combine with NH$_3$ to form NH$_4$SH particles in the 1--2 bar region, the gaseous absorption bands of H$_2$S (taken from HITRAN2022) coincide with those of NH$_3$ and CH$_4$ in the MUSE/VIMS spectral range (Fig. \ref{fig:absorption}) and cannot be detected from these observations. The H$_2$S profile was thus left fixed at that determined from the ECCM.

\subsection{Linedata and Collision-induced absorption}

The absorption coefficients for different atmospheric species were taken from the following databases.

\textit{Methane:} For the main isotopologue of methane at wavelengths less than 0.93 $\mu$m we used the low-resolution coefficients of \citet{kark10}, while at longer wavelengths we used the TheoReTS line database of \citet{rey17}. For other methane isotopologues we used just the TheoReTS line database. The absorption data of \citet{kark10} is actually for all isotopes, but we assumed that the first isotopologue dominates at visible wavelengths. For the methane line data we used the sub-Lorentzian line shape of \citet{hartmann02}, and at each output wavelength considered lines up to 1000 cm$^{-1}$ away. 

\textit{Ammonia:} At wavelengths less than 0.977 $\mu$m we used the ExoMOL line data of \citet{coles19}, which we processed as described by \citet{irwin19jupiter}, combining the lines with a strength less than $10^{-28}$ cm$^{-1}$/(molecule/cm$^3$) at 296K into a `pseudo-continuum'. At wavelengths longer than 0.977 $\mu$m we used the HITRAN2022 database \citep{gordon20}, using their hydrogen-broadened parameters. We assumed a Voigt line shape and imposed a line wing cut-off of 35 cm$^{-1}$.

\textit{Other gases:} For all other constituents we used the line data of HITRAN2022 \citep{gordon20}, using their hydrogen-broadened parameters. We assumed a Voigt line shape and imposed a line-wing cut-off of 35 cm$^{-1}$.

\textit{Collision-induced absorption - CIA:} For collision-induced absorption of H$_2$--H$_2$ and H$_2$--He, we used formulations from \citet{borysow85,borysow89a,borysow89b,borysow91}.

\subsection{Radiative Transfer Modelling} \label{sec:radtrans}

To simulate the observed spectra of Jupiter, the atmospheric model must first be split up into a number of discrete levels and then the mean pressure, temperature and gas abundances of each layer calculated. Then, since Jupiter's atmosphere is highly scattering, a multiple-scattering calculation must be performed. For these calculations we used our NEMESIS radiative transfer and retrieval model \citep{irwin08}, using the Matrix-operator method (otherwise known as doubling and adding) of \citet{plass73}.

The most accurate method to compute the scattered radiances is using the line-by-line (LBL) method, where the contribution from each line of each gas is calculated explicitly and the calculation is performed at sufficient spectral resolution to sample the various gaseous absorption features. However, since LBL calculations are very slow, most radiative transfer studies of giant planet atmospheres take the available line data and from them construct a number of `$k$-tables' for each gas, for a range of pressures and temperatures likely to be found in the atmosphere of interest. The $k$-tables are computed at the spectral resolution and sampling of the instrument being simulated. For each pressure and temperature in the $k$-table, and for each tabulated wavelength bin, a line-by-line absorption-coefficient spectrum is first calculated and then multiplied by the instrument function. The resulting coefficients are then ranked in ascending order and sampled at 10 -- 20 ordinates to capture how the mean absorption varies as a fraction of the tabulated wavelength bin \citep{lacis91, irwin08}. Once the $k$-tables are compiled, the mean absorption coefficients of each gas are interpolated to the pressure and temperature of each atmospheric layer and then combined using the `overlapping line approximation' \citep{lacis91}, where it is assumed that the line positions of different gases are not correlated with each other. To do this, the $k$-coefficients of each gas are multiplied by their fractional abundance in the layer, ranked in order and then a new, mean $k$-distribution for the layer calculated. Finally, although the absorption coefficients have been heavily processed, it is found that since the high- and low- absorbing spectral regions within a calculation spectral bin are correlated, it is a reasonable approximation to assume that the $k$-coefficients are correlated also. This is known as the `correlated-$k$ approximation'.

Although they are simplifications, the overlapping line and correlated-$k$ approximations are both found to work well in practice with typical errors of the order of 5 -- 10\%. While measurements and absorption coefficient data were also measured to this level of precision, and while there were great uncertainties in the atmospheric models used to simulate these data, the methodology was satisfactory. However, as measurements, absorption coefficient data and spectral resolution have continued to improve, the shortcomings of this scheme have become apparent, especially when analysing higher spectral resolution observations, where the random overlap of lines from different gases is more difficult to justify. In addition, for sufficient accuracy the $k$-tables must be constructed with 10 -- 20 coefficients for each wavelength bin, which reduces calculation speed, and the process of combining the $k$-coefficients for the different gases in each layer can become computationally expensive if there is a large number of gases to consider. Finally, $k$-tables are relatively inflexible with respect to wavelength shifts or variations of resolving power with wavelength and so need to be individually recomputed for different instrument configurations, which is computationally expensive and also expensive in terms of data storage.

In light of these difficulties, especially with regards to this study where we wish to analyse data from instruments with very different and variable spectral characteristics, we instead simulated the spectra with NEMESIS using line-by-line (LBL) look-up tables rather than $k$-tables. For each gas, and for a range of pressures and temperatures suitable to model Jupiter's atmosphere, we computed high-resolution LBL spectra covering the combined spectral range of 0.4 to 5.2 $\mu$m. We sampled the LBL absorption coefficient spectra with a step of 0.0002 $\mu$m, which we found to be sufficient. To save on data storage, rather than compute the tables over a wide grid of both pressures \textbf{and} temperatures, we instead calculated the coefficients for 15 pressures spanning 1.063 mb to 9.74 bar (equally spaced in log pressure), but for each pressure considered just two temperatures -- one extracted at that pressure from our reference Jupiter temperature-pressure profile, and one 10 K warmer, to roughly capture how the absorption varies with temperature. For this study, where the temperature profile was assumed to remain unchanged, there was no temperature interpolation/extrapolation error at all. 

Once these LBL-lookup tables had been calculated we then computed our simulated Jovian spectra at the same sample step of 0.0002 $\mu$m and then averaged the spectra over the MUSE and VIMS instrument functions. This meant that we could easily model both datasets simultaneously and we could assess the effect of different resolutions and different wavelength calibrations. In addition, LBL-lookup-table calculations are not affected at all by the overlapping line and correlated-$k$ approximations and so are more accurate and better able to simulate the MUSE/VIMS data. Finally, although it might be expected that such calculations are much slower than correlated-$k$ calculations, we found that the increase in computation time was not substantial and our LBL-table calculations were only  roughly twice as slow as our correlated-$k$ calculations. Hence, this method gives greatly improved accuracy with only a modest increase of computation time and is much more flexible and able to cope with revised or variable instrument calibrations.

\subsection{Particle-scattering model} \label{sec:part-scatter}

 Since we do not know what the particulates in the Jovian cloud decks are composed of, we need to determine their scattering properties from the spectra themselves. To do this, as originally developed by \citet{irwin15}, we determine the imaginary refractive index $n_\mathrm{imag}$ spectra of the particles, and reconstruct the real refractive index spectrum $n_\mathrm{real}$ using the Kramers-Kronig relation for an assumed value $n_\mathrm{real}$ at a specific wavelength. We then determine the extinction cross-section, single-scattering, and phase-angle spectra using Mie theory, so that all these properties are self-consistent. Of course, Mie scattering assumes spherical particles, which are highly unlikely for the ice and solid particles that make up the particulates in Jupiter's cold atmosphere. However, if we assume that particles are not aligned with respect to each other then this method gives useful first-order approximations to the extinction and scattering cross-section spectra. The phase functions calculated with Mie theory have features specific to spherical particles such as the `rainbow' and the back-scattering `glory', which are not appropriate for solid particles. Hence, to eliminate these features, we fit the Mie-calculated phase functions with combined Henyey-Greenstein functions that approximate the forward- and back-scattering behaviour. We assumed Gamma size distributions \citep{hansen71,hansen74} for all our particulates 

\begin{figure}
	\includegraphics[width=\columnwidth]{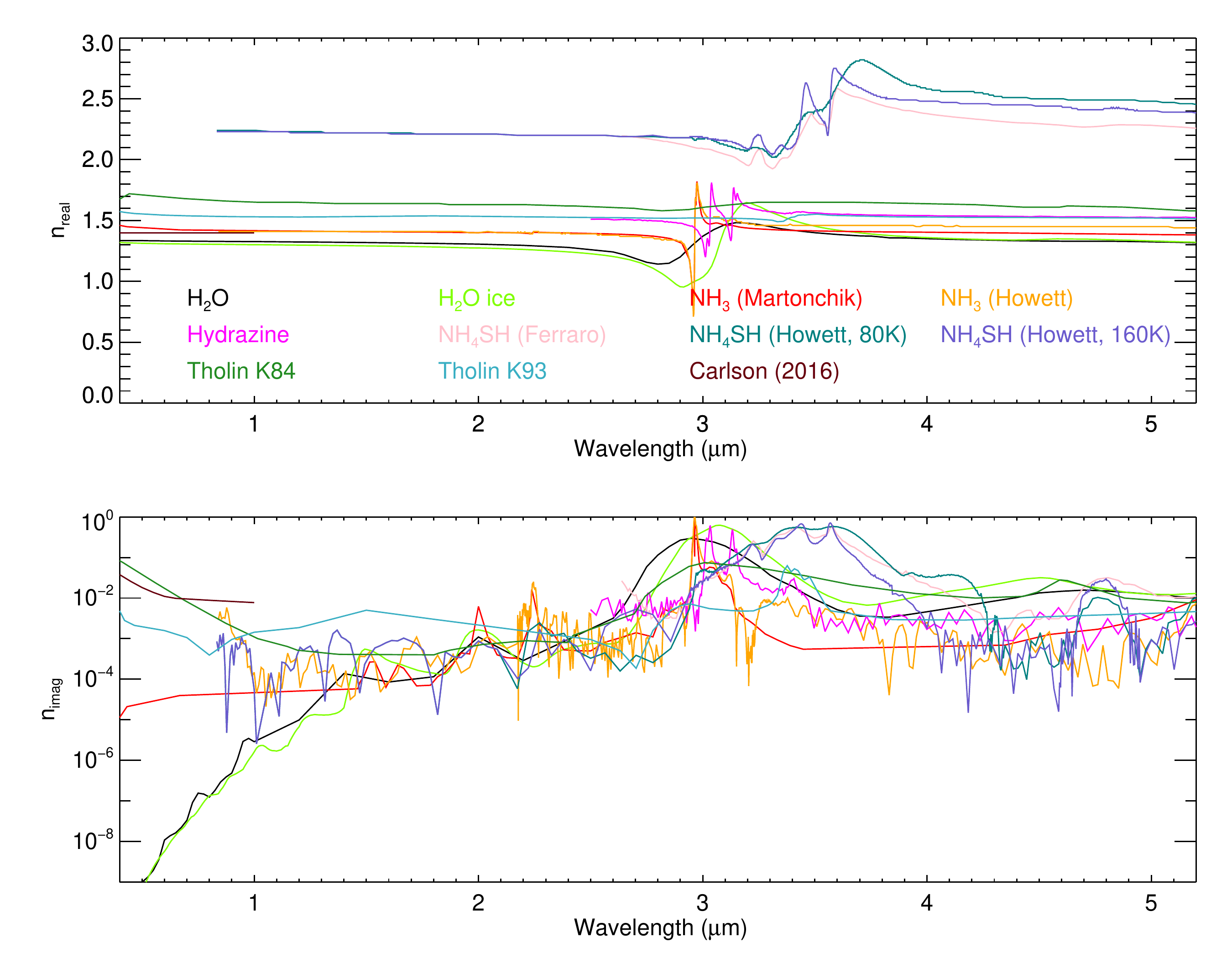}
    \caption{Complex refractive index spectra of candidate Jovian condensates from various sources: Water \citep{hale73}, Water ice \citep{warren08}, Ammonia \citep{martonchik84,howett07}, NH$_4$SH \citep{ferraro80,howett07}, `Tholin K84' \citep{khare84}, `Tholin K93' \citep{khare93}, hydrazine \citep{clapp96}, and the proposed Jovian chromophore of \citet{carlson16}.}
    \label{fig:refractive_indices}
\end{figure}

The published complex refractive index spectra of several Jovian condensate candidates are shown in Fig. \ref{fig:refractive_indices}. We will refer back to this figure on several occasions during this study.

\begin{figure}[!h]
	\includegraphics[width=\columnwidth]{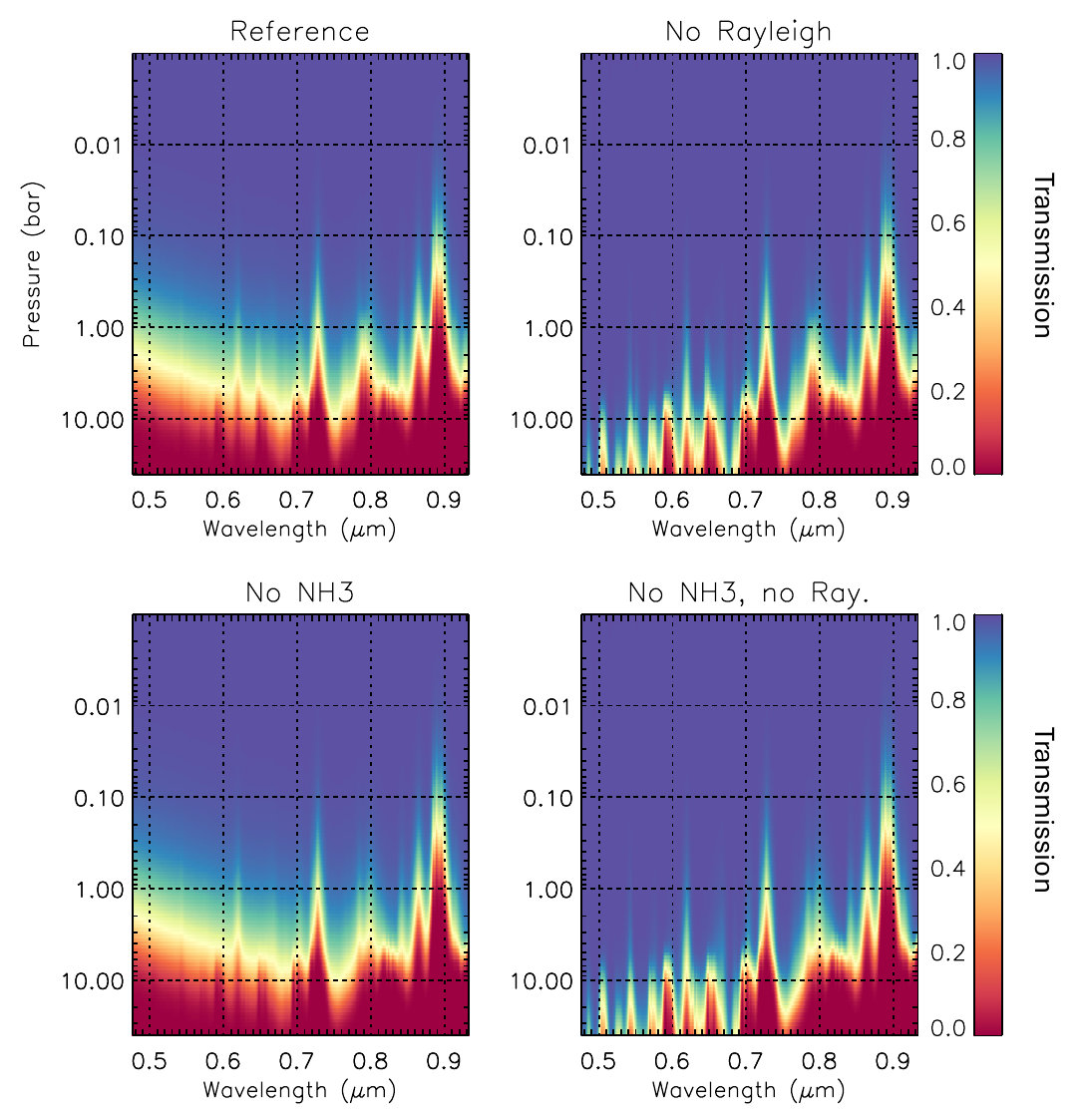}
    \caption{Calculations of nadir two-way cloud-free transmission down to different pressure levels in Jupiter's atmosphere at MUSE wavelengths. Calculations are shown for ammonia absorption included, or not included, and also for Rayleigh scattering included, or not included. }
    \label{fig:jupiter_muse_transmission}
\end{figure}

\begin{figure}[!h]
	\includegraphics[width=\columnwidth]{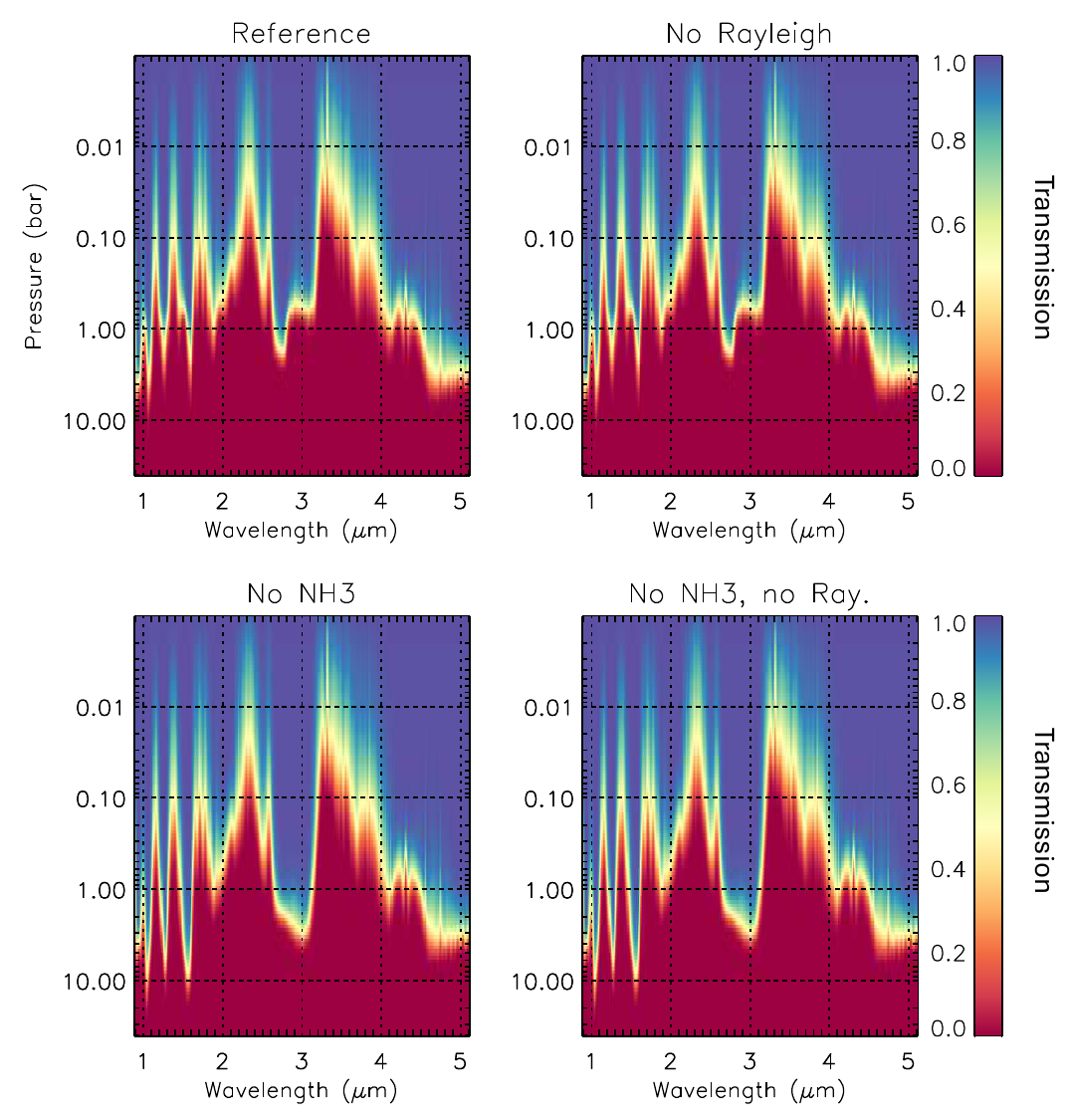}
    \caption{Calculations of nadir two-way cloud-free transmission down to different pressure levels in Jupiter's atmosphere at VIMS-IR wavelengths. Calculations are shown for ammonia absorption included, or not included, and also for Rayleigh scattering included, or not included. At these wavelengths, omitting Rayleigh scattering opacity makes little difference, but large changes are seen at short-wave continuum wavelengths if ammonia absorption is neglected, especially at 3 $\mu$m. }
    \label{fig:jupiter_vims_transmission}
\end{figure}

\section{Preliminary Modelling}\label{Section:Initial}

Before presenting our preferred retrieval model for simultaneously fitting the entire VLT/MUSE--Cassini/VIMS-IR spectral range in Sections \ref{Section:EZ-NEDF-NEB} and \ref{Section:EZ-NEB-Minnaert}, in this section we present initial modelling studies that were used to refine our understanding of the necessary cloud and ammonia profiles needed to match the observed Jovian spectra.

In Section \ref{sec:probe_level} we assess which wavelengths are sensitive to reflection from which pressure levels, and which wavelengths are sensitive to ammonia variations.  In Section \ref{Section:NEDF} we analyse difference spectra between the EZ, NEB, and a North Equatorial Dark Feature (NEDF) to localise the main reflectivity changes between these locations to be in the 1.5 to 1.7 bar region. We also outline evidence that suggests the main opacity of the chromophore is situated at $p>1$ bar. In Section \ref{sec:band-depth} we present evidence that a cloud at 1 -- 2 bar must be composed of large, highly scattering particles to be consistent with the observed band strength of the 619-nm methane band and in Section \ref{sec:onion} we report the results of simple `onion peeling' and reflecting layer models, which establish that there must be at least two distinct cloud layers, with a thin upper cloud at $p\sim0.5$ bar and a lower cloud at 1--2 bar responsible for the bulk of the spatial variations seen. In Section \ref{sec:vims-5micron} we examine the 5-$\mu$m window spectrum of the EZ, observed by Cassini/VIMS-IR during its flyby, to differentiate between the thermal emission and reflected sunlight components, showing that the main cloud top must be at 1.1 -- 1.2 bar, and that the reflected sunlight spectrum can be approximated mostly by reflection from two thin layers, one at $\sim$1.1 bar and one at $\sim$0.75 bar (dependent on the assumed PH$_3$ profile). Finally, in Section \ref{Section:JIRAM-EZ} we fit JIRAM-SPE spectra (2.40 -- 5.05 $\mu$m) of the same area at the northern edge of the EZ measured on both the day and night side and show that the optimal fit is achieved with clouds at 1.2, 0.4 and 0.09 bar, and show that the 0.4-bar cloud needs a strong absorption feature at 3 $\mu$m.

\subsection{Probed pressure levels}\label{sec:probe_level}

To understand the observed spectra of Jupiter (Fig. \ref{fig:VIMS_spx_comparison}) we need to understand to what pressure level in the atmosphere we can see at each wavelength. Fig. \ref{fig:jupiter_muse_transmission} shows the calculated two-way nadir cloud-free transmission spectra to different pressure levels in our reference Jupiter atmosphere at VLT/MUSE wavelengths, while Fig. \ref{fig:jupiter_vims_transmission} shows the corresponding calculation at Cassini/VIMS-IR wavelengths. In both examples, we show the transmission in the nominal case and also in three other cases: 1) no Rayleigh-scattering cross-section; 2) no ammonia absorption; and 3) no Rayleigh-scattering cross-section or ammonia absorption. For the nominal case we include the Rayleigh-scattering cross-section as if each Rayleigh-scattering event permanently removes a photon from the beam. It can be seen that Rayleigh scattering is important to consider at MUSE wavelengths, but becomes negligible at VIMS-IR wavelengths. However, since Rayleigh-scattering is conservative this means that even in regions of strong Rayleigh-scattering it is quite possible for light to scatter to much deeper levels and then be scattered back to the observer. Hence,  we should use the transmissions calculated with no Rayleigh-scattering when considering to what depths we might actually be able to see.  The conservative nature of Rayleigh scattering also means that when considering the back-scatter of clouds in the atmosphere, at short wavelengths the back-scatter of the air itself can become at least as important as that from the clouds. Hence, counter-intuitively, the addition of clouds that might not be conservative scatterers could in some cases reduce the overall reflectivity at short wavelengths, not increase it. 

While the inclusion or exclusion of ammonia absorption can be seen to make only minor changes in the calculated transmissions at MUSE wavelengths (with the exception of features at 647 and 770 nm), there are substantial changes seen at VIMS-IR wavelengths from 0.9 to 1.5 $\mu$m,  near 2 $\mu$m, and also near 3 $\mu$m. Since the vertical profile of ammonia and its spatial distribution is very ill-constrained, and since the identity and scattering properties of Jupiter's aerosols are also unknown this presents a significant challenge to decoupling ammonia variations from aerosol variations.

\subsection{MUSE -- VIMS-IR Comparison of North Equatorial Dark Feature with the EZ and NEB}\label{Section:NEDF}
From the MUSE and VIMS-IR datasets we can identify regions that lie within North Equatorial Dark Features (NEDF). Such features were analysed by \citet{irwin25}, and are also known as 5-$\mu$m hotspots. While the cloud structure measured by the Galileo probe nephelometer \citep{ragent98} has for many years been considered to be anomalous, it was measured from within just such an NEDF and so it is sensible to expect that we might be able to use the nephelometer profile as a prior estimate of the cloud structure in an NEDF. Indeed, although it is sometimes thought that these NEDFs are substantially `cloud-free', clouds were detected here by the Galileo probe, and \citet{irwin25} demonstrated that an entirely cloud-free atmosphere would have a reflection spectrum that looks nothing like the spectra of NEDFs observed by MUSE \citep[see Fig. S4 of][]{irwin25}.

\begin{figure*}
	\includegraphics[width=\textwidth]{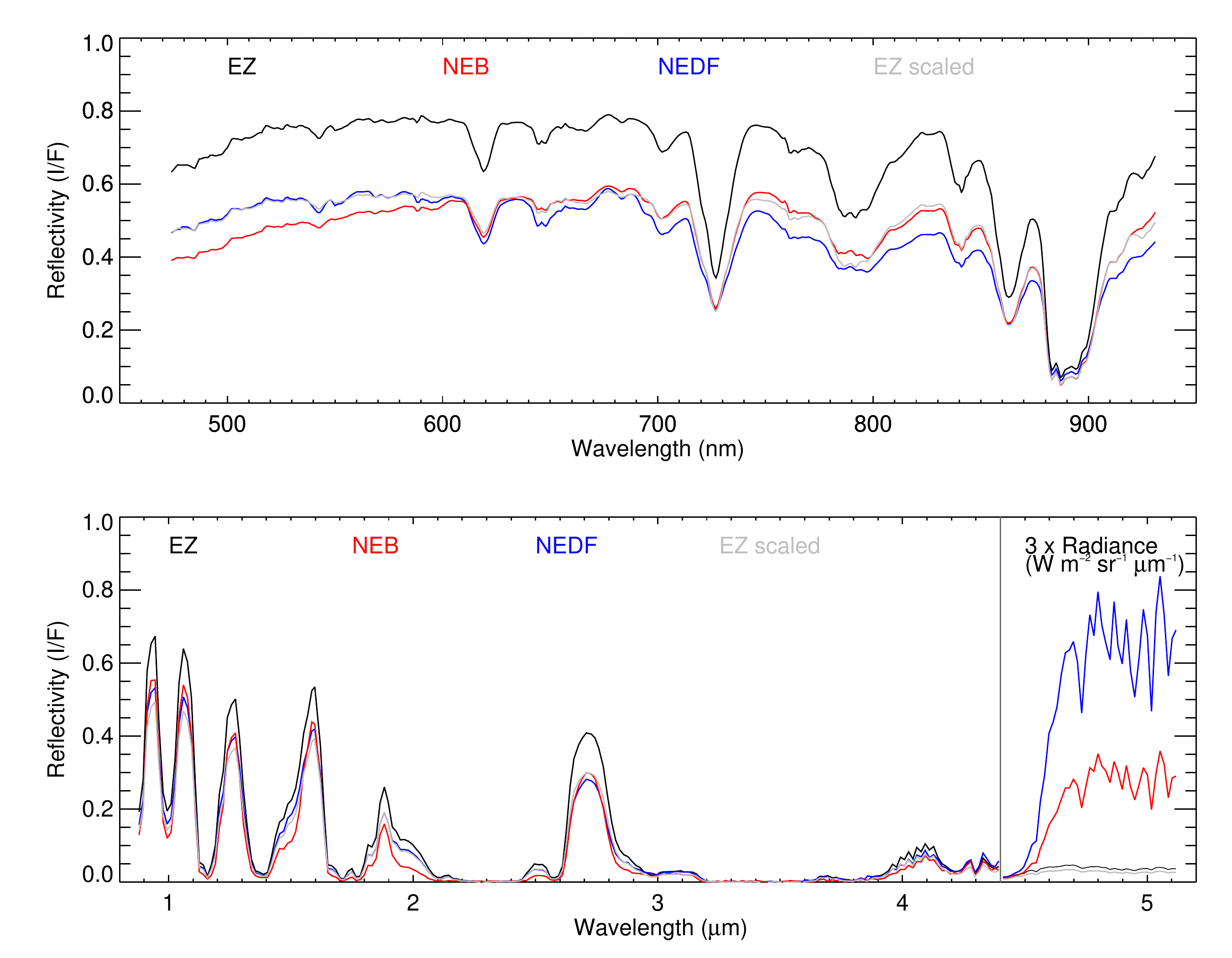}
    \caption{Typical spectra from our 2018 VLT/MUSE (top) and 2000 Cassini/VIMS-IR (bottom) observations, centred on the Equatorial Zone (EZ), the North Equatorial Belt (NEB) and in the centre of a North Equatorial Dark Feature (NEDF). Both panels also show the EZ spectrum, roughly scaled to match the NEDF spectrum at short wavelengths to aid comparison. In the bottom panel, showing the VIMS-IR spectra, the reflectivity is shown for wavelengths less than 4.4 $\mu$m, while the radiance is shown for longer wavelengths, scaled by a factor of 3 to aid comparison with the reflectivity spectra.}
    \label{fig:spot_compare}
\end{figure*}

To compare the spectrum and cloud structure of a typical NEDF with those of the more representative EZ and NEB regions, we extracted spectra from $3\times3$ pixel boxes from our MUSE observation. For the NEDF we selected a region in the centre of the NEDF seen above and to the right of the centre of disc in  Fig. \ref{fig:jupiter_muse}, while for the EZ and NEB we selected typical regions near the central meridian. We compared these spectra with those extracted from single pixels from corresponding regions in our approach VIMS-IR observation (top row of Fig. \ref{fig:jupiter_vims}). The spectra from these regions are shown in Fig. \ref{fig:spot_compare}. In the MUSE spectral range (top panel of Fig. \ref{fig:spot_compare}) we can see that in addition to being less reflective than the EZ in general, the NEDF spectrum becomes progressively darker with respect to the EZ at longer wavelengths. This can best be seen by comparing to the EZ spectrum scaled to roughly match the NEDF spectrum at short wavelengths. In contrast, the scaled-EZ and NEB spectra are reasonably similar at longer wavelengths, but become progressively different at short wavelengths, due to the absorption of the chromophore that gives the belts their characteristic brown/red colour. At the longer wavelengths probed by VIMS-IR (bottom panel of Fig. \ref{fig:spot_compare}) the differences between the spectra are less easily interpreted, which is partly because at these wavelengths the variation in ammonia abundance also has a strong effect, which is much less significant for MUSE. In general, though, the NEDF and scaled EZ spectra (using the same scaling factor as was applied to the MUSE spectrum) agree reasonably well in the VIMS-IR spectral range at wavelengths dominated by reflected sunlight, suggesting that such regions have similar cloud and ammonia profiles, although the clouds must be thinner (or less reflective) in the NEDF. However, in the 5-$\mu$m window we can see very different spectra between the EZ and NEDF, with high 5-$\mu$m emission seen in the NEDF, but very low emission seen in the EZ. The NEB is similarly different from the EZ in this range, but to a lesser extent than the NEDF. 

\begin{figure}
	\includegraphics[width=\columnwidth]{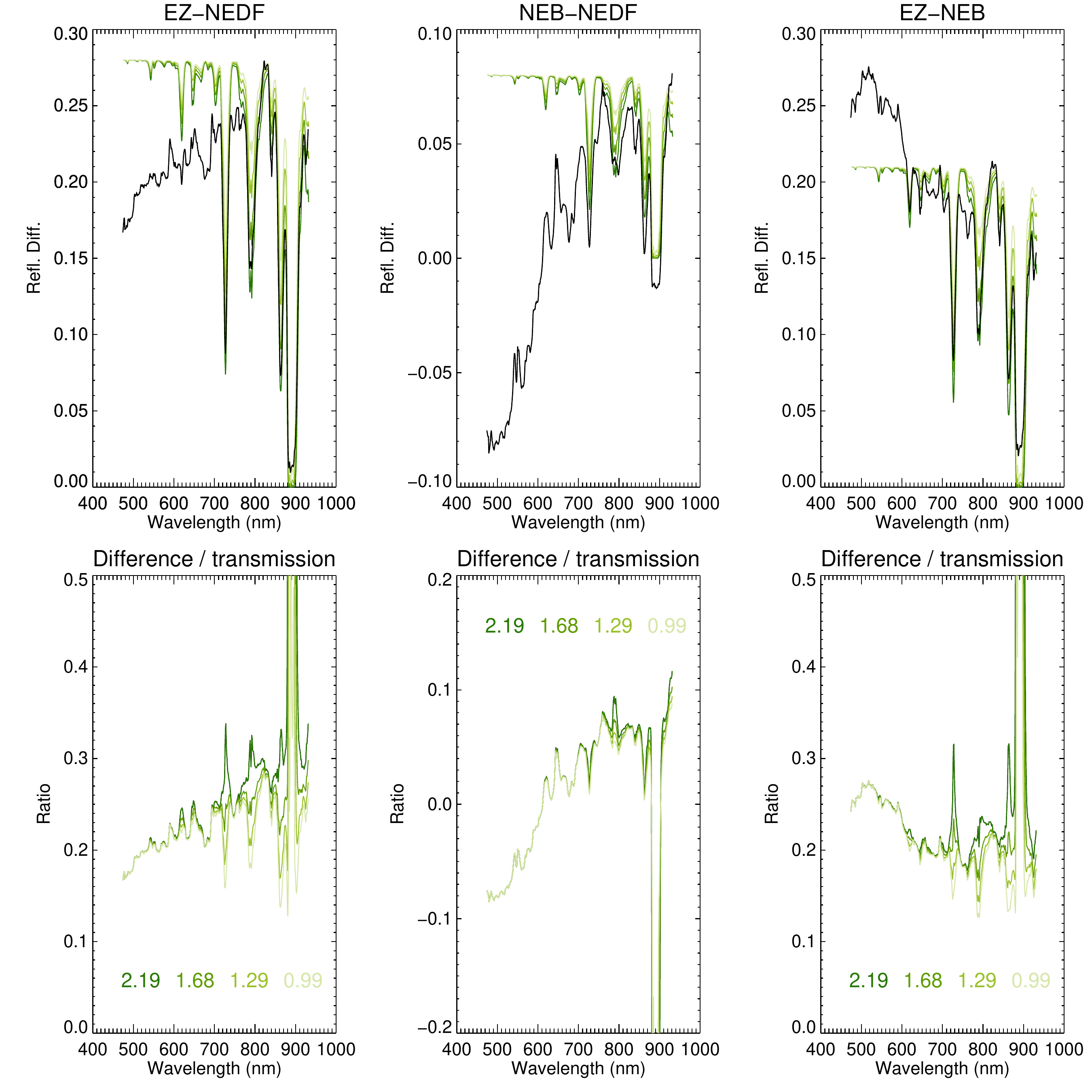}
    \caption{Analysis of VLT/MUSE spot spectra. The top row shows the difference between the observed EZ, NEB and NEDF spectra (in black) compared with calculated two-way transmissions of our standard cloud-free atmosphere (omitting Rayleigh scattering) to various pressure levels, roughly scaled to be consistent with the peak EZ -- NEDF difference near 820 nm. The transmissions are colour-coded, as defined in the panel below, and the pressures are in bar.  The bottom row shows the difference between the EZ, NEB and NEDF spectra divided by the calculated transmission spectra. }
    \label{fig:spot_muse_analyse}
\end{figure}

To analyse these differences more quantitatively, in Fig. \ref{fig:spot_muse_analyse} we compare the differences between EZ, NEDF, and NEB spectra at MUSE wavelengths (i.e., EZ -- NEDF, NEB -- NEDF, and EZ -- NEB) and the two-way transmission to different pressure levels calculated for conditions of zero Rayleigh scattering opacity and shown earlier in Fig. \ref{fig:jupiter_muse_transmission}. In the top row of Fig. \ref{fig:spot_muse_analyse} we compare the difference spectra with the calculated transmissions to several levels roughly scaled to match the difference spectra near 820 nm. The transmission spectra are colour-coded and the chosen base pressure values are shown in the bottom panels. As we go to deeper pressures the absorption bands in the calculated transmission spectra become deeper and by comparing the relative depths of the calculated absorptions with those seen in the observed difference spectra we can gain insight into the pressure level in Jupiter's atmosphere where the reflectivity is changing between the three locations. In the bottom row of Fig. \ref{fig:spot_muse_analyse} we show the observed difference spectra divided by the calculated two-way transmission spectra to the different pressure levels, which we will call `ratio plots'. Considering first the EZ -- NEDF difference spectrum we can see that for the transmission spectra calculated at high base pressures, the ratio plot has peaks at absorption band wavelengths, indicating that the computed absorption bands are deeper than those seen in the difference spectrum. Conversely, at low calculation pressures, the ratio plot shows troughs at absorption band wavelengths, indicating that the computed absorption bands are shallower than those seen in the observed difference spectrum. However, at intermediate pressures (1.5 -- 1.7 bar) we can see that the features in the ratio spectrum are to first order `flattened-out'. Hence, this simple analysis shows that the bulk of the reflectivity changes responsible for the difference between the EZ and NEDF spectra take place in the 1.5 -- 1.7 bar pressure region. Aside from the gaseous absorption wavelengths, we can see that the difference between the EZ and NEDF spectra generally increases with wavelength, which is probably due to the particles near 1.5 -- 1.7 bar, responsible for this extra reflectivity, being more reflecting at longer wavelengths. 

Turning to the EZ -- NEB difference spectrum we see very similar behaviour in the absorption bands and again find that having reflectivity differences in the 1.5 -- 1.7 bar region mostly `flattens out' the gaseous absorption features in the ratio spectrum. However, here we see very different behaviour at shorter wavelengths, where the NEB is increasingly dark at shorter wavelengths, due to the chromophore. It seems likely that the  chromophore is also in the 1.5 -- 1.7 bar region: even though increased absorption in any cloud above this level might lead to a smooth reduction in reflectivity at shorter wavelengths, the increase in absorption would have to \textbf{exactly} mirror the increase in reflectivity from the 1.5 -- 1.7 bar region, requiring great correlation of the Jovian aerosol properties over a wide range of pressures. In addition, the limb-darkening seen by VLT/MUSE increases with wavelength for all latitudes (e.g., Fig. \ref{fig:jupiter_spectra_comparison} for the EZ, and \ref{fig:jupiter_spectra_comparison1} for the NEB), while the absorption of chromophores such as tholins and the Carlson chromophore decreases with wavelength (Fig. \ref{fig:refractive_indices}). From this we again deduce that the chromophore is most likely in the lower layer and its darkening is more masked by scattering from particles in the overlying clouds as we move towards the limb. The NEB -- NEDF difference spectrum  has much smaller features in the methane bands at 727 and 790 nm than the EZ -- NEDF spectrum, suggesting the vertical cloud structure is more similar. The only region where the observed-difference/calculated-transmission spectra change significantly with pressure is at 790 nm, and again is consistent with changes at 1.5 -- 1.7 bar.

We performed a similar analysis of the VIMS-IR difference spectra, which we show in Fig. \ref{fig:spot_vims_analyse}.  In this spectral region the analysis is far less simple because absorption of ammonia becomes very strong and there are also several strong absorption bands of methane in which we cannot see the deeper clouds at all. Hence, at VIMS-IR wavelengths we need to consider both changes in the opacity in the upper haze/cloud layer \textbf{and} at deeper levels, and also ammonia abundance differences, which makes it difficult to draw reliable conclusions.  

\subsection{MUSE Cloud-top pressures and cloud scattering properties}\label{sec:band-depth}

In our band-depth analysis of the same VLT/MUSE observations \citep{irwin25}, the strength or `equivalent width' of the 619-nm methane band feature was used to infer cloud-top pressures of 2 -- 3 bar. This is significantly deeper than the cloud-top pressures determined from almost all other previous observations to be in the range 1 -- 2 bar, and from the previous section analysing the same MUSE data, where we conclude that cloud changes are at 1.5 -- 1.7 bar. How can we explain this discrepancy? Section \ref{sec:probe_level} showed that Rayleigh scattering has very significant effects at MUSE wavelengths and indeed \citet{irwin25} showed that if we remove clouds altogether from our model, we still see significant reflection from the atmosphere and strong methane absorption at 619 nm \citep[See Fig. S4 of][]{irwin25}. According to Mie theory, particles that are large compared to the wavelength are predominantly forward-scattering, but this asymmetry decreases as the particle size reduces, with the particles tending to Rayleigh scattering as their size becomes comparable to, or smaller than, the wavelength. We realised that we could have our main cloud layer at  1 -- 2 bar and \textbf{still} infer a cloud-top pressure of 2--3 bar from the 619-nm methane band if the particles are composed of large, highly scattering particles that  only partially back-scatter sunlight at visible wavelengths, and allow the remaining light to forward-scatter down to deeper levels before being Rayleigh-scattered back. Looking at the known scattering properties of candidate aerosols (Fig. \ref{fig:refractive_indices}), it can be seen that water liquid/ice has the lowest imaginary refractive index (and thus highest single-scattering albedo) at visible wavelengths, thus making a water condensate an attractive potential candidate for the required particles at 1 -- 2 bar. However, it is speculation to conclude that the lower cloud \textbf{is} composed of water ice -- all we can conclude here is that the presence of large, highly scattering particles in the 1--2 bar region is an attractive potential solution to the cloud height discrepancy. We will return to the question of the identity of this cloud in the Discussion section.

\subsection{MUSE `Onion-peeling' and reflecting layer models} \label{sec:onion}
The distribution of clouds seen in methane absorption bands (e.g., at 889 nm, Fig. \ref{fig:jupiter_muse}) is very different from that seen at other wavelengths, where the atmospheric gases are more transparent, with enhanced reflectivity seen over the EZ, GRS and the North and South Tropical Zones. This reflection is from particles high on the atmosphere ($p < 0.5$ bar), and is sometimes referred to as a haze. As we move from 889 nm to adjacent wavelengths where the atmosphere is more transparent, the reflection steadily increases as we see deeper into the atmosphere and the lower, and more spatially varied cloud structure becomes apparent. The rate of increase of reflectivity with depth varies between the EZ, NEB and NEDFs and can be used to make first-order inferences of the vertical cloud structure, without having to resort to a full multiple-scattering retrieval model. In Appendix \ref{app:onion} the method of `Onion-peeling' was applied to MUSE observations from 875 to 910 nm and the results support the preliminary conclusions that in addition to the upper layer of aerosol at $p\sim0.5$ bar, a second optically-thicker cloud is present at 1 -- 2 bar, whose variations are mainly responsible for the differences seen between these regions (Figs.  \ref{fig:oniontest1} -- \ref{fig:onionfigure}).

The `Onion-peeling' method works best when the contributions of overlying layers can be cleanly peeled away to reveal the effects of lower layers. The approach is interesting and simple, but does not account for the fact that the transmission to different pressure levels increases gradually with pressure rather than sharply. Hence, the assigned pressures of the inferred clouds depend strongly on an assumed limiting two-way opacity.

\begin{figure*}[!h]
	\includegraphics[width=\textwidth]{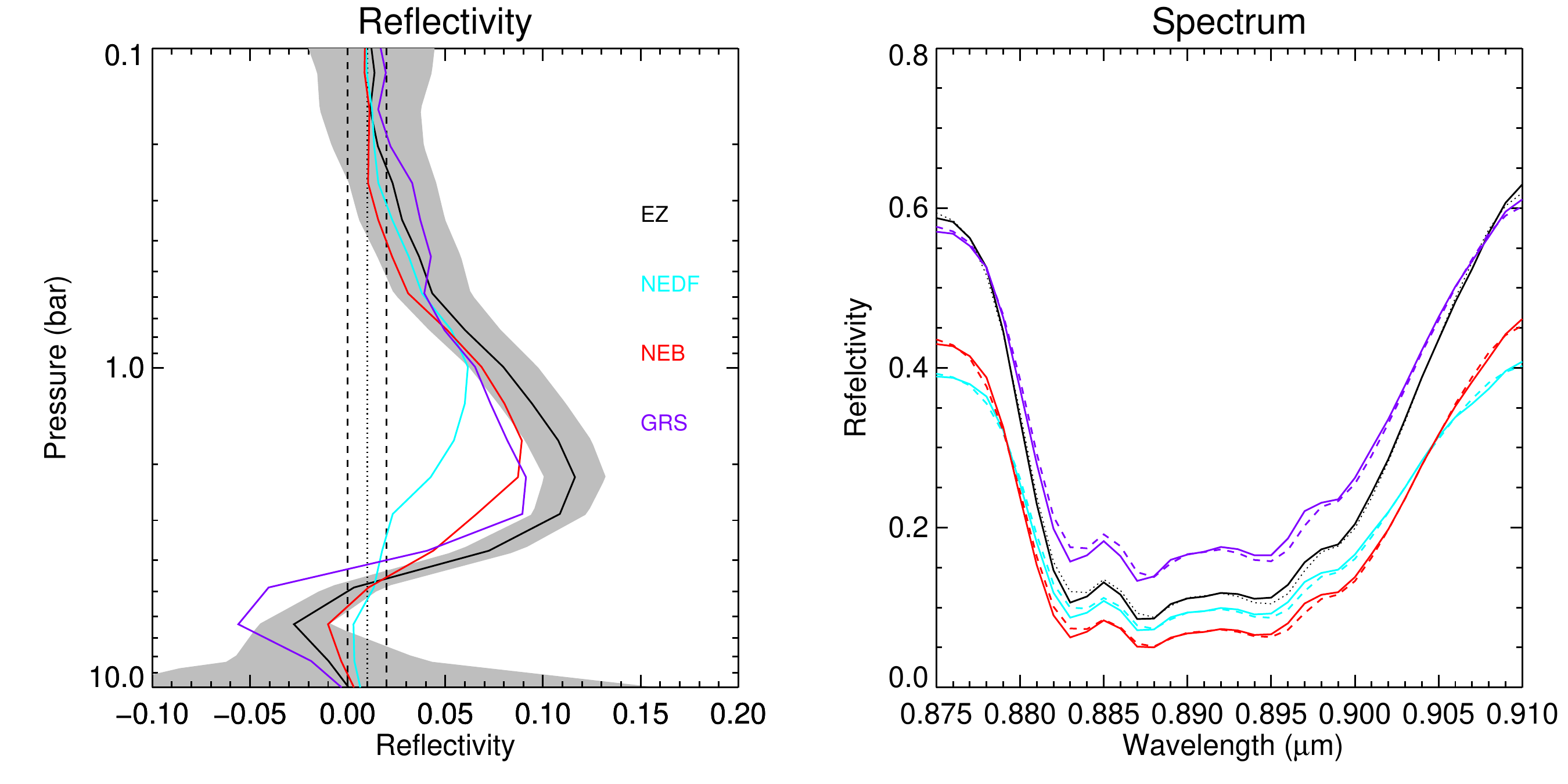}
    \caption{Reflecting layer model retrievals of MUSE spectra of four representative regions near Jupiter's equator: EZ (Equatorial Zone), NEB (North Equatorial Belt), NEDF (North Equatorial Dark Feature) and the GRS (Great Red Spot). The left hand panel shows the reflectance profiles retrieved from the sample MUSE spectra in the 875 -- 910 nm band in the these regions, and the right hand panel shows the observed spectra in these region (solid lines) and the fits to them (dashed lines), showing that the reflecting layer model fits the observed spectra well. In the left hand panel, the prior reflectivity profile and error limits are shown as the dotted and dashed lines. The modified retrieval uncertainty, $\sigma_m$, is shown as the grey region, but only for the EZ for clarity since the error profiles are very similar for all locations; in all cases the retrieval error reduces as we go from 0.1 to about 5 bar, but the retrieval becomes unconstrained at higher pressures as the atmosphere `blacks out'. }
    \label{fig:onionretrieval}
\end{figure*}

An alternative approach is to consider that the total reflected spectrum is a linear combination of light reflected from particulates at different levels, taking no account of the absorption of overlying clouds. In this approach, the observed reflectivity $R(\lambda)$ at a particular wavelength $\lambda$ can be written as a sum of the reflections from $N$ different pressure levels multiplied by the two-way transmissions to those levels, i.e.,

\begin{equation}\label{eq:refllayer}
R(\lambda) = \sum_{i=1}^N T(p_i,\lambda)R(p_i),
\end{equation}

where $p_i$ is the pressure of each level, $R(p_i)$ is the reflectivity of the level (assumed to be constant over the wavelength range considered), and $T(p_i,\lambda)$ is the two-way pre-calculated nadir transmission spectrum to that level, presented earlier in Section \ref{sec:probe_level}. The observed MUSE reflectivity spectrum over the 875 -- 910 nm region probes a wide range of pressure levels in a narrow spectral range and so can be simply modelled in this way. The observed reflectivity spectrum can be easily inverted using the same optimal estimation algorithm that is employed by NEMESIS to determine $R(p)$, and in this case the model is linear and so the retrieval is a single-step process. The reflectivity spectrum of four representative locations (EZ, NEB, NEDF and GRS) were modelled in this way and the retrieved reflectivity profiles $R(p)$ are shown in Fig. \ref{fig:onionretrieval}. The optimal estimator returns the retrieved reflectivity profile and covariance matrix $\mathbf{S_x}$. If we take the diagonal values of this matrix, $\sigma^2$, then we can make a modified estimate of the retrieved measurement error $\sigma_m = 1/\sqrt{1/\sigma^2-1/\sigma_0^2}$,
where $\sigma_0$ are the assumed prior errors. The calculated values of $\sigma$ approach $\sigma_0$ where there is no information in the retrieval, while the calculated values of  $\sigma_m$ approach infinity and so gives a better pictorial representation of the information content. 

In Fig. \ref{fig:onionretrieval}, which also shows the uncertainty using $\sigma_m$, we can again see that the bulk of the reflectance difference between these regions occur at pressures greater than 1 bar. At lower pressures we can see smaller, but significant variation in the 0.2 -- 0.6 bar region, which accounts for the differences in reflectivity in the centre of the 889-nm methane absorption band. Hence, the evidence again supports the existence of two independent clouds -- the main one based at 1 -- 3 bar, and a less reflecting cloud in the 0.2 -- 0.6 bar region. 

Finally, we should note that the `weighting functions' of these retrievals are rather broad as the transmission spectra vary smoothly with pressure. Hence, the underlying reflectivity profiles may not be as broad and smooth as are retrieved here. Indeed, we were able to achieve similar fits to the EZ spectrum, for example, with two thin layers, based at 0.45 and 1.69 bar, respectively. Hence, the `true' reflectivity profiles may be much more sharply peaked than those retrieved here. We will return to this point in the Discussion section.

\subsection{VIMS-IR Equatorial Zone 5-$\mu$m window} \label{sec:vims-5micron}

Having found a cloud structure consistent with the visible/near-IR MUSE and VIMS spectra, we then sought to find what vertical cloud structure is also consistent with the variations seen in the `5-$\mu$m window' of our observations. The observed spectrum in the range 4.5 to 5.2 $\mu$m comes mostly from thermal emission from levels at pressures of 5 -- 10 bar in Jupiter's atmosphere. However, on the dayside a proportion of the observed radiance will be reflected sunlight. Figure \ref{fig:VIMS_spx_comparison} shows the $0^\circ$ zenith angle spectra of the Equatorial Zone (EZ) and North Equatorial Belt (NEB) extracted from our approach VIMS-VIS and VIMS-IR cubes using our Minnaert model. As can be seen, in the EZ, where the thermal emission is very low, the observed radiance is consistent with a reflectivity of roughly 0.15, and hence reflected sunlight is likely to be a significant component of the EZ 5-$\mu$m spectrum, and must be properly accounted for. Figure \ref{fig:VIMS_spx_comparison}  also underlines that while the 5-$\mu$m spectrum changes by an order of magnitude between the EZ and NEB, the radiance at shorter continuum wavelengths changes much less dramatically. Hence, whatever is causing the 5-$\mu$m brightness modulation has rapidly decreasing effects at shorter wavelengths. The difference in the EZ and NEB at other wavelengths is dependent on the variation of upper level cloud/haze, the ammonia abundance and the phosphine abundance.

Discriminating between reflected solar and thermal radiation at 5 $\mu$m is difficult in observations from the Earth or from the VIMS approach observations as the two components are entangled. However, VIMS also observed Jupiter during its flyby and our flyby VIMS-IR cube  can be used to extract spectra of the EZ from both the dayside and nightside, which can be used to differentiate between the two components. We extracted two spectra from the flyby VIMS-IR cube along the horizontal line of the cube in the EZ, one from either side of the central meridian at zenith angles as close as possible to 30$^\circ$. The two spectra extracted had viewing zenith angles of 30.8$^\circ$ and 31.9$^\circ$, and solar zenith angles of 37.4$^\circ$ (i.e., dayside) and 99.0$^\circ$ (i.e., nightside), respectively.

\begin{figure}
	\includegraphics[width=\columnwidth]{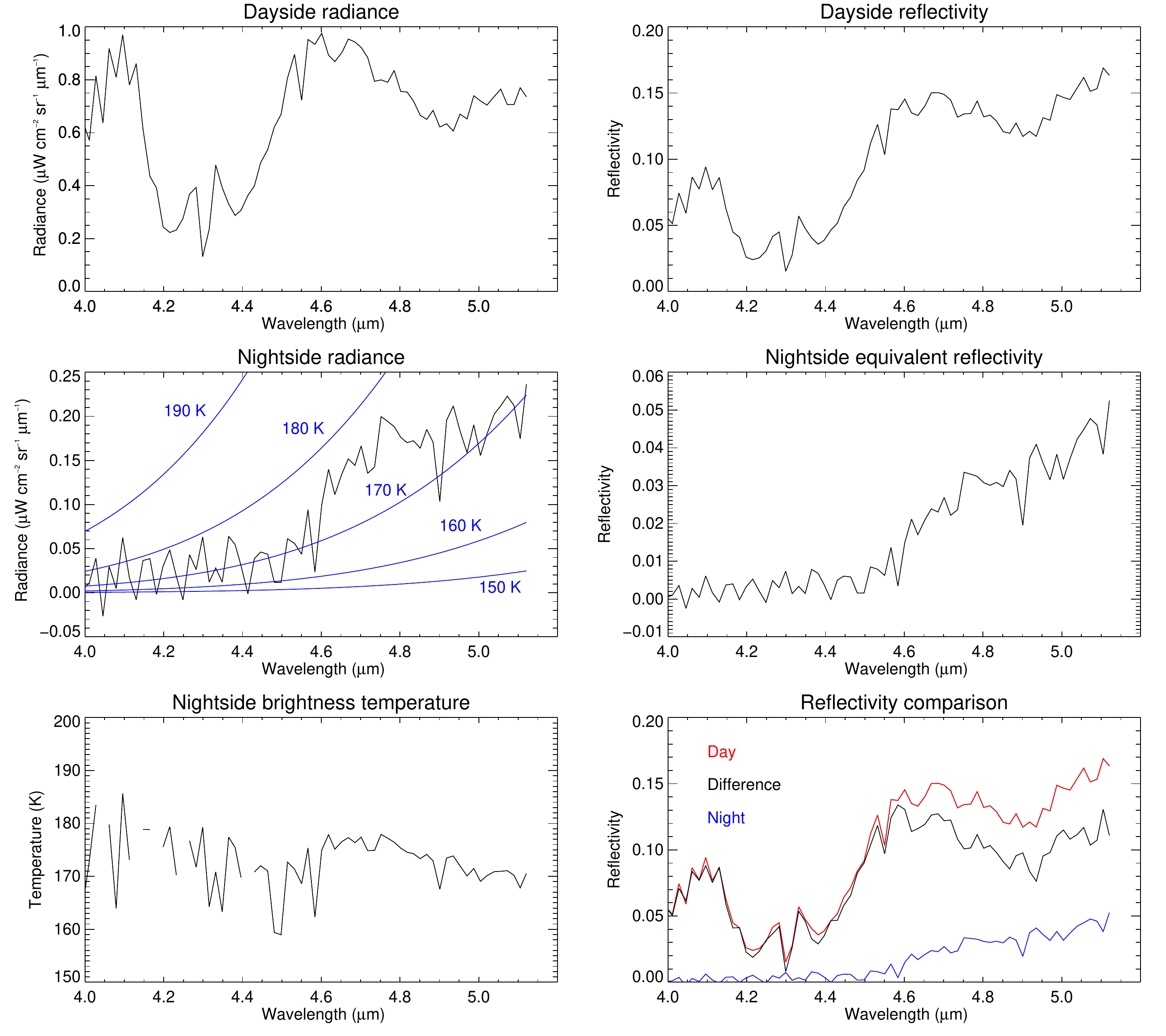}
    \caption{5-$\mu$m spectra extracted from the EZ of the VIMS flyby observation V1356976257\_3 on the dayside (top row) and nightside (middle row) at a viewing zenith angle of $\sim$30$^\circ$, in terms of both radiance and equivalent reflectivity (N.B., on the nightside there is clearly no actual reflectivity, so the spectrum should be considered as an \textbf{equivalent} dayside reflectivity). The nightside spectrum is compared to black-body curves for surfaces at 150, 160, 170, 180 and 190 K. In the bottom left panel the nightside spectrum is converted to equivalent black-body temperature. The bottom right panel compares the day and night spectra in terms of reflectivity, and also their difference. }
    \label{fig:VIMS_compare_refl1}
\end{figure}

In Fig. \ref{fig:VIMS_compare_refl1} we show these dayside and nightside EZ spectra plotted in terms of radiance and equivalent reflectivity (i.e., we divided the dayside and nightside radiances by the expected reflected sunlight spectrum from a Lambertian surface at 5.2 AU from the Sun at normal incidence). The dayside spectrum shows a considerable component of sunlight as can be seen from the 4 to 4.6 $\mu$m range, where we can see sunlight reflected from the clouds modulated by the strong absorption band of PH$_3$, centred at 4.3 $\mu$m. In the nightside spectrum, this component is completely absent. Instead, we see very low emission, which increases with wavelength and appears approximately consistent with the spectrum of a black body at a temperature of $\sim$172 K. This observation gives us a considerable insight into the vertical cloud structure of Jupiter and suggests that the tops of the clouds in the EZ as seen at 5 $\mu$m lie at a point in the atmosphere where the temperature is $\sim$172 K, which from our assumed temperature-pressure profile corresponds to a pressure of 1.1--1.2 bar. 

Since the EZ appears almost featureless, we can assume that the cloud opacity is very thick all along the EZ and thus the thermal emission component in the nightside spectrum is likely to be exactly the same as that on the dayside. Hence, we can subtract it to determine the reflected solar component on the dayside (also shown in Fig. \ref{fig:VIMS_compare_refl1}), where we can see that the shape of the reflected solar component is very dissimilar to the spectrum we usually see at 5 $\mu$m. Indeed, there appears to be a broad absorption band centred at 4.9 $\mu$m, which to our knowledge has not previously been noted.

\begin{figure}
	\includegraphics[width=\columnwidth]{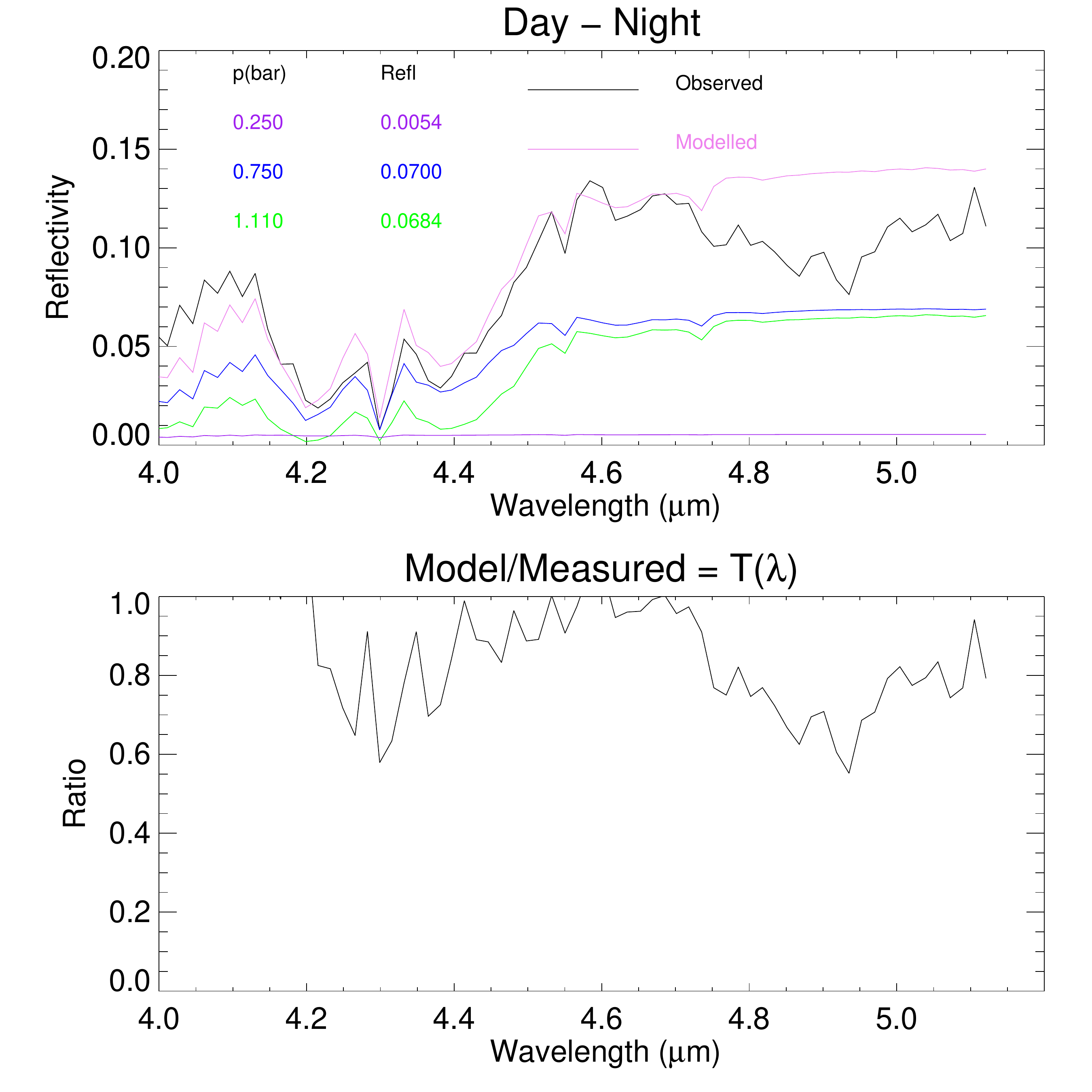}
    \caption{Top panel shows the Day -- Night spectrum extracted for the EZ from the VIMS-IR flyby cube (V1356976257\_3) at a solar zenith angle of 37.4$^\circ$ and a viewing zenith angle of $\sim$ 30$^\circ$. Also shown are the predicted spectra from a simple three-layer reflectance model (violet line) and the contributions from each layer (coloured green, blue and purple). The pressure levels of these layers and the estimated reflectivities are shown in the panel. Bottom panel shows the estimated reflectance spectrum (i.e., Day -- Night) divided by the modelled spectrum, showing the increased absorption needed between 4.6 and 5.2 $\mu$m to explain the observed reflectance spectrum. Differences at wavelengths less than 4.6 $\mu$m can be attributed to discrepancies between between the modelled and true phosphine (PH$_3$) abundance profiles.} 
    \label{fig:VIMS_compare_refl2}
\end{figure}

What cloud structure might account for these spectra? We have found from the nightside EZ spectrum that there must be a cloud with a top at 1.1--1.2 bar. But we also know that there is an upper tropospheric component of cloud/haze, seen in the methane bands of both the MUSE and VIMS observations at $p < 0.5$ bar. Hence, using our calculated two-way transmission spectra for our cloud-free reference atmosphere down to these two pressure levels, we explored whether we might be able to combine these components in a simple reflecting layer model to reproduce the Day-Night reflectivity spectrum. In practice, this simple two-layer model gave a poor fit, but by adding a third reflecting layer at a pressure of $\sim$0.7 bar we were able to fit the observed reflectance spectrum $R_{obs}$ with a simple, linear reflectance model $R_{obs}=\sum_{i=1}^3 R_i T_i(\lambda)$, where $R_i$ are the fitted layer reflectivities and $T_i(\lambda)$ are the calculated two-way transmission spectra down these levels at pressures $p_i$. For the deep cloud we kept $p_1$ fixed to the level determined from the nightside spectrum (1.1 bar) and found that we could achieve a reasonable fit to the Day--Night reflectivity spectrum for the remaining reflecting layers at pressures of $p_2 \sim$ 0.75 bar and $p_3 \sim$ 0.25 bar as shown in Fig. \ref{fig:VIMS_compare_refl2}. It can be seen that the additional layer near 0.75 bar is necessary to fit the shape of the PH$_3$ absorption band at 4.3 $\mu$m (assuming our adopted PH$_3$ profile is correct) and that the modelled reflectivity from the uppermost `haze' layer is very low, suggesting that only the lower two clouds are actually important in this wavelength range. At the longer wavelengths  we see that the calculated two-way transmission spectrum down to even the deepest layer at 1.1--1.2 bar is effectively unity for wavelengths longer than 4.5 $\mu$m, and does not model at all the broad absorption band seen centred at 4.9 $\mu$m. Since we can discount this band as being caused by gaseous absorption, this would appear to be a cloud absorption feature. Whether this absorption feature is in the cloud at 0.75 bar or at 1.1 bar is not possible to discern from this analysis. 

\subsection{JIRAM-SPE EZ spectrum}\label{Section:JIRAM-EZ}
\begin{figure}
	\includegraphics[width=\columnwidth]{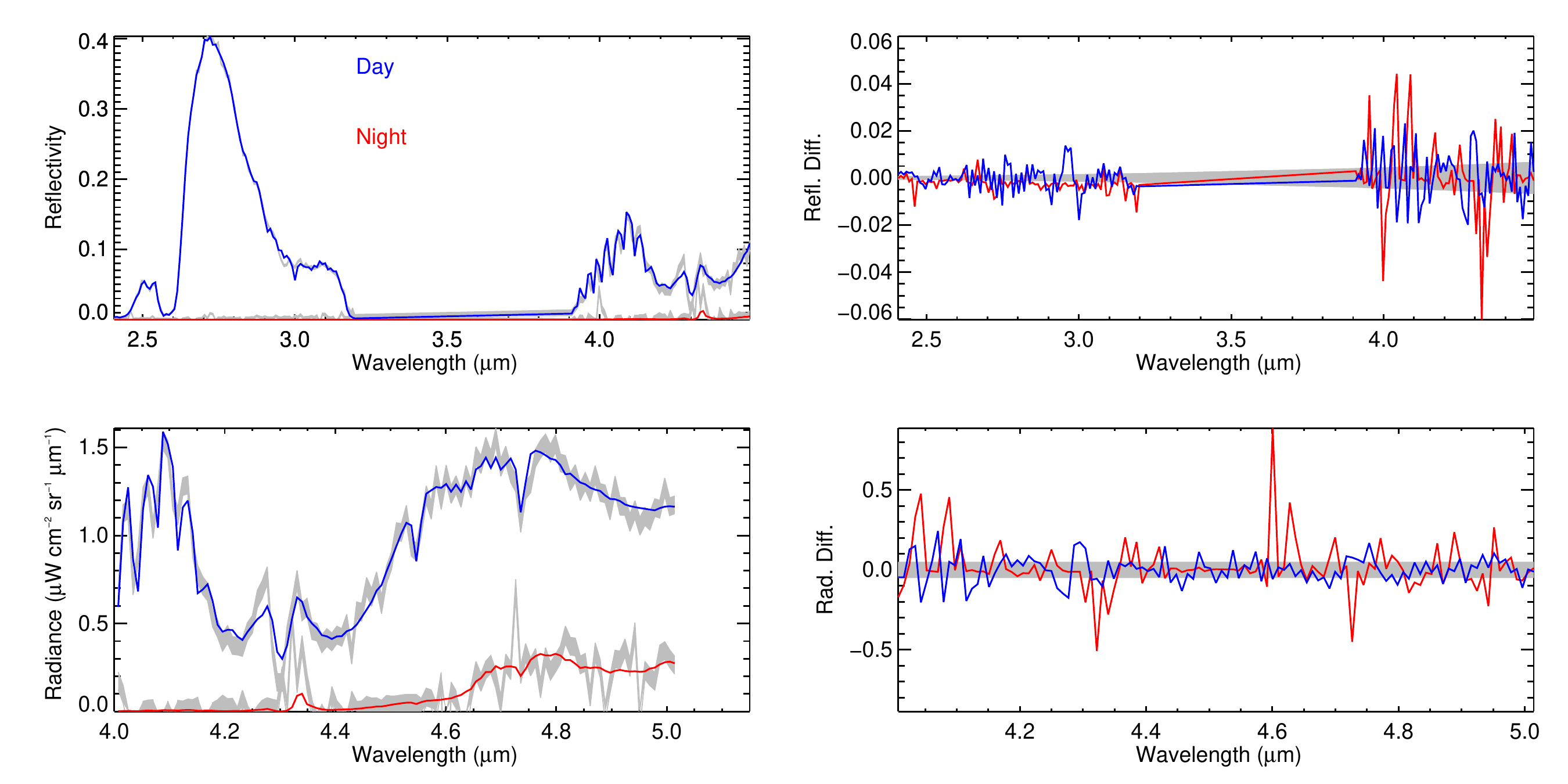}
    \caption{Simultaneous fit to two JIRAM-SPE spectra observed near the northern edge of the EZ in 2016 at 4.75$^\circ$N and 11.75$^\circ$E, one on nightside (viewing zenith angle 31.5$^\circ$) and one on dayside (viewing zenith angle of 75.1$^\circ$, solar zenith angle of 30.1$^\circ$, azimuth angle 94$^\circ$). The spectrum is split into two regions: 2.4 -- 4.4 $\mu$m reflectivity (top row); and 4.0 -- 5.0 $\mu$m radiance (bottom row). The observed spectra and error limits are shown in grey, while the fits to them using a model with three vertically-thin layers at 1.2, 0.4 and 0.09 bars are shown as red for the nightside spectrum and blue for the dayside. The left-hand column compares the measured and fitted spectra, while the right-hand column shows the differences between the measured and fitted spectra.} 
    \label{fig:JIRAM_NEB_fit}
\end{figure}

Having used the VIMS data to find a cloud model consistent with VIMS 5-$\mu$m spectra we then extended this analysis to selected JIRAM-SPE spectra. We first tested our model against a typical JIRAM-SPE 5-$\mu$m spectrum, covering just the 5-$\mu$m region and neglecting reflected sunlight. Using a model setup identical to that of \citet{grassi20} we were able to fit the spectrum to a similar level of accuracy and also verified that our wavelength calibration of the VIMS spectra is accurate (Figs.  \ref{fig:JIRAM_comparison} -- \ref{fig:jiram_vims_comparison}). We then tested our model against JIRAM-SPE spectra extracted over a wider wavelength range, including regions controlled by reflected sunlight. From the binned JM0003 JIRAM-SPE spectra, described in Section \ref{sec:JIRAMBIN} and summarised in Fig. \ref{fig:jiram_images}, we extracted spectra along a line running north-south at the northern edge of the EZ, and to the north west of the GRS. In these spectra, we neglected wavelengths below 2.4 $\mu$m where the calibration is less certain \citep{grassi21}. One bin in this range, at 4.75$^\circ$N and 11.75$^\circ$E, had a good measurement of both the nightside (viewing zenith angle 31.5$^\circ$) and dayside (viewing zenith angle of 75.1$^\circ$, solar zenith angle of 30.1$^\circ$, azimuth angle 94$^\circ$), and very low 5-$\mu$m emission, making it a good sample EZ observation. We fitted these two spectra simultaneously, using our full multiple-scattering NEMESIS model, assuming a three-cloud atmosphere with thin layers centred at $p_1 = 1.2$ bar (fixed), $p_2 = $ 0.7, 0.55 or 0.4 bar, and $p_3 = $0.25, 0.15, or 0.09 bar, respectively. We assumed the upper layer was composed of small particles with a gamma size distribution of mean radius $r=0.4$ $\mu$m, variance $\sigma = 0.05$ and we assumed the spectral properties were fixed with the imaginary refractive index set to 0.001 at all wavelengths. We assumed the lowest cloud was composed of very large particles ($r=40$ $\mu$m, $\sigma = 0.05$), in line with expectations from modelling at visible wavelengths, and arbitrarily set the size distribution of middle cloud to be  $r=0.4$ $\mu$m, $\sigma=0.3$. We retrieved the $n_\mathrm{imag}$ spectra for both the lower two clouds and also allowed the gas abundances to vary. We did not add any forward-modelling error in these retrievals and so the model was under-constrained, meaning that the retrieved parameters were not held closely to their priors and the retrieved $n_\mathrm{imag}$ spectra could vary freely, and perhaps unphysically with wavelength. However, this approach enabled us to quickly eliminate solutions that provided poor fits to the observations. The best fit was achieved with the middle cloud and haze at the lowest pressures sampled ($p_2 = 0.4$ bar  and $p_3 = 0.09$ bar, respectively) and is shown in Fig. \ref{fig:JIRAM_NEB_fit}; the retrieved cloud opacities (at 1.5 $\mu$m) in this case were 40.0, 2.05 and 0.02, respectively. 

\begin{figure}
	\includegraphics[width=\columnwidth]{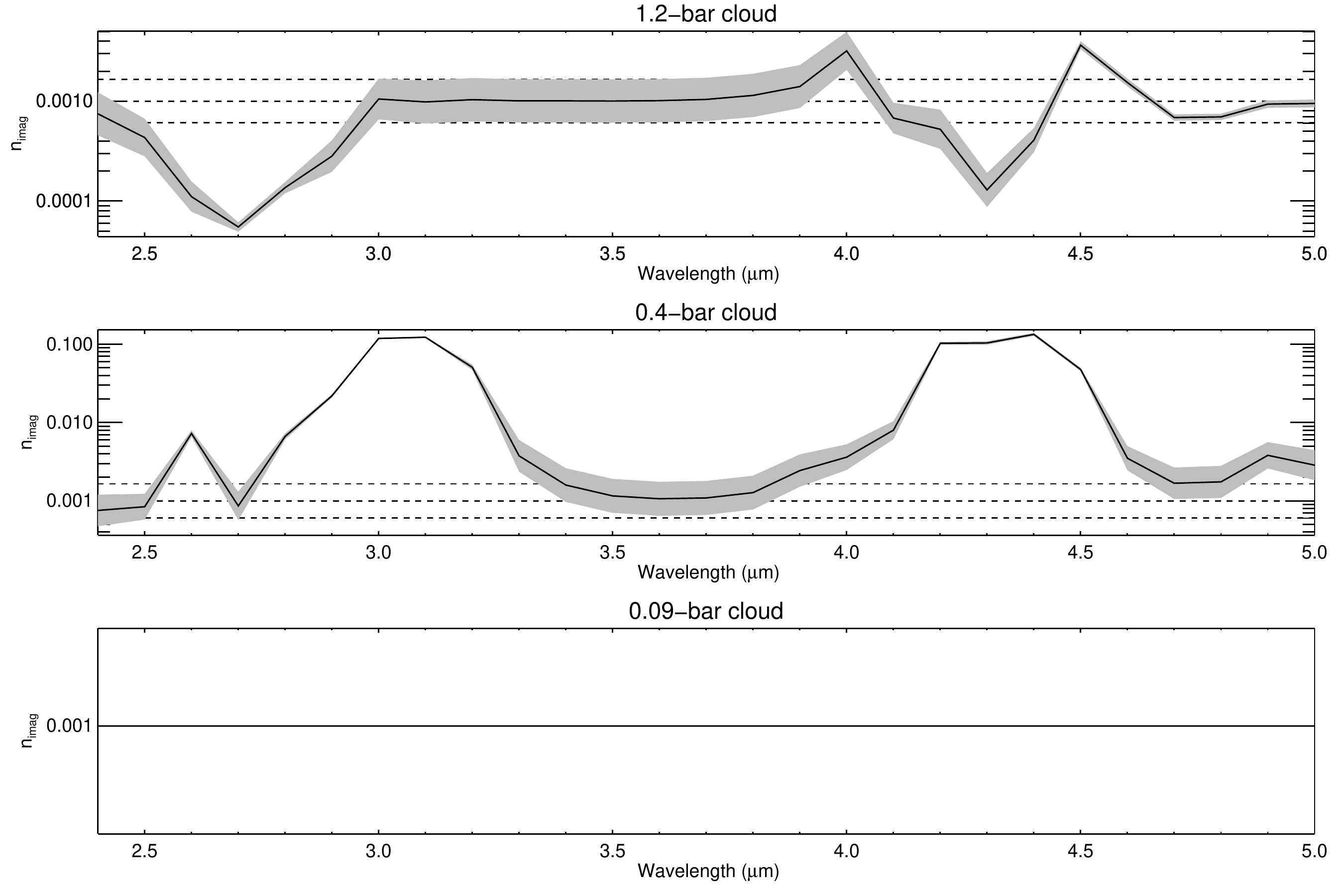}
    \caption{Prior and retrieved spectral variation of $n_\mathrm{imag}$ from the fit to the two JIRAM-SPE spectra shown in Fig. \ref{fig:JIRAM_NEB_fit}, with clouds assumed at 1.2 bar (top), 0.4 bar (middle) and 0.09 bar (bottom). Note these retrievals are very underconstrained and should be taken to represent indicative spectral variations, rather quantitative ones. The \textit{a priori} imaginary refractive indices and errors are shown as dotted lines, while the fitted values are indicated by solid lines within grey error envelopes. The $n_\mathrm{imag}$ spectrum of the uppermost haze (bottom row) was assumed to be invariant. } 
    \label{fig:JIRAM_NEB_nimag}
\end{figure}

The excellent fit to the dayside spectrum is partly due to NEMESIS finding that the aerosols at 0.4 bar have a strong absorption feature at 3 $\mu$m (Fig. \ref{fig:JIRAM_NEB_nimag}), which is consistent with the possible presence of large ammonia ice particles as noted by \citet{brooke98}, or another condensate that strongly absorbs at 3 $\mu$m \citep{sromovsky10a,sromovsky10b,biagiotti25}. However, as this fit is unconstrained, the good fit to the
PH$_3$-absorption region (4.1 -- 4.6 $\mu$m) has been achieved by also increasing the $n_\mathrm{imag}$ of the 0.4-bar cloud here, resulting in a very low retrieved PH$_3$ abundance, which we know from other studies at mid-infrared wavelengths is unphysical. Similarly, the retrieved abundance of NH$_3$ is lower than expected, since much of the absorption has been aliased by the imaginary refractive index spectrum of the 0.4-bar cloud. This analysis underlines how misleading it can sometimes be to run unconstrained retrievals over limited spectral regions. In the 5-$\mu$m region, the main absorption comes from the relatively flat retrieved $n_\mathrm{imag} \sim 0.001$ spectrum of the 1.2-bar cloud, combined with its very large opacity. Most of the radiation we see on the dayside in this region is due to reflection from the 0.4-bar cloud, and it can be seen that this has increased $n_\mathrm{imag}$, and thus lower single-scattering albedo from 4.7 to 5.0 $\mu$m, noted earlier to be necessary to match the VIMS-IR dayside spectrum in this range. Looking at the known scattering properties of candidate aerosols (Fig. \ref{fig:refractive_indices}), it can be seen that water liquid/ice has the highest absorption at 5-$\mu$m wavelengths of any of the usually considered condensate candidates. Combined with the very high single-scattering albedo of water ice/liquid at visible wavelengths, a condensate with properties similar to water ice/liquid again emerges as an attractive candidate for the condensate in the deeper layer at $p \sim 1.2$ bar, while the presence of a strong absorption at 3 $\mu$m makes a condensate with properties similar to ammonia ice an attractive candidate for the condensate in the upper cloud layer. Whether this upper cloud is ammonia ice, or something like tholins \citep{khare93}, or indeed anything showing this characteristic N--H feature \citep{biagiotti25} is unclear, although the previously noted scarcity of SIACs suggests pure ammonia ice may be unlikely.

\section{Final retrieval model}\label{sec:final_retrieval}

\subsection{Preliminary Modelling Conclusions}

Based on the preliminary analyses presented in Section \ref{Section:Initial} we can see that there is very good evidence to conclude that there are three main cloud components necessary to model the Jovian spectra:

\begin{enumerate}
    \item A cloud at 1 -- 2 bar, composed of large particles ($r \sim 10 \mu$m) that are highly scattering at visible wavelengths and rather absorbing at 5-$\mu$m wavelengths, possibly consistent with water ice or another unknown condensate with these properties;
    \item A second optically thinner cloud at 0.4 -- 0.7 bar, perhaps composed of large particles ($r \sim 10 \mu$m), which have a strong absorption feature at 3 $\mu$m, possibly consistent with ammonia ice or another condensate that has a strong N--H absorption band \citep{biagiotti25}.
    \item A chromophore layer to account for the absorption seen at blue visible wavelengths, which our initial modelling suggests is most likely to be concentrated in the lower layer.
\end{enumerate}

As noted earlier, when analysing ISO 3-$\mu$m spectra, \citet{brooke98} modelled the upper clouds as moderately large ($r \sim 10$ $\mu$m) ammonia ice particles. In addition, \citet{fletcher09} converged on a 800-mbar layer of 10-$\mu$m sized particles, possibly NH$_3$ ice, to match Cassini/CIRS observations and \citet{harkett24} placed a single cloud at 1 bar, with 5--10 $\mu$m radii particles, which allowed them to achieve good fits to JWST/MIRI data. Hence, the conclusion of the presence of large radii particles at $p < 1$ bar seems to be a growing consensus. These particles are unlikely to be pure ammonia ice, however, otherwise SIACs (spectrally-identifiable ammonia clouds) would be widespread and not, in fact, rather uncommon. Such large particles have a cross-section that will not drop too rapidly across the combined MUSE/VIMS range, making these particles also an attractive candidate for what has historically been known as the upper level `haze', which can clearly be seen in MUSE, VIMS and JIRAM-SPE spectra in all strongly-absorbing methane bands from 0.8 to 4.0 $\mu$m. The similar reflectivity seen over a wide wavelength range suggests these particles have a radius larger than $\sim$ 1 $\mu$m (the opacity of small particles would fall rapidly with wavelength). In addition, at methane-absorbing wavelengths Jupiter's disc is still seen to be limb-darkened, not limb-brightened, which suggests that the `haze' is vertically confined with a density that drops quickly with altitude, making it very dissimilar from the photochemical haze layers seen in other planets, which are typically vertically extended and show limb-brightening. A large size for the particles present in a putative photochemical haze created at low pressures also seems unlikely. Could this `haze' layer instead perhaps be merely the top of condensate with properties similar to ammonia ice, seen not as SIACs, but combined with something including a `ubiquitous' 3-$\mu$m absorption \citep{sromovsky10b, biagiotti25}? The aerosol may then be more visible in the EZ because this cloud is optically thicker and/or more vertically extended. This interpretation is supported by the fact that from JWST/NIRCam \citet{hueso23} distinguished discrete features in this layer, which they used for cloud tracking. Such discrete clouds are suggestive of a convective origin for these particles, rather than they being an extended photochemical haze. Finally, as can be seen in Figs. \ref{fig:jupiter_muse} and \ref{fig:jiram_images}, the upper level `haze' shows remarkable correspondence with convective features at lower levels, making it more likely to be generated from below, rather than settling down from above. Hence, we believe the `haze' may actually be the top of an optically-thin cloud, with spectral properties similar to ammonia ice, and perhaps analogous to the upper tropospheric water ice cirrus clouds seen in Earth's atmosphere.

Assuming either the ammonia hydrosulphide or `mushballs' hypotheses for cloud formation in the 1 -- 2 bar pressure region, we would expect the mole fraction of ammonia to decrease sharply in this pressure region and then remain constant until reaching its condensation level at a lower pressure. We thus placed two main clouds in our final model: one coinciding with an initial reduction of ammonia mole fraction at 1 -- 2 bar; and one coinciding with the ammonia condensation level at lower pressures and ran simulations to determine the required spectral properties of these two clouds, and also a chromophore, which we placed within the lower cloud, and various other parameters as we shall describe. The resulting simulations gave good results and are described in the following Sections \ref{Section:EZ-NEDF-NEB} and  \ref{Section:EZ-NEB-Minnaert}.

\subsection{Retrievals of typical combined EZ,  NEB and NEDF spectra}\label{Section:EZ-NEDF-NEB}

We first applied our retrieval model to typical single-angle spectra extracted from the Equatorial Zone (EZ), North Equatorial Belt (NEB) and a North Equatorial Dark Feature (NEDF). For the EZ and NEB spectra we used our Minnaert analyses of the VLT/MUSE and Cassini/VIMS-IR cubes to generate the spectra seen with both the Sun and observer at zenith. We used VLT/MUSE for wavelengths less than 0.93 $\mu$m and Cassini/VIMS-IR above. Known problematic wavelengths in the VIMS-IR spectra \citep[listed in ][and noted in Fig.  \ref{fig:vims_calibration}]{clark18}  were omitted and the remaining spectra sampled at 180 wavelengths. We sampled the VLT/MUSE spectra at wavelengths less than 0.93 $\mu$m with the same number of wavelengths to give equal weight to the MUSE and VIMS-IR spectra, giving combined spectra with 360 wavelengths in total. The spectral radiance errors were set to either 0.9\% of the reflected sunlight from a perfect Lambertian scatterer, or a brightness temperature error of 0.75 K, whichever was larger (N.B., the brightness temperature errors dominate at 5 $\mu$m). These error bars were chosen mostly to incorporate the `forward-modelling' uncertainty of our retrieval model, rather than being purely random measurement noise. The measurement errors are very small and we found that our model could not fit to this level of uncertainty. However, the optimal estimation algorithm we use is predicated on being able to fit the data to within some estimated error otherwise the constraints used are ineffective. Hence, we add this `forward-modelling error' to enable NEMESIS to achieve an optimal, constrained fit. For the NEDF spectrum, we used the same MUSE and VIMS-IR spectra described in Section \ref{Section:NEDF}. The NEDF spectrum was sampled at the same wavelengths as our EZ and NEB spectra and assigned errors in the same way. The assumed geometry of the combined NEDF spectrum was also set to zenith for both observer and Sun since the NEDF regions were located near the centre of the disc for both datasets.

Since the Galileo Entry probe found the ammonia abundance to increase with pressure \citep{sromovsky98,folkner98} and since in the 1--2 bar range we suspect that ammonia combines with either H$_2$S to form NH$_4$SH, or perhaps with H$_2$O to eventually form `mushballs', we adopted the following combined cloud/ammonia model. The ammonia mole fraction profile was assumed to have a deep value, then drop to an intermediate value at pressure $\sim 1-2$ bar. This pressure was determined by the retrieval model and a lower cloud, `Cloud-1', based at this level, with a fixed fractional scale height of 0.25 above. The ammonia mole fraction was then held at a fitted intermediate value above this cloud until the pressure was low enough that the partial pressure exceeded the saturated vapour pressure, calculated from our reference temperature-pressure profile. The base of a second upper cloud, `Cloud-2', was set to this condensation pressure and had a variable fractional scale height above, with an \textit{a priori} fractional scale height of $0.25 \pm 0.05$. Above this level, the ammonia mole fraction was determined by a variable condensation-level ($z_0$) relative humidity of $R_H(z_0) = 0.5 \pm 0.1$, which was assumed to decrease with altitude above the condensation level as $R_H(z)=R_H(z_0)\exp({-(z-z_0)/H})$, where $H$ was set to 10 km. This was used to account for photolytic destruction of ammonia in the upper troposphere and ensured that ammonia abundance decreased rapidly above the condensation level, which was found necessary to suppress some of its absorption features in the VIMS-IR spectral range. Finally, we added a chromophore profile which we assumed to be based at 1.25 bar (fixed), with a fixed fractional scale height of 0.25. The opacities of lower (Cloud-1), upper (Cloud-2) and the chromophore layers were all fitted by NEMESIS, as were several other parameters, listed in Table \ref{reftable1}. 

As noted in Section \ref{sec:part-scatter} we assumed Gamma size distributions for all our particulates and settled on the following values of the mean particle radius $r$ and the variance $\sigma$ of the Gamma size distribution: 1) lower cloud (Cloud-1): $r = 10$ $\mu$m, $\sigma=0.05$; 2) upper cloud (Cloud-2): $r = 10$ $\mu$m, $\sigma=0.05$; and 3) chromophore: $r = 0.2$ $\mu$m, $\sigma=0.1$. The prior imaginary refractive index spectra were set to those of water ice \citep{warren08} for the lower cloud (Cloud-1), ammonia ice \citep{martonchik84} for the upper cloud (Cloud-2), and for the chromophore, the chromophore spectrum of \citet{alexander24,Alexander2026}, which is closely related to the `Carlson chromophore' \citep{carlson16}. For Cloud-1 and Cloud-2 we limited the minimum imaginary refractive indices to $10^{-4}$ to ensure that the observations led our model to \textbf{require} small visible imaginary indices at visible wavelengths, as we expect from our MUSE cloud-top pressure analysis (Section \ref{sec:band-depth}), rather than prescribing these at the start. The real refractive indices were fixed at 1.5 $\mu$m to values of 1.296, 1.407 and 1.4 for Cloud-1, Cloud-2, and the chromophore, respectively, and computed from the fitted $n_\mathrm{imag}$ spectra at other wavelengths using the Kramers-Kronig relation. The \textit{a priori} $n_\mathrm{imag}$ errors for all three cloud components were set to 50\%. The cross-section, single-scattering albedo, and phase function spectra were all self-consistently calculated from the fitted $n_\mathrm{imag}$ spectra as outlined in Section \ref{sec:part-scatter}. Varying the $n_\mathrm{imag}$ spectra of all three aerosol layers simultaneously makes the retrieval potentially prone to degeneracy between these parameters, and also between these $n_\mathrm{imag}$ parameters and retrieved gaseous abundances. We hope to explore issue this more carefully in a future paper utilising the `Nested-sampling' capability of the new archNEMESIS model (see Conclusions). In the meantime, we believe the $n_\mathrm{imag}$ spectra to be mostly decoupled as these clouds affect very different parts of the spectrum. The chromophore, made of small particles, only significantly affects the spectrum at wavelengths less than 600 nm. The lower and upper clouds (Cloud-1 and Cloud-2) have significant reflectivity over a wide range of wavelengths, but the lower cloud (Cloud-1) can only be seen at continuum wavelengths, making the effects of these two layers easier to decouple. 

The mole fraction profile of phosphine was assumed to be fixed up to a `knee' pressure of 1.0 bar and decrease at lower pressures with a fractional scale height (FSH) of 0.3. This is an assumption that has on onwards effect on the retrieved aerosol opacity at pressures less than 1 bar and also on the retrieved $n_\mathrm{imag}$ spectrum of Cloud-2, but these effects are non-trivial to decouple. \citet{harkett24} faced similar difficulties in their analysis of JWST/MIRI spectra and eventually chose a `knee' pressure of 0.8 bar, although at these longer wavelengths it was possible to retrieve spatial variations of FSH. Again, a more thorough exploration of possible degeneracies will be attempted in a future study using Nested Sampling. The water profile was assumed here to be constant at depth, and limited to a maximum relative humidity of 100\% at lower pressures. Interestingly, for the cases (NEB, NEDF) where we are able to see through the lower cloud layer (Cloud-1) to detect appreciable thermal emission from the deep atmosphere, we retrieve low deep water abundances, with the profiles only reaching saturation at $\sim$2 bar, close to the fitted Cloud-1 base pressure. The abundance of all other gases was assumed to not vary with height. 

\begin{figure}
	\includegraphics[width=\columnwidth]{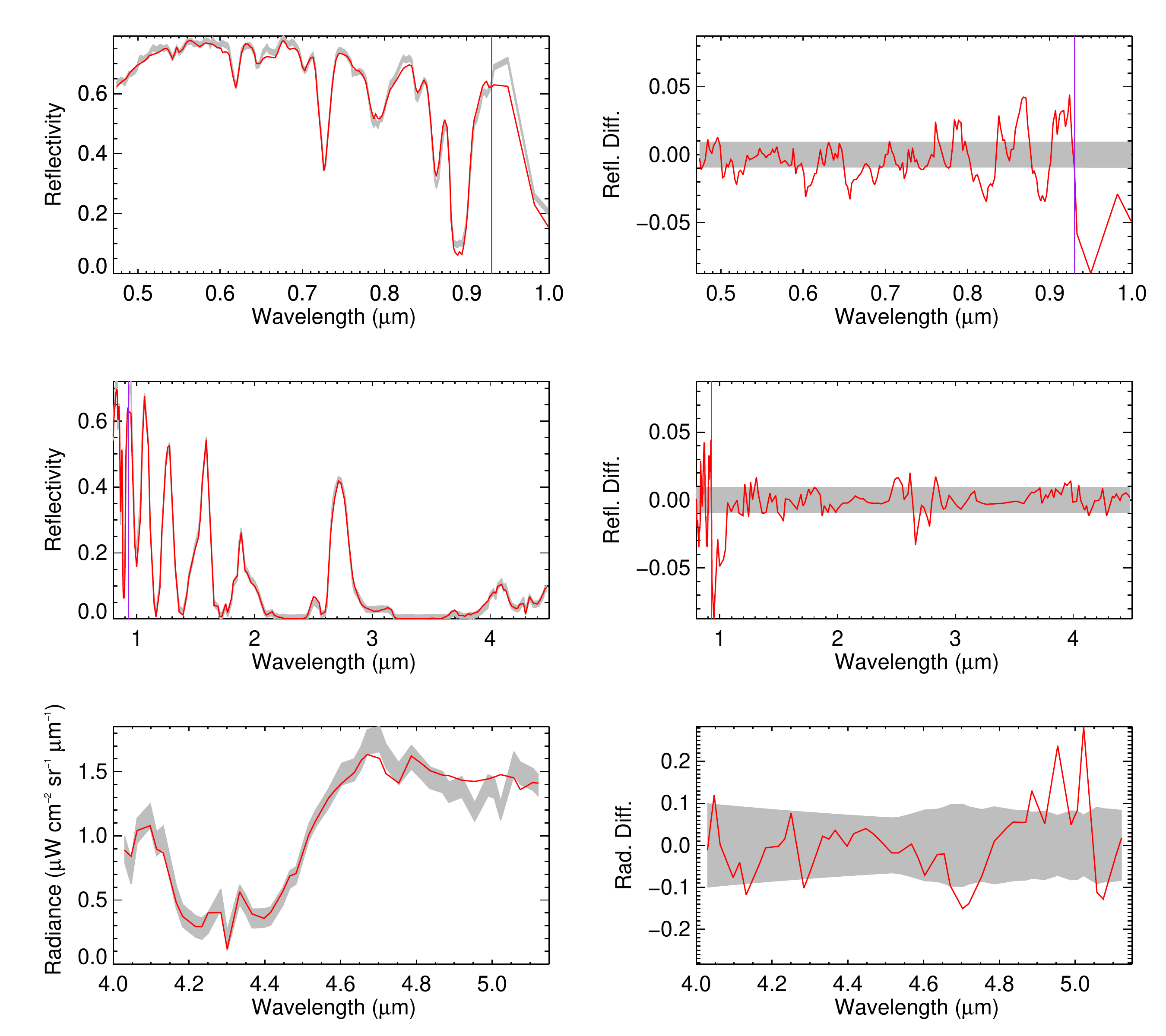}
    \caption{Fit to combined MUSE/VIMS EZ $0^\circ -$zenith angle spectrum ($\chi^2/n = 2.61$). The combined spectral range has been split into separate sections to aid comparison. Top row shows the measured and fitted reflectivity spectra (left) and difference (right) for the range 0.475 to 1.0 $\mu$m, which covers the MUSE range and the lower part of VIMS-IR. The transition between the MUSE and VIMS-IR data is indicated by the vertical purple line at 0.93 $\mu$m. The measured spectrum and assumed error range is shown in grey, and the fitted spectrum is shown in red. The middle row compares the reflectivity spectra from 0.8 to 4.5 $\mu$m, while the bottom row compares the \textbf{radiance} spectra from 4 to 5.15 $\mu$m.}    \label{fig:fitspecEZ}
\end{figure}

\begin{figure}
	\includegraphics[width=\columnwidth]{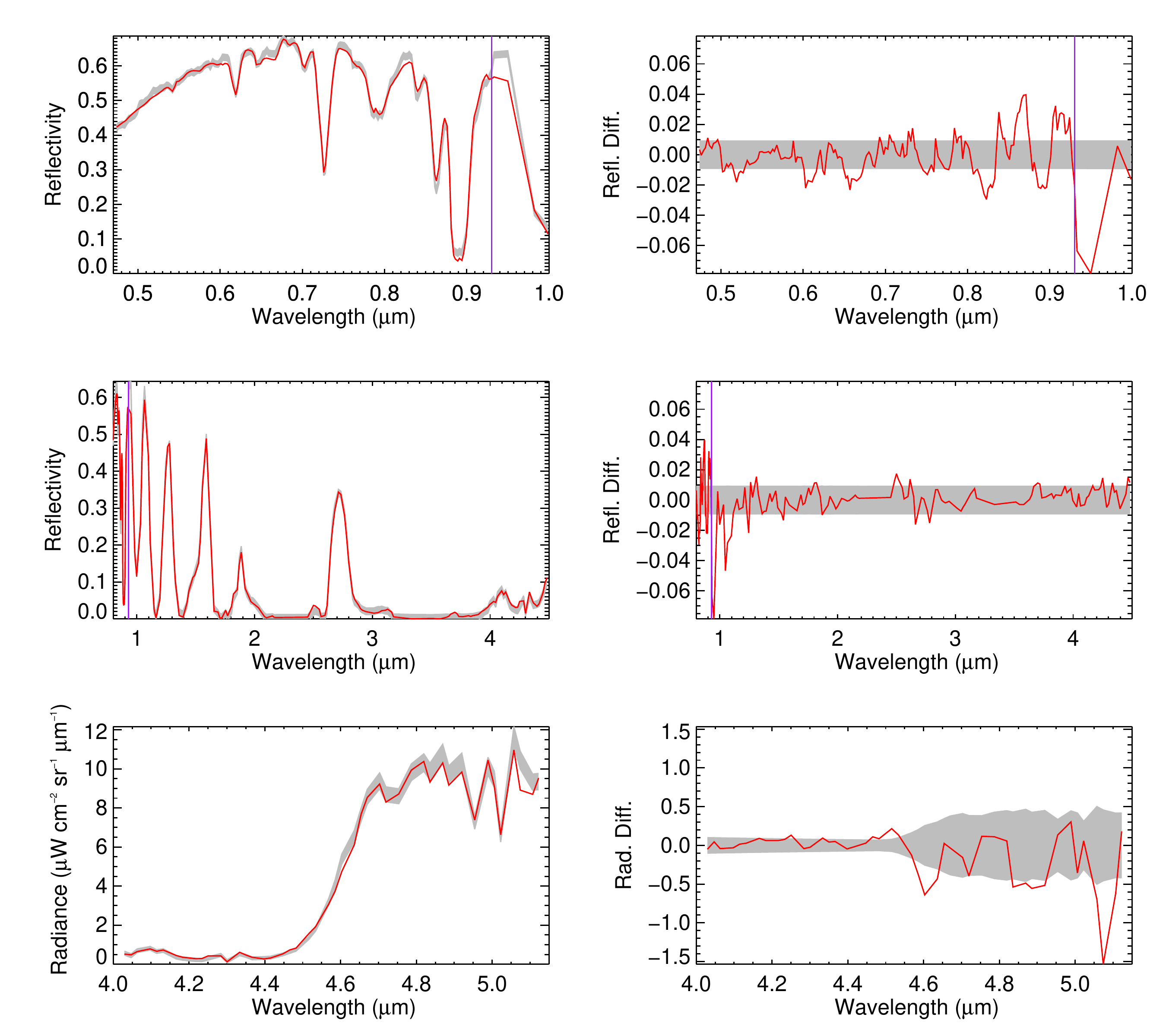}
    \caption{As Fig. \ref{fig:fitspecEZ}, but showing fit ($\chi^2/n = 1.88$) to the combined MUSE/VIMS NEB $0^\circ -$zenith angle spectrum.}    \label{fig:fitspecNEB}
\end{figure}

\begin{figure}
	\includegraphics[width=\columnwidth]{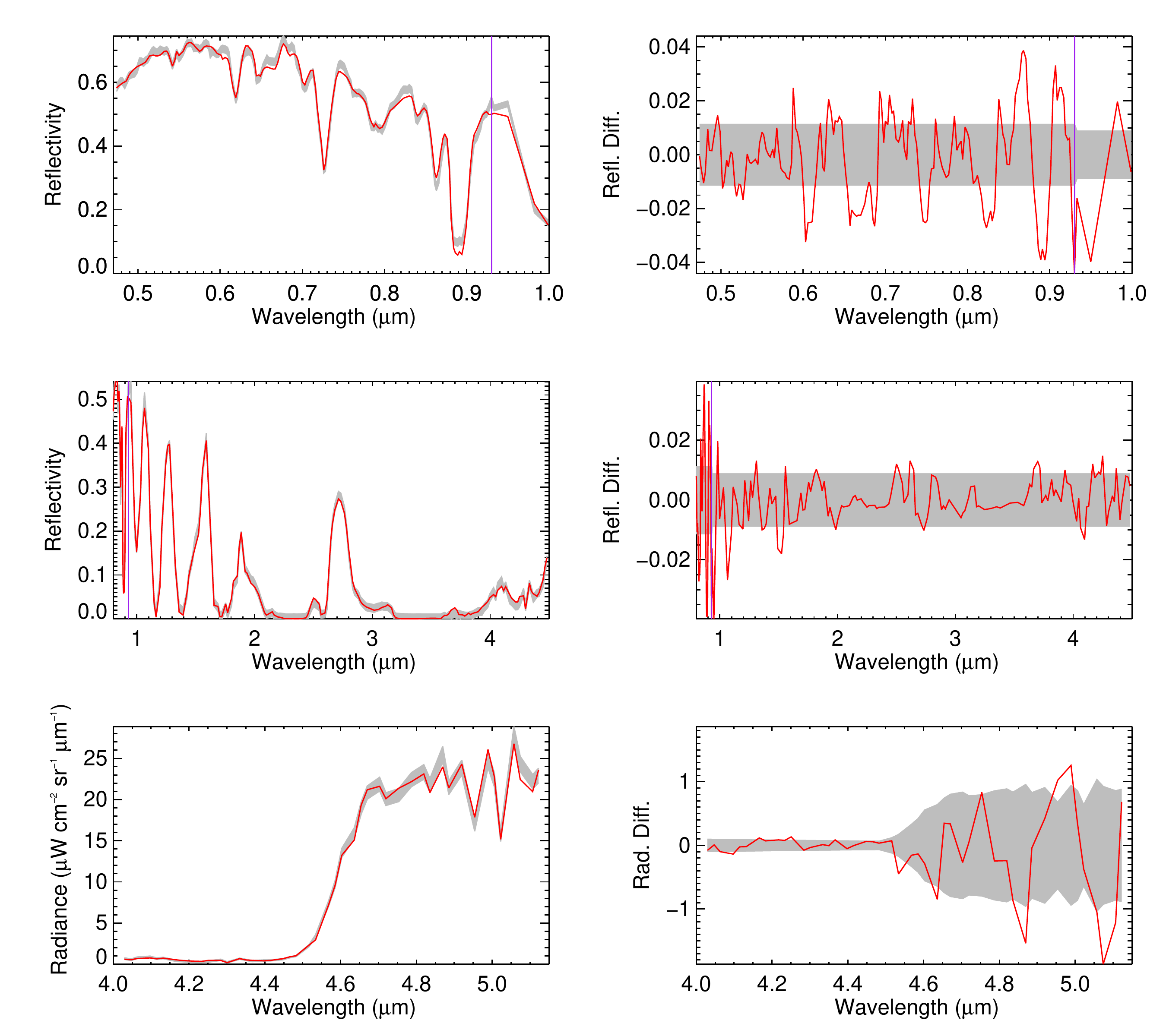}
    \caption{As Fig. \ref{fig:fitspecEZ}, but showing fit ($\chi^2/n = 1.44$) to the combined MUSE/VIMS NEDF $0^\circ -$zenith angle spectrum.}    \label{fig:fitspecNEDF}
\end{figure}

\begin{table*}
\caption{Retrieved quantities from our fits to EZ, NEB and NEDF spectra}
\begin{tabular}{llllll}
\hline
 & EZ  & NEB   & NEDF  & EZ (2-ang) & NEB (2-ang)  \\
\hline
Lower cloud opacity $\tau_1$ & 38.17(2.97) & 23.14(1.32) &  6.18(0.48) & 121.15(5.14) & 56.26(1.62) \\
Lower cloud pressure (bar) $p_1$ & 2.12(0.04) & 1.96(0.04) & 1.55(0.04) & 2.15(0.02) & 1.85(0.02) \\
Upper cloud opacity $\tau_2$ & 1.61(0.05) & 0.83(0.04) & 1.12(0.08) & 2.15(0.08) & 1.05(0.04) \\
Upper cloud pressure (bar) $p_2$ & 0.57 & 0.56 & 0.54 & 0.54 & 0.51 \\
Upper cloud FSH $f_2$ & 0.39(0.05) & 0.24(0.04) & 0.37(0.05) & 0.32(0.03) & 0.28(0.04) \\
Chromophore opacity $\tau_3$ & 0.22(0.05) & 0.32(0.04) & 0.38(0.02) & 0.11(0.02) & 0.20(0.02) \\
Chromophore pressure (bar) & 1.25 & 1.25 & 1.25 & 1.25 & 1.25  \\
NH$_3$(d) (ppm) & 497(117) & 160(42) & 191(26) & 1847(1067) & 134(38) \\
NH$_3$(m) (ppm) & 61.4(5.1) & 45.4(3.8) & 50.4(9.3) & 37.8(2.6) & 27.1(2.0) \\
NH$_3$ RH(\%) & 47(5) & 49(5) & 46(5) & 47(5) & 48(5) \\
PH$_3$(d) (ppm) & 6.91(1.39) & 0.51(0.08) & 0.26(0.04) & 10.42(1.32) & 0.48(0.05) \\
H$_2$O(d) (ppm) & 3.30(3.30) & 8.88(2.39) & 4.53(0.74) & 3.42(3.42) & 8.74(2.01) \\
AsH$_3$ (ppb) & 0.30(0.29) & 0.23(0.18) & 0.28(0.20) & 0.24(0.22) & 0.15(0.11) \\
GeH$_4$ (ppb) & 0.07(0.07) & 0.06(0.05) & 0.08(0.05) & 0.06(0.06) & 0.03(0.02) \\
CO (ppb) & 0.73(0.70) & 0.31(0.29) & 0.64(0.42) & 0.48(0.46) & 0.15(0.15) \\
CH$_3$D/CH$_4$ ($\times 10^{-5}$) & 3.15(2.90) & 3.08(1.99) & 6.74(3.00) & 3.08(2.68) & 1.26(0.94) \\
Cloud-1 $n_\mathrm{imag}$(4.8 $\mu$m) $(\times 10^{-3})$ & 8.3 & 0.6 & 7.0 & 2.6 & 0.06 \\
Cloud-1 $\varpi$ (4.8 $\mu$m)  & 0.83 & 0.98 & 0.85 & 0.94 & 0.99 \\
\hline
\end{tabular}
\begin{tablenotes}
\small
\item Formal retrieved errors are indicated in  (brackets).
\item All opacities are those at a reference wavelength of 1.5 $\mu$m.
\item All \textit{a priori} errors were set to 100\%, except for $p_1$ $(0.25 \pm 0.05)$, NH$_3$ RH ($50\pm 5$\%), and $f_2$ $(0.25 \pm 0.05)$.
\item There is no listed error for $p_2$ as this is computed from NH$_3$(m) and the assumed temperature-pressure profile.
\item There is no listed error for $p_3$ as this was fixed at the assumed chomophore layer base presure of 1.25 bar.
\item The cases where the model was fitted simultaneously to observations at zenith angles of 0$^\circ$ and 42.37$^\circ$ are indicated by (2-ang) in the heading.
\end{tablenotes}
\label{reftable1}
\end{table*}

Our fits to the EZ, NEB and NEDF spectra using this model are shown in Figs. \ref{fig:fitspecEZ}, \ref{fig:fitspecNEB}, and Fig. \ref{fig:fitspecNEDF}, respectively. As can be seen the agreement between the measured and modelled spectra is generally very good. There seems to be a slight discontinuity between the MUSE and VIMS data at 0.93 $\mu$m for all three cases, where the two datasets have been combined,  but the model reproduces the observed peaks and troughs mostly very well.  The fitted parameters from our retrievals are shown in Table 1 and the fitted cloud/ammonia profiles and $n_\mathrm{imag}$ spectra are shown in Figs. \ref{fig:gas_cloud_oneang} and \ref{fig:nimag_oneang}, respectively. 

\subsubsection{EZ and NEB}
\begin{figure*}
	\includegraphics[width=\textwidth]{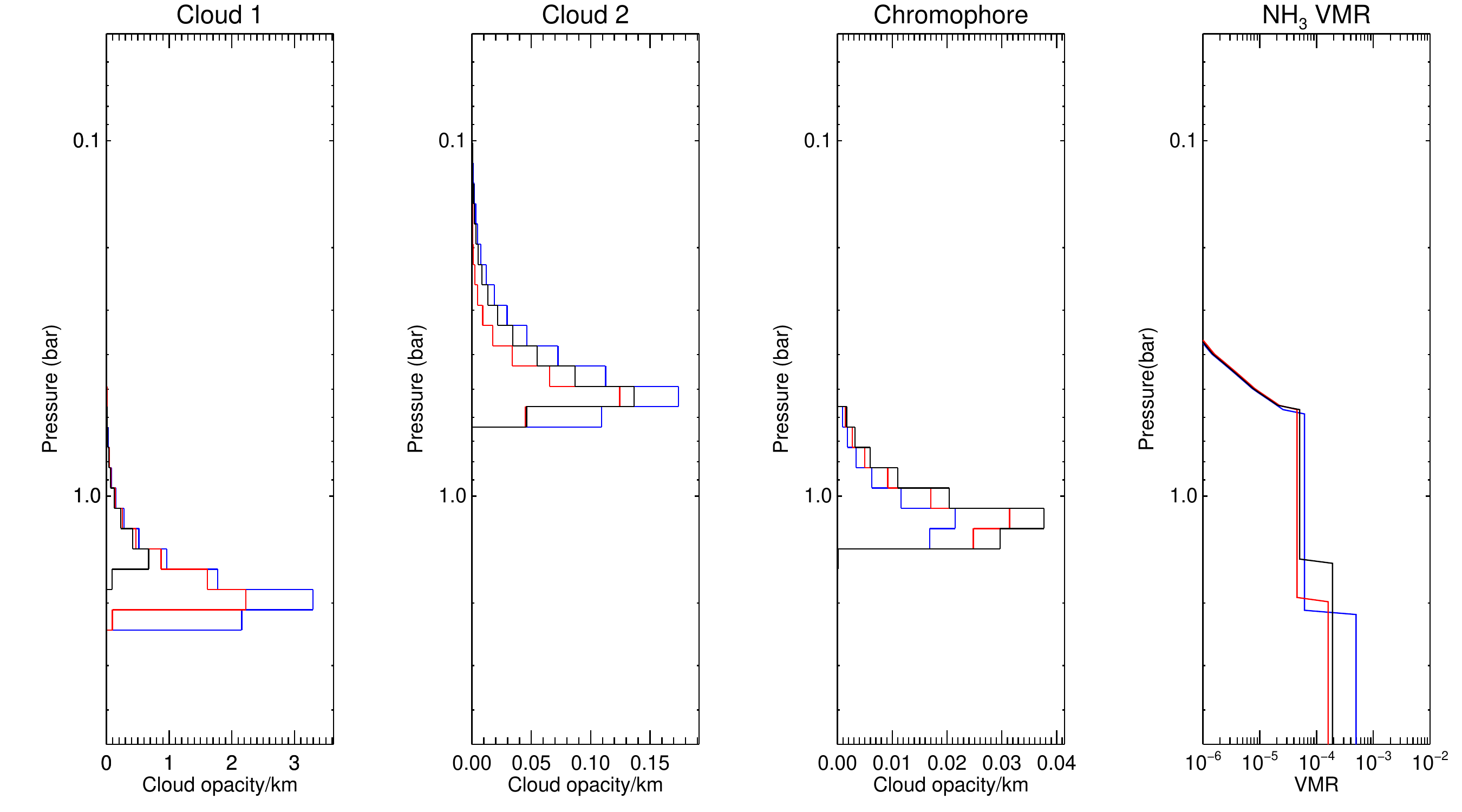}
    \caption{Cloud and ammonia profiles fitted to observed $0^\circ -$zenith angle spectra in the EZ (blue), NEB (red) and NEDF (black). The left hand panel shows the lower cloud profiles (Cloud-1), the second panel shows the upper cloud profiles (Cloud-2), while the third panel shows the chromophore profiles. The right hand panel shows the ammonia mole fraction profiles. }
    \label{fig:gas_cloud_oneang}
\end{figure*}

\begin{figure}
	\includegraphics[width=\columnwidth]{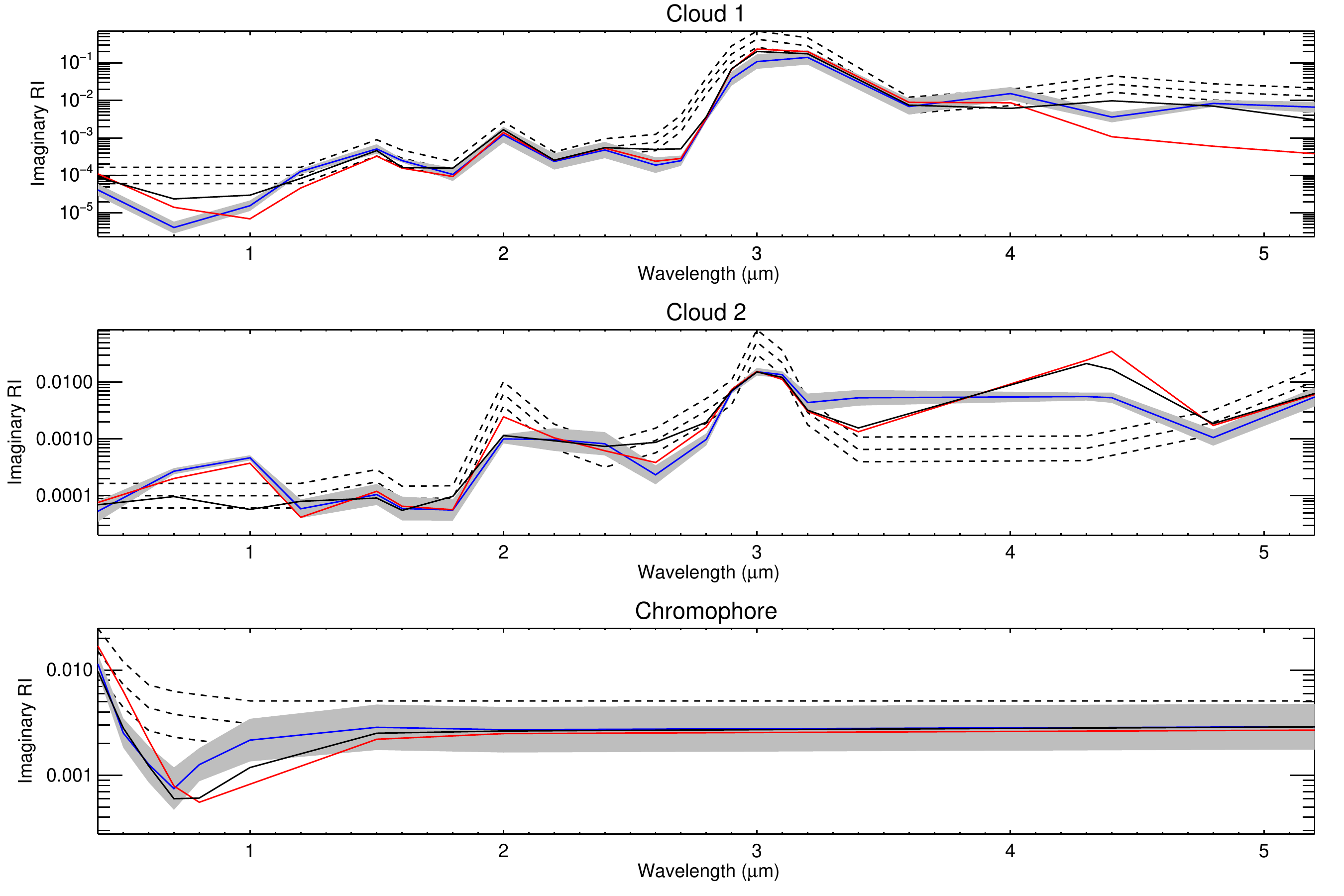}
    \caption{Spectral variations of $n_\mathrm{imag}$ fitted to the combined nadir MUSE/VIMS $0^\circ -$zenith angle spectra in the  EZ (blue), NEB (red) and NEDF (black). The \textit{a priori} spectra and errors are indicated by the dashed lines. For clarity, only the retrieved error range on the EZ spectrum is shown in grey. The top panel shows the fitted lower cloud (Cloud-1) $n_\mathrm{imag}$ spectra, the middle panel shows the fitted upper cloud (Cloud-2) $n_\mathrm{imag}$ spectra, and the bottom panel shows the fitted chromophore spectra. }
    \label{fig:nimag_oneang}
\end{figure}

Considering first the EZ and NEB, we can see that the solutions favour a cloud base of $\sim$ 2 bar for the lower cloud (Cloud-1) with large opacity for the EZ $(\tau = 38)$ reducing to $\tau = 23$ in the NEB. The vastly different 5-$\mu$m brightness is then accounted for by the strong reduction in $n_\mathrm{imag}$ at 5 $\mu$m for Cloud-1 from $8.3\times 10^{-3}$ in the EZ to $6.0\times 10^{-4}$ in the NEB, increasing the single-scattering albedo from $\varpi = 0.83$ to $\varpi = 0.98$ and thus allowing much more thermal radiation to escape from deep levels. The observed spectrum from 4 to 4.5 $\mu$m in the EZ is found to come entirely from reflection from the overlying upper cloud layer (Cloud-2), with increasing contribution at wavelengths $> 4.5$ $\mu$m from reflection from the lower cloud (Cloud-1) and thermal emission. Longwards of 4.5 $\mu$m the upper cloud (Cloud-2) reflection is modelled as decreasing to negligible amounts at 5.15 $\mu$m, whereas the reflection from lower cloud (Cloud-1) increases over the same range. Hence, in this retrieval the mysterious `absorption' at 4.9 $\mu$m, identified in Section \ref{sec:vims-5micron}, cannot be attributed to the absorption of a single layer, but instead results from the interplay between the reflection spectra of the two cloud layers. However, since this retrieval is trying to fit all wavelengths similarly, the match to the observed 4.9 $\mu$m feature is not as convincing as the underconstrained JIRAM-SPE case presented in Fig. \ref{fig:JIRAM_NEB_fit}, where a cloud at a pressure similar to our Cloud-2 was clearly found to be responsible for this feature.  Isolating the location of the 4.9 $\mu$m feature will be addressed further in future work using Nested Sampling. The opacity of the upper cloud (Cloud-2) is retrieved to be $\tau = 1.61$ in the EZ and  $\tau = 0.83$ in the NEB, and can be seen in Fig. \ref{fig:gas_cloud_oneang} to extend higher in the atmosphere in the EZ. The retrieved opacity of the upper cloud is much less than that of the lower cloud, consistent with the hypothesis that the upper cloud (Cloud-2) is a thin `cirrus'-like layer, overlying the main, deeper, and optically thicker Cloud-1 layer. 

In the visible wavelength range, the solution favours very low $n_\mathrm{imag}$ for the lower cloud (Cloud-1), especially in the EZ, which confirms our suspicions that large, highly scattering particles are needed in this wavelength range to allow light to scatter to deep levels and be Rayleigh-scattered back, increasing the equivalent widths of the observed visible-wavelength methane and ammonia bands. The retrieved $n_\mathrm{imag}$ spectra of both main clouds (Cloud-1 and Cloud-2) have large absorptions near 3 $\mu$m, but with the constraint much stronger on the the upper cloud (Cloud-2) $n_\mathrm{imag}$ spectrum. This reduces the reflectivity of the upper cloud in this range, which is necessary to match the observed spectra as identified in the ISO analysis of \citet{brooke98}. This spectral feature is consistent with the particles in the upper cloud (Cloud-2) containing a significant opacity of something like ammonia ice, but without the narrow features at shorter wavelengths of SIACs. 

Finally, the chromophore layer is retrieved to have an opacity of  $\tau = 0.22$ in the EZ and $\tau = 0.32$ in the NEB. Hence, the increase in redness of the NEB appears to be mostly caused by a  reduction in the opacity of the bright particles in the lower cloud (Cloud-1) rather than an increase in the opacity of the chromophore particles. However, the retrieved $n_\mathrm{imag}$ chromophore spectrum for the NEB is considerably higher than for the EZ and is well constrained at shorter wavelengths. Hence, even if the chromophore particles are not especially more abundant in the NEB, they are darker and more blue-absorbing. The calculation is largely insensitive to these small chromophore particles at longer wavelengths with retrieved errors identical to \textit{a priori}. 

It can be seen that the dominant cloud opacity is that of the lower cloud (Cloud-1), based at 1--2 bar, in agreement with previous observations and analyses, and is approximately 10 times more optically thick than the upper cloud (Cloud-2) at $\sim 0.5$ bar, which as we suggested earlier may be something akin to an ammonia ice `cirrus' cloud, and 100 times thicker than the chromophore layer. It is for this reason that we show the cloud profiles for the three cloud types on separate panels in Fig. \ref{fig:gas_cloud_oneang}. 

For ammonia, we see large deep abundances retrieved below the lower cloud (Cloud-1), albeit with large retrieval errors, with enhanced abundance in the EZ $(497\pm117)$ compared with $160\pm42$ in the NEB, values which are consistent with previous Juno/JIRAM-SPE determinations \citep{grassi20} and more consistent with Juno/MWR \citep{guillot20li} and VLA \citep{moeckel23} values (discussed in section \ref{sec:ammonia_comp}). The mid-level ammonia abundance, between Cloud-1 and Cloud-2 is much better constrained and shows lower values in the NEB of $45.4\pm3.8$ ppm, compared with $61.4\pm5.1$ in the EZ, both leading to similar ammonia ice condensation pressures of $\sim$0.56 bar. The profiles for both latitudes are very similar above the condensation level and it can be seen that the relative humidity is barely changed from \textit{a priori}, both in terms of value and error, showing that our model is largely insensitive to this value. 

The deep abundances of trace gases, shown in Table \ref{reftable1}, can be seen to be mostly poorly constrained in the EZ (where sensitivity to deep gas abundances is very low) and only marginally well constrained in the NEB, although the PH$_3$ abundance is reasonably well constrained for both locations. We hope to improve our constraints on these deep abundances in future work.

\subsubsection{NEDF}

The lower cloud (Cloud-1) profiles in the EZ and NEB are very different from that retrieved in the NEDF, with the NEDF cloud having much lower opacity $(\tau=6.18)$ and much lower cloud base pressure (1.55 bar). This is entirely consistent with our preliminary findings in Section \ref{Section:NEDF},  where we found that the NEDF region has considerable cloud-clearing at these pressure levels. However, the NEDF still has considerable opacity in Cloud-1 and cannot be considered as completely cloud-free, as indeed was seen by the Galileo probe nephelometer \citep{ragent98}. The opacities of the upper cloud (Cloud-2) and the chromophore are more similar to those determined in the  EZ and NEB, with the Cloud-2 opacity  $(\tau=1.12)$ and  chromophore opacity $(\tau=0.38)$ lying part-way between those determined in the EZ and NEB. The computed 5-$\mu$m single-scattering albedo of the lower cloud (Cloud-1) particles are similar to those determined in NEDFs from JIRAM-SPE observations \citep{grassi24} for a single cloud whose vertical location is broadly consistent with our lower cloud (Cloud-1). 

The ammonia profile of the NEDF is most similar to that of the NEB, with similar retrieved values of the mid-level mole fraction of $50.4\pm9.3$ ppm, leading to an ammonia condensation level of 0.54 bar. The deep abundance is retrieved to be lower than both the EZ and NEB with a value of $191 \pm 26$ ppm. 

\subsubsection{Ammonia mole fraction variations}\label{sec:ammonia_comp}

Our adoption of a two-stage ammonia profile, with an intermediate value between the two cloud pressure levels (Cloud-1 and Cloud-2) leads to high retrieved deep abundances, which are more in line with values determined from Juno/MWR and VLA (Section \ref{Section:Introduction}) and consistent with those determined from ISO observations by \citet{brooke98}, who adopted a similar ammonia profile parameterisation. \cite{irwin25} compared the latitudinal variation previously retrieved from VLT/MUSE observations with a one-stage ammonia profile (i.e., ammonia profile constant below the condensation level) and an extended cloud, which returned abundances of typically 190 ppm in the EZ and 120 ppm in the NEB, compared with MWR/VLA values of 400 ppm and 200 ppm respectively. \citet{irwin25} noted that having an ammonia abundance that increases with pressure below the ammonia condensation level might resolve this discrepancy, and this new study suggests that this is indeed the case. The ammonia profiles retrieved here have a similar form to those derived from the VLA by \citet{depater16, depater19} and also, away from the EZ, with the follow-on VLA/MWR analysis of \citet{moeckel23}. In the EZ, \citet{moeckel23} and also \citet{guillot20li} (from MWR alone) find the ammonia abundance decreases with depth at $p>2$ bar. Unfortunately, we do not have sufficient vertical resolution or sensitivity in our observations to confirm this apparent behaviour.

\subsection{Retrievals of typical combined EZ,  NEB two-angle spectra}\label{Section:EZ-NEB-Minnaert}

The Minnaert analysis of spectral imaging observations allows NEMESIS to fit simultaneously to both the $0^\circ$ zenith angle spectrum of a latitude band and also a second spectrum reconstructed at higher zenith angles to constrain the limb-darkening. Fitting both spectra simultaneously places much stronger constraints on the particle opacities and scattering properties and was key to the development of the `holistic' cloud model of Uranus and Neptune \citep{irwin22}. However, such double-angle retrievals are much more computationally expensive than $0^\circ$ zenith angle retrievals, and it was not possible to apply this method to our NEDF observations. Hence, in the previous section we only considered $0^\circ$ zenith angle spectra of the EZ, NEB and NEDF and used these observations and retrievals to refine our model parameterisations. 

\begin{figure}
	\includegraphics[width=\columnwidth]{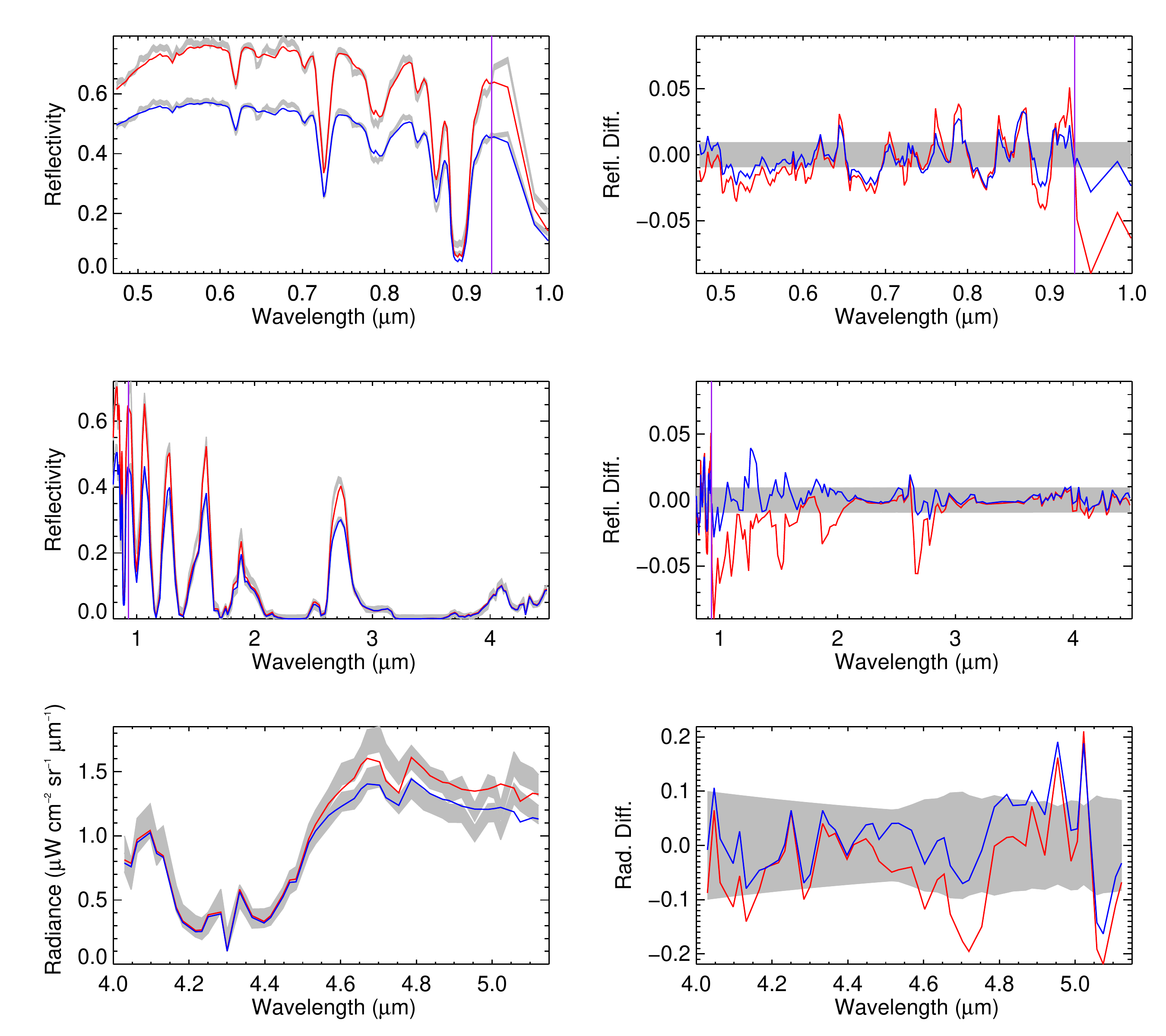}
    \caption{As Fig. \ref{fig:fitspecEZ}, but showing two-angle fit ($\chi^2/n = 3.47$) to the combined MUSE/VIMS EZ spectra, reconstructed at zenith angles of 0$^\circ$ (modelled spectrum in red) and 42.37$^\circ$ (modelled spectrum in blue). The measured spectra and assumed error ranges are again shown in grey.}    \label{fig:fitspecEZtwoang}
\end{figure}

\begin{figure}
	\includegraphics[width=\columnwidth]{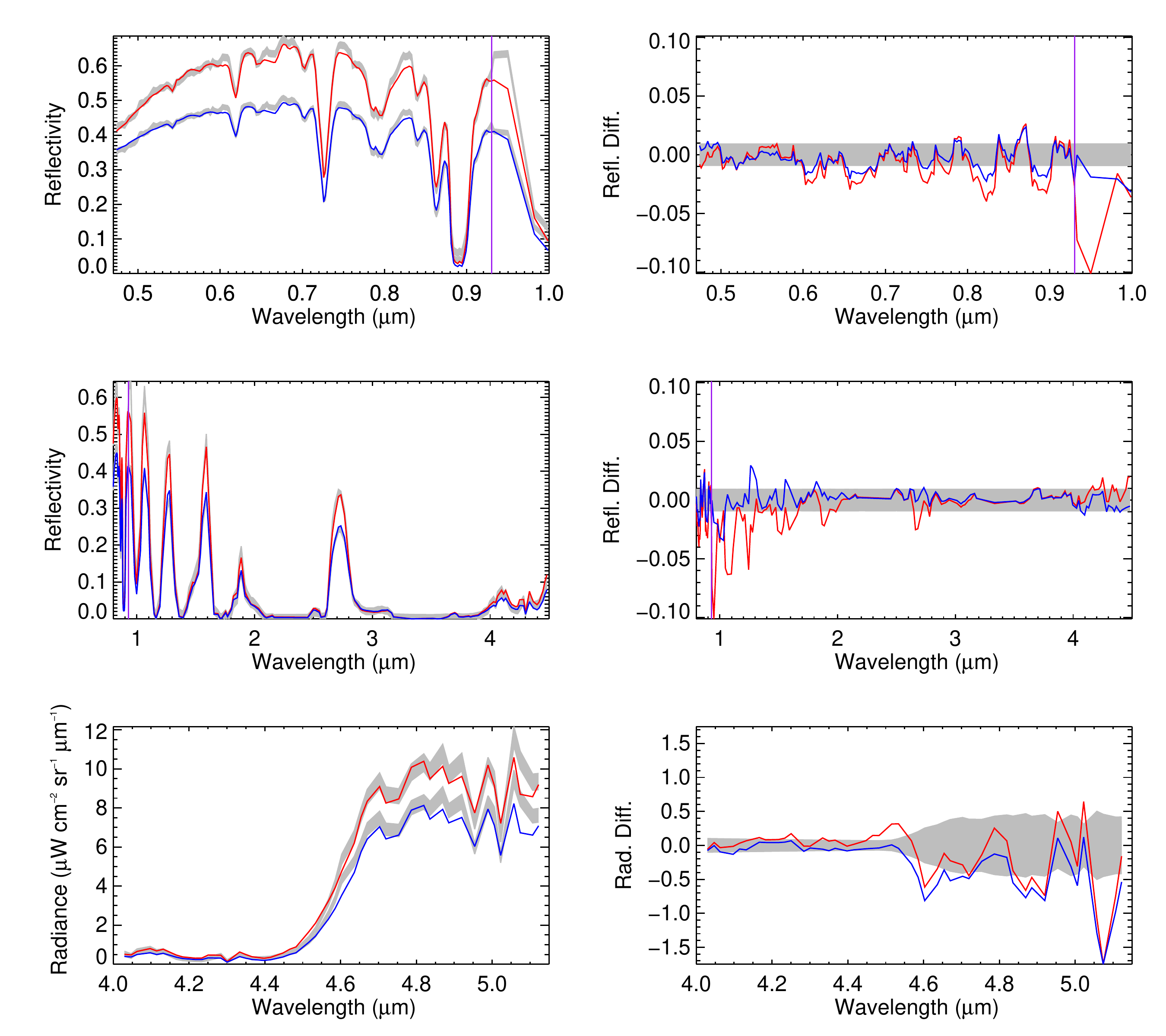}
    \caption{As Fig. \ref{fig:fitspecEZ}, but showing two-angle fit ($\chi^2/n = 2.52$) to the combined MUSE/VIMS NEB spectra, reconstructed at zenith angles of 0$^\circ$ (modelled spectrum in red) and 42.37$^\circ$ (modelled spectrum in blue). The measured spectra and assumed error ranges are again shown in grey.}    \label{fig:fitspecNEBtwoang}
\end{figure}

\begin{figure*}
	\includegraphics[width=\textwidth]{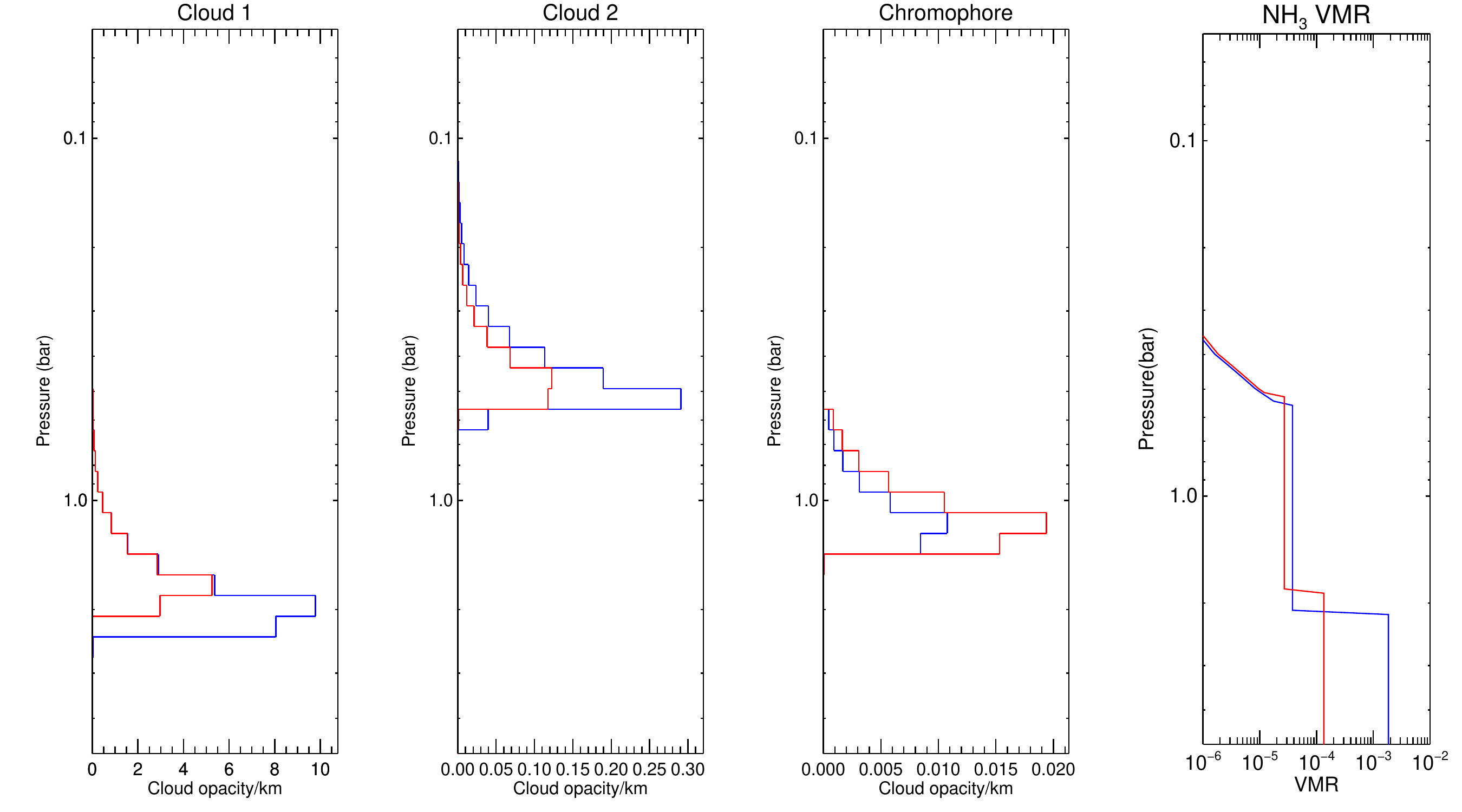}
    \caption{Cloud and ammonia profiles fitted to the observed two-angle spectra in the EZ (blue), NEB (red) and NEDF (black). The left hand panel shows the lower cloud profiles (Cloud-1), the second panel shows the upper cloud profiles (Cloud-2), while the third panel shows the chromophore profiles. The right hand panel shows the ammonia mole fraction profiles.}
    \label{fig:gas_cloud_twoang}
\end{figure*}

\begin{figure}
	\includegraphics[width=\columnwidth]{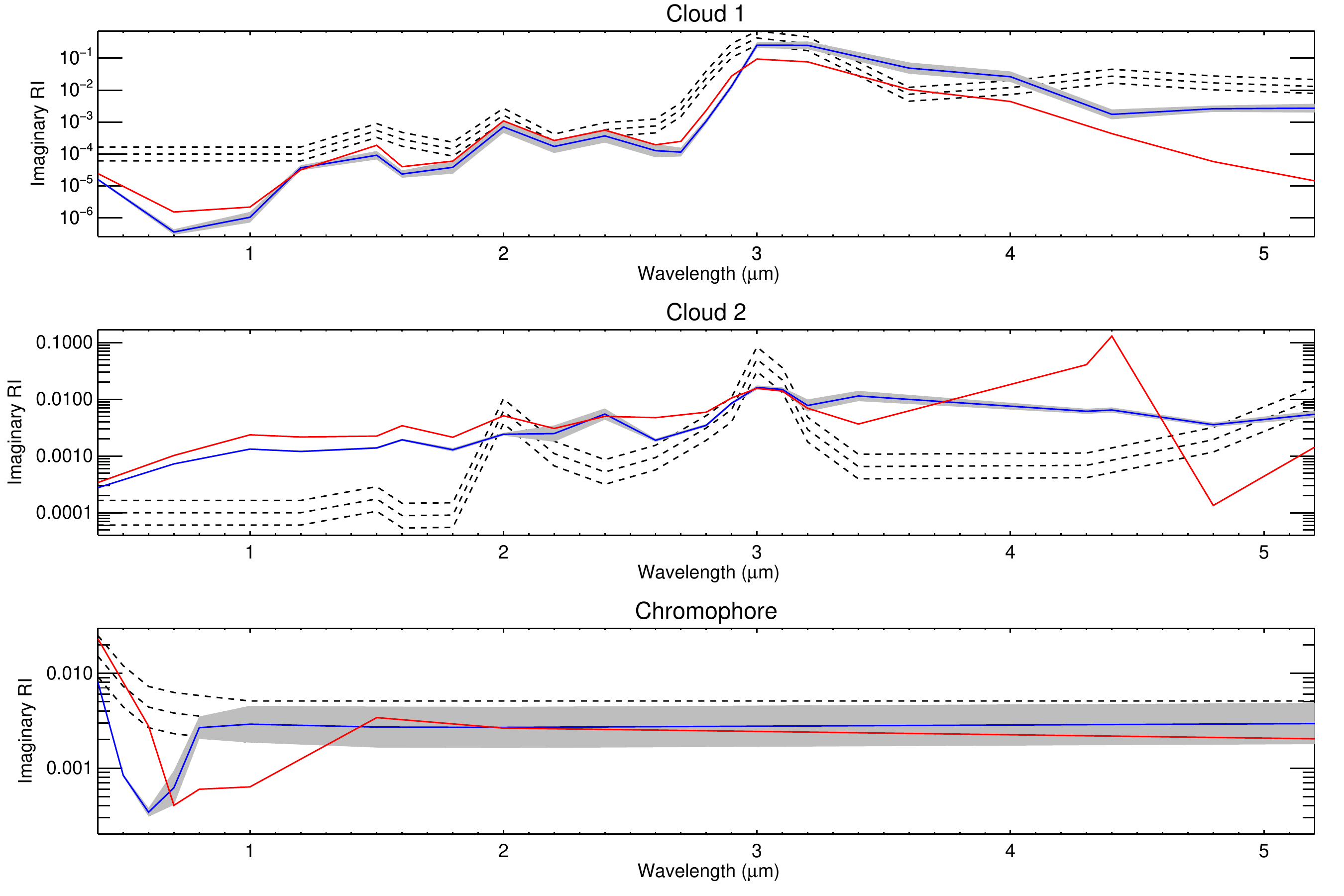}
    \caption{As Fig. \ref{fig:nimag_oneang}, but showing the retrieved $n_\mathrm{imag}$ spectra determined from the combined two-angle MUSE/VIMS spectra in the EZ (blue) and NEB (red). The top panel shows the fitted lower cloud (Cloud-1) $n_\mathrm{imag}$ spectra, the middle panel shows the fitted upper cloud (Cloud-2) $n_\mathrm{imag}$ spectra, and the bottom panel shows the fitted chromophore spectra.}
    \label{fig:nimag_twoang}
\end{figure}

Having validated our retrieval model in this way, we then tested how our retrieved quantities in the EZ and NEB might change if we fitted to two zenith angles. From our Minnaert analyses of our VLT/MUSE and VIMS-IR data in the EZ and NEB we fitted NEMESIS to two spectra simultaneously: one with the Sun and observer at zenith and one with the solar and viewing zenith angles set to 42.37$^\circ$ in the back-scatter direction.  Both angles were chosen to coincide with angles used in our 5-point Gauss-Lobatto zenith angle quadrature scheme in our Matrix operator multiple scattering model (Section \ref{sec:radtrans}), to minimise any possible interpolation errors. The same wavelengths were chosen as in the previous section, and the same errors. 
Our fits to the double-angle EZ and NEB  spectra are shown in Figs. \ref{fig:fitspecEZtwoang} and \ref{fig:fitspecNEBtwoang}. Figures \ref{fig:gas_cloud_twoang} and \ref{fig:nimag_twoang} compare the retrieved cloud/ammonia profiles and $n_\mathrm{imag}$ spectra.

It can be seen that our model parameterisation is able to fit double-angle spectra just as accurately as the single-angle spectra shown in Section \ref{Section:EZ-NEDF-NEB}. However, by fitting to two observation angles simultaneously we are able to constrain the cloud properties less degenerately and more robustly, in exactly the same was we did for Uranus and Neptune \citep{irwin22}. The retrieved solutions for the EZ and NEB are qualitatively similar to those determined from the $0^\circ$ zenith angle spectra alone (Section \ref{Section:EZ-NEDF-NEB}), but there are some key differences. The two-angle retrievals are summarised below and also in Table \ref{reftable1}.

\begin{enumerate}
    \item The retrieved opacity of the lower cloud (Cloud-1) is found to be higher than the single-angle retrievals by a factor of 2--3, but with similar base pressures. However, the opacities of the upper cloud (Cloud-2) and the chromophore are more similar to those determined in the single-angle case.
    
    \item The mid-level ammonia mole fraction is found to be slightly lower than the single-angle case, again with slightly higher values in the EZ.

    \item The deep ammonia abundance in the EZ is found to be very high, but very poorly constrained, while that in the NEB is similar to the single-angle case.

    \item The retrieved phosphine abundances are similar for the two-angle and single-angle cases, in both the EZ and NEB.

    \item The deep gaseous abundances are again unconstrained in the EZ, but have similar values to the single-angle case in the NEB.

    \item The retrieved $n_\mathrm{imag}$ spectra are similar to the single-angle case, but more constrained for the two-angle case. Again, the lower cloud (Cloud-1) is found to have higher $n_\mathrm{imag}$ in the 5-$\mu$m window in the EZ of $2.6\times10^{-3}$ at 4.8 $\mu$m, reducing to $6\times10^{-5}$ in the NEB. This increases the single-scattering albedo at 4.8 $\mu$m from $\varpi=0.94$ in the EZ to $\varpi=0.99$ in the NEB.
    
\end{enumerate}

The ability of our atmospheric model to match the observed spectra of the EZ and NEB at two angles simultaneously adds considerable confidence to the effectiveness of this model to explain the observed spectra of Jupiter from 0.48 -- 5.20 $\mu$m.

\section{Discussion}\label{Section:Discussion}

\subsection{Comparison of retrieval strategies}

We have found that we can fit very well to EZ, NEB and NEDF spectra with a single-angle multiple-scattering retrieval (Section \ref{Section:EZ-NEDF-NEB}), and that including limb-darkening information for the EZ and NEB leads to better constrained parameters for the two-angle retrievals (Section \ref{Section:EZ-NEB-Minnaert}). In these retrievals we retrieve not just the cloud opacities, but also the gaseous abundances together with cloud $n_\mathrm{imag}$ spectra that allow us to simultaneously fit from visible wavelengths through to \
5 $\mu$m. Before we discuss the findings of these analyses further, in this section we compare the retrieved cloud profiles of the EZ, NEB and NEDF with previous determinations.

\begin{figure}
	\includegraphics[width=\columnwidth]{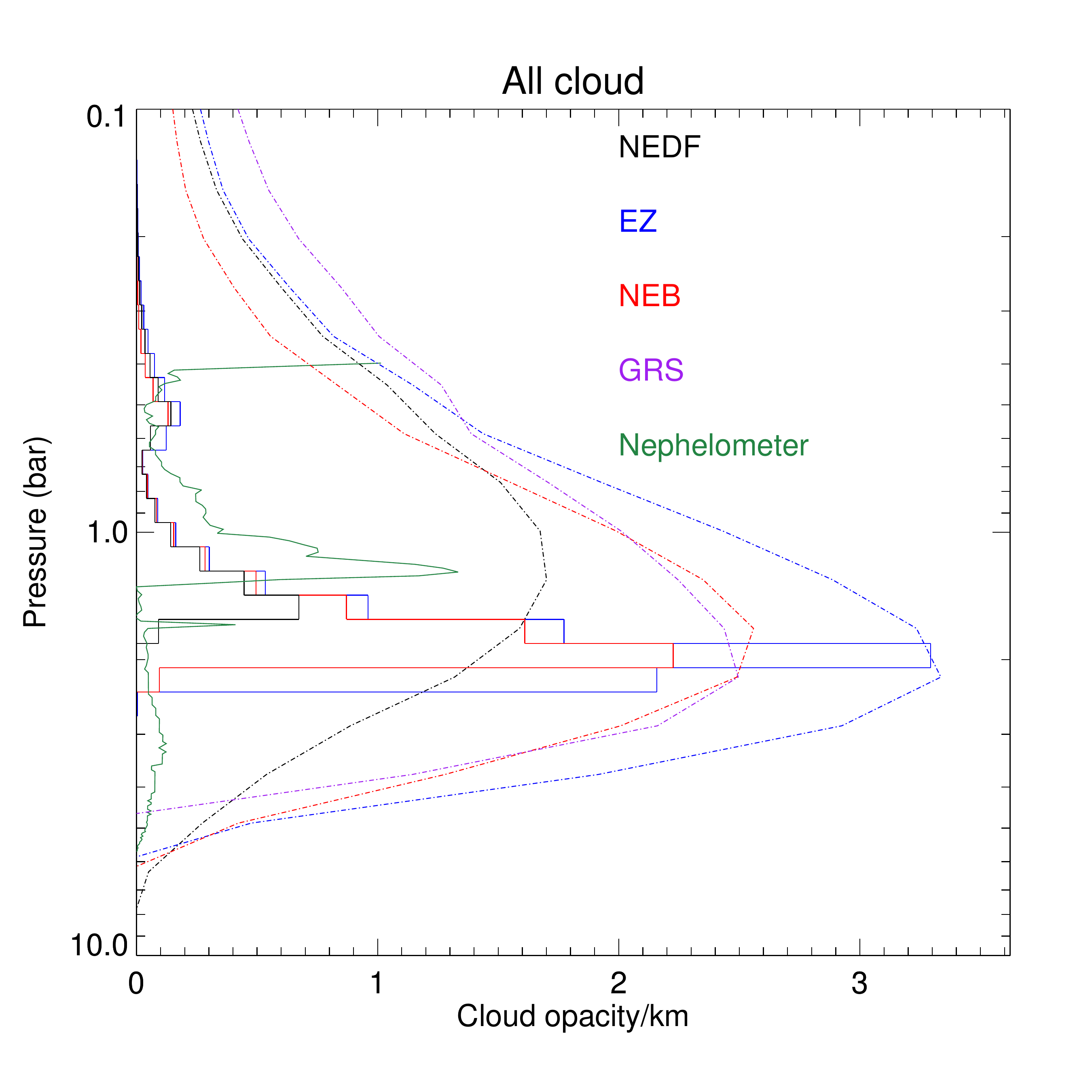}
    \caption{Cloud retrieval comparison. Solid lines show the cloud opacity/km profiles retrieved earlier for single-angle ($0^\circ$ zenith angle) retrievals of the EZ, NEB and NEDF, and shown separately in Fig. \ref{fig:gas_cloud_oneang}, but here summed over all cloud components. Also shown is the cloud profile determined from the Galileo Probe Nephelometer (previously shown in Fig. \ref{fig:nephelometer} and here arbitrarily scaled) and the reflectivity profiles (dash-dot lines, again arbitrarily scaled) retrieved with a reflecting-layer model from the MUSE 875 -- 910 nm reflectivity spectra, and shown previously in Fig. \ref{fig:onionretrieval}. }
    \label{fig:cloud_comparison}
\end{figure}

Figure \ref{fig:cloud_comparison} shows the cloud opacity/km profiles retrieved from the single-angle retrievals of Section \ref{Section:EZ-NEDF-NEB} and shown previously in Fig. \ref{fig:gas_cloud_oneang}. Here the opacity (at 1.5 $\mu$m) of all three components have been added together. Also plotted on this figure is the Galileo Probe Nephelometer profile, shown previously in Fig. \ref{fig:nephelometer}, and arbitrarily scaled  for ease of comparison.  The Galileo probe fell through an NEDF and so we might expect the measured profile to be most like that retrieved from our NEDF observations, which we can see is indeed the case. However, the base pressure of the main nephelometer cloud is at a slightly lower pressure than the base of our retrieved lower cloud (Cloud-1), and the reflectivity from the upper component at 0.4 bar is more significant and also at a lower pressure than our upper cloud (Cloud-2). Of course, the NEDF observed by the Galileo probe is not the same as the NEDF locations viewed in our VLT/MUSE and Cassini/VIMS-IR datasets, and the probe viewed a single location, whereas the remote observations are averaged over a small spatial area. Hence, we should not be surprised that there is imperfect correspondence. It may just be that the NEDF seen by the Galileo probe was even more cloud-free than that observed in our datasets. This might also make sense with regards to the relative back-scatter of the nephelometer profile at 1.2 and 0.4 bar, which seem roughly equal in the nephelometer profile, but are very dissimilar for our the retrieved case. 

Also shown in Fig. \ref{fig:cloud_comparison} are the reflectivity profiles for these three locations (and also the GRS) retrieved from the VLT/MUSE spectra from 875 to 910 nm with our simple reflectivity model in Section \ref{sec:onion}, and shown previously in Fig. \ref{fig:onionretrieval}. These profiles have again been scaled for ease of comparison. Putting aside the difference in vertical resolution, with the continuous reflectivity profiles smoothed by the width of the transmission weighting functions, the correspondence between the pressure levels of the main features is remarkable, with the MUSE reflectivity profile peaks corresponding closely with the retrieved lower cloud (Cloud-1) pressure levels, and the variations at lower pressures  consistent with the differing upper cloud (Cloud-2) opacities. For the NEDF we can see that the reflectivity profile peaks at significantly lower pressures than the other regions, while at 0.2 -- 0.6 bar the reflectivity of the NEDF falls half way between that of the EZ and NEB, exactly as retrieved in the full multiple-scattering retrievals. We can also see that the reflectivity profile of the GRS is closest in shape to that of the EZ at depth, but has much more reflectivity at 0.2 -- 0.6 bar, suggesting that the GRS is mostly a feature at the upper cloud (Cloud-2) pressure level. Overall, it is apparent that if we are just interested in the vertical profile of aerosols, a simple reflectivity profile retrieval of the observed spectrum from 875 to 910 nm provides a very effective indicator of the vertical cloud structure and can be computed at a tiny fraction of the computational cost of a full multiple-scattering retrieval. However, if we want to understand the scattering properties of these clouds over a wide spectral range and also determine the abundances of different gases, the multiple-scattering retrieval route is required.

\subsection{Belt/Zone differences}

For both the single-angle and double-angle retrievals we find that the lower cloud (Cloud-1) layer has approximately twice the opacity in the EZ as in the NEB. However, although these opacities are large, we find the Cloud-1 particles are highly scattering at visible wavelengths and so this does not give a huge change of visible reflectivity as the reflection is already almost `saturated'. The visible reddening of the NEB compared with the EZ is accounted for by a reduction in opacity of the highly-scattering Cloud-1 particles, a slight increase in the opacity of the chromophore, and a significant increase in $n_\mathrm{imag}$ of the chromophores particles  at blue visible wavelengths. At longer wavelengths, the large reduction in 5-$\mu$m emission going from the NEB to the EZ is mostly caused by an increase in $n_\mathrm{imag}$ at 5 $\mu$m of the lower cloud (Cloud-1) particles in our retrievals. 

If we believe our retrieved vertical cloud structure and our assumed particle sizes to be accurate then these changes must be caused by compositional changes in the lower cloud (Cloud-1) and chromophore particles. In the EZ it may be that upwelling of moist, volatile-rich air and the formation of the Cloud-1 layer leads to reduced single-scattering albedo of the Cloud-1 particles as well as higher overall opacity, thus reducing the 5-$\mu$m thermal emission to close to zero. Conversely, reduced upwelling in the NEB and a thinning of the Cloud-1 layer may be associated with photochemical processing in the chromophore particles, making them darker. It may also be that these Cloud-1 (and indeed Cloud-2 particles at higher altitudes) are hybrid particles, formed by condensible species condensing on pre-existing aerosols (chromophore or otherwise). 

\begin{figure}
	\includegraphics[width=\columnwidth]{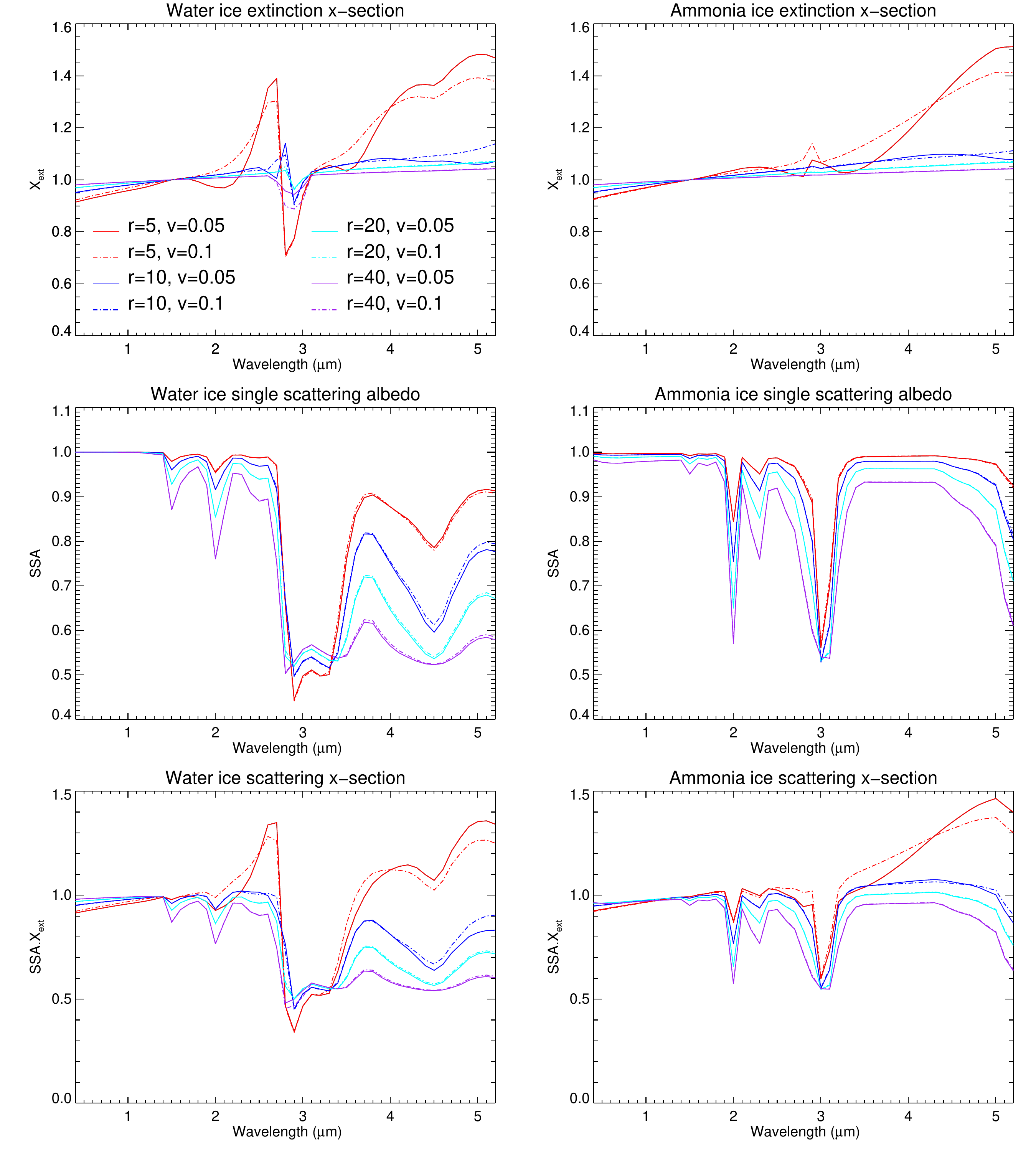}
    \caption{Comparison of cross-section and single-scattering albedo spectra calculated for water ice (left) and ammonia ice (right) using gamma size distributions of various mean radii $r$ = 5, 10, 20 and 40 $\mu$m, and variances $\sigma =  0.05, 0.1$. The calculations use the complex refractive index data of \citet{warren08} for water ice, and \citet{martonchik84} for ammonia ice. Top row shows the extinction cross-section spectra, the middle row shows the single-scattering albedo spectra, and the bottom row shows the scattering cross-section spectra. It can be seen that the main variation is with mean radius, and that the assumed variance has a secondary effect on the calculations.}
    \label{fig:x-section-ssa}
\end{figure}

Alternatively, the change in the 5-$\mu$m single-scattering albedo (SSA) of the lower cloud (Cloud-1) particles may be due to a change in the size of the particles. In Fig. \ref{fig:x-section-ssa} we compare calculations of the cross-section and SSA spectra of water ice and ammonia ice using the complex refractive indices of \citet{warren08} and \citet{martonchik84}, respectively. We have assumed gamma size distributions and present results for a range of mean radii ($r= 5$, 10, 20 and 40 $\mu$m) and variances ($\sigma = 0.5$, 1). It can be seen that ammonia ice has strong absorption features at 2 $\mu$m, 2.2 $\mu$m, 3 $\mu$m and wavelengths longer than 5 $\mu$m, but the absorption in the 5-$\mu$m window varies significantly with wavelength and is also not strong enough to account for the 5-$\mu$m absorber. However, it can be seen that water ice is highly scattering at short wavelengths and has a broad absorption longwards of 4 $\mu$m, with the SSA decreasing rapidly as the mean radius increases and also becoming less varying with wavelength. This is precisely what we need to explain belt/zone differences, with small variations at visible wavelengths, but large differences at 5 $\mu$m. How water ice might actually be present at such pressure levels is unclear as its expected condensation level from an ECCM is 5 -- 7 bar (Fig. \ref{fig:jupiter_atmosphere_model}). Is it the case that small water ice particles are perhaps lofted up to these levels by vigorous convection? Or is it perhaps that the air is close to water saturation and so local cooling at 1 -- 2 bar can lead to the formation of a higher water ice cloud, analogous to `cirrus' clouds seen in the Earth's atmosphere? In addition, we do not know the shape of the deep H$_2$O profile. Does it follow the SVP curve from its deep condensation level as predicted by the ECCM? Or does it perhaps combine with NH$_3$ more readily than expected at depth, reducing its abundance and then allowing water vapour to condense again at lower pressures? It is notable that the water vapour abundances at $\sim 5$ bar we retrieve in the NEB and an NEDF of  just a few ppm, would lead to water ice condensing near our retrieved lower cloud (Cloud-1) base. However, test retrievals we made with variable-radius water ice used as Cloud-1 particles had insufficient reflectivity at 2.7 $\mu$m due to the strong absorption of water ice at this wavelength and hence it seems unlikely that the Cloud-1 particles are pure water ice condensates. Since we find that the ammonia abundance also drops rapidly at the Cloud-1 condensation level, it is possible that water ice may be combined with ammonia to form an H$_2$O--NH$_3$ complex that leads to the `mushballs' proposed by \citet{guillot20}. Unfortunately, the scattering properties of such condensates are unknown, but to a first approximation it would seem likely that they are a combination of those of water and ammonia ice, and we know that water ice is a strong 5-$\mu$m absorber. Hence, a very plausible interpretation of our results is that the Cloud-1 layer is composed of water-ammonia mixture of a composition similar to that leading to the proposed theoretical mushballs of \citet{guillot20}, which have larger opacities and \textbf{larger radii} in the zones compared with the belts, rather than there being compositional differences of the Cloud-1 particles between the belts and zones. Whatever the belt/zone changes may be, they must also make the absorption of the chromophore at visible wavelengths less strong in the zones, with the increased lower cloud (Cloud-1) opacity in the zones masking the chromophore absorption. Whichever interpretation is correct, since the bulk properties of the NEB and EZ are very similar to those of the other belts and zones, these changes are presumed to apply to other belts and zones also. 

To explore this question further requires a more thorough investigation of the solution space. In this study we used our traditional NEMESIS model \citep{irwin08}, which is built on Fortran and is not parallelisable. NEMESIS reads and writes many files to the disk during a calculation and the multiple-scattering calculations necessary to model the observation geometry are slow, especially at the higher zenith angles. As a result, the double-angle retrieval runs presented here took many days to complete on a high performance computer. In addition, NEMESIS uses Optimal Estimation to arrive at its solution, which although relatively fast can become trapped in local minima and may not necessarily arrive at the globally optimal solution. Hence, although we have found a good solution to the required cloud and ammonia profiles, there may be other combinations of particle sizes and properties that fit the observations equally well. We will return to this point in our conclusions.

\subsection{The `Ammonia Ice Cloud'}

The latitudinal distribution of particulates seen at strongly methane-absorbing wavelengths in the MUSE (Fig. \ref{fig:jupiter_muse}), VIMS (Fig. \ref{fig:jupiter_vims}) and JIRAM-SPE (Fig. \ref{fig:jiram_images}) observations is rather different from that seen at wavelengths where methane is less absorbing and points to a different latitudinal distribution of particles at pressures of $p < 0.5$ bar. In many previous studies this has been taken as evidence for the presence of a detached photochemical haze. However, we have shown here that the similar reflectivities seen from this pressure level across the combined spectral range indicate that the particles cannot be small photochemical haze products, but must be somewhat larger and perhaps more likely to be condensed particles. In our final retrieval model it will be noted that we have not had to assume the presence of an additional haze layer at all to match our observations. Instead, the particles in our model responsible for reflection from these pressure levels are the upper reaches of the upper cloud (Cloud-2) layer, based at $\sim$0.55 bar, which is optically thicker and extends higher in the EZ than the NEB, and whose spectral properties are possibly consistent with a significant component of ammonia ice, although other species with a strong N--H absorption band may be responsible \citep{biagiotti25}. 

For many years the accepted wisdom published in most text books and websites was that the clouds of Jupiter we see in telescope and spacecraft images are composed of ammonia ice. However, it has been shown, both here and in a large number of previous studies, that the dominant reflection comes from clouds based at 1--2 bar (Cloud-1 in our study), where it is so warm that pure ammonia ice would sublime. Although, we conclude (since the ammonia gas abundance drops near this level) that the main cloud layer of Jupiter has a significant component of ammonia, either in the form of H$_2$O--NH$_3$ that leads to `mushball' precipitation or ammonium hydrosulphide (NH$_4$SH), or both, the main clouds we see cannot be pure ammonia ice. Instead, we find that ammonia is combined both in the lower cloud (Cloud-1) at 1--2 bar and also in a second, optically thin `cirrus-like' Cloud-2 layer at low pressures ($p\sim 0.55$ bar). Although this well-known upper layer has historically been described as a detached haze,  a high-level haze would be unlikely to exhibit the variations in opacity seen at the northern edge of the NEB, where a wave-like variation in reflectivity is seen in the 890-nm image of MUSE (Fig. \ref{fig:jupiter_muse}). But if this component is instead the top of deeper dynamical upwellings, then such a correlation is perfectly understandable and indeed to be expected. Interestingly, thermal waves detected \citep{fletcher17} in the stably-stratified upper troposphere of the NEB were found to be anti-correlated with `haze' opacity, suggesting again that these are condensates rather than photochemical haze particles. In addition, we know from JWST/NIRCam observations \citep{hueso23} that the particles in this layer are formed into discrete cloud-like `clumps'. Such a formation is again much more consistent with the particles being more like an upper-level `cirrus' cloud than an extended photochemical haze.

The fact that SIACs (Spectrally-Identifiable Ammonia Clouds) are not more widely seen suggests that it is likely that the upper cloud (Cloud-2) particles are generally `hybrid' particles, composed of several possible constituents, perhaps condensing on existing aerosols such as photochemically-produced chromophores. Although the strong 3-$\mu$m absorption needed by the Cloud-2 particles to match the observed spectra is consistent with larger radius ammonia ice particles, as previously concluded by \citet{brooke98}, retrieval tests where we assumed the Cloud-2 particle were made of pure ammonia ice did not match the observed spectra well. Pure ammonia ice also has absorption features at 2 $\mu$m and although we still fitted the 3-$\mu$m absorption well, we then underestimated the reflectivity at 2 $\mu$m. In contrast, the retrieved Cloud-2 $n_\mathrm{imag}$ spectra of our preferred model (Figs. \ref{fig:nimag_oneang} and \ref{fig:nimag_twoang}) can be seen to suppress the 2-$\mu$m absorption peak of the ammonia ice prior. However, in certain locations of significant convection, it may be that ammonia condenses at higher pressures and into optically thicker clouds that give rise to the sparsely distributed SIACs discussed in the Introduction. 

\subsection{Chromophore Layer}

The vertical position and nature of Jupiter's visible colouring agent, the `chromophore' has long been a mystery in terms of composition, spectral properties and vertical location. Previous analysis of  MUSE observations \citep{alexander24,Alexander2026} strongly suggests that the main chromophore resides in the lower cloud at 1--2 bar, rather than higher in the atmosphere, and this is also our conclusion. Attempts were made to add chromophore in the upper cloud, but this made little difference in how well we could fit the data over the  wavelength range considered (0.48 -- 5.20 $\mu$m). Since the overwhelming bulk of the opacity is found to be in the lower cloud layer (Cloud-1), it makes sense that any darkening by a chromophore would be most effective here also. It is, however, quite possible, and indeed likely, that chromophore particles are present throughout the vertical column if they are generated, as is widely thought, in the upper troposphere/lower stratosphere by photochemical reactions such as those between acetylene and ammonia \citep{carlson16}. Indeed, we find that the upper cloud (Cloud-2) particle scattering properties are inconsistent with pure ammonia ice, suggesting the presence of another component. In addition, in preliminary analysis of observations at shorter wavelengths covered by HST/WFC3, Cassini/ISS and HST/FOS, extending into the UV, where the atmosphere becomes increasingly Rayleigh-scattering, we find the need for a significant presence of chromophore absorption at lower pressures. These are wavelengths where the chromophore absorption is greatest, so small opacities are needed and if we assume the particles are small, as expected, the effect of these upper level chromophores would be difficult to discern in the wavelength range we have considered. The extension of our model to shorter, UV, wavelengths will be the subject of a future study. 

In the `Cr\`eme Br\^ul\'ee' model of \citet{baines19} the chromophores colour the upper layers in locations such as the Great Red Spot, while \citet{sromovsky17} find an absorber with the same spectral properties may be generally responsible for Jupiter's colouration. However, since we find that the opacity of the chromophores in any but the main lower cloud (Cloud-1) layer is very small, our inferred vertical distribution is the opposite to that of the  `Cr\`eme Br\^ul\'ee' model and perhaps more analogous to a `Chocolat Li\'egeois' distribution, with most of the chromophore lying deep in the Cloud-1 layer and partly obscured by it, lying well below the upper, thinner Cloud-2 hybrid-ammonia-ice layer. 

A second reason to suspect that chromophore particles are present in the lower cloud is that we find we need a component of small particles at 1--2 bar to give enough reflection to reproduce the observed spectrum at MUSE wavelengths. When the small chromophore particles were removed from our model, we found that although the retrieved $n_\mathrm{imag}$ values of Cloud-1 increased at MUSE blue wavelengths to provide the additional absorption, the fitted spectra were not as good as those presented from our preferred model. In addition, it was only the blue $n_\mathrm{imag}$ values of the Cloud-1 particles that needed to be increased to match the observations, not those in upper cloud (Cloud-2), adding further confirmation that the darkening needed is of the lower cloud (Cloud-1) layer, not the upper cloud (Cloud-2). However, these findings might also point to the Cloud-1 layer being bimodal, and composed of both the large particles we have assumed here, together with a component of smaller particles. 

Finally, in this study we have considered only the EZ, NEB and and NEDF, and have deliberately avoided considering special regions such as the Great Red Spot (GRS). The colouration of the GRS does to the eye look rather different than the EZ, and from maps of the upper cloud (Cloud-2) `hybrid-ammonia-ice' distribution in regions of strong methane absorption in MUSE, VIMS and JIRAM-SPE images (Figs. \ref{fig:jupiter_muse}, \ref{fig:jupiter_vims} and \ref{fig:jiram_images}) we can see that the Cloud-2 opacity is much higher in the GRS than it is in the EZ. Hence, future studies may find that the `Cr\`eme Br\^ul\'ee' model is more appropriate in the GRS, for which it was originally proposed, although this scenario is not favoured by \citet{asier23, asier26}. The fact that the `Cr\`eme Br\^ul\`ee' model does not seem to be applicable to the regions we have considered here represents a problem to this model. It may be that the chromophore particles are indeed generated in the upper troposphere by a photolytic reaction between C$_2$H$_2$ and NH$_3$, but are then vertically mixed downwards before accumulating in a region where the air is more statically stable at 1--2 bar. Indeed, this is very similar to our `holistic' model that successfully matched Uranus and Neptune observations \citep{irwin22}. In the atmosphere of Uranus and Neptune, the condensation of methane causes a rapid drop in mean atmospheric molecular weight that stabilises the atmosphere at similar pressures and inhibits convective overturning. For Jupiter, the 1--2 bar region is where we find the abundance of NH$_3$ decreases, perhaps through combination with H$_2$S, or with H$_2$O. This condensation will then similarly reduce the mean molecular weight -- not as much as for Uranus and Neptune, but perhaps enough to concentrate the chromophore particles mixing down from above. Alternatively, the identity of the chromophore may not be the `Carlson' chromophore \citep{carlson16} at all, but something derived from processing of H$_2$O, H$_2$S, PH$_3$ and NH$_3$ that happens to have similar qualities. \citet{harkett24} recently reported a very interesting correlation between enhanced phosphine and aerosol abundance in the GRS, which may perhaps indicate that phosphine is involved in aerosol production.

\section{Conclusions}\label{Section:Conclusions}

\begin{figure*}
	\includegraphics[width=\textwidth]{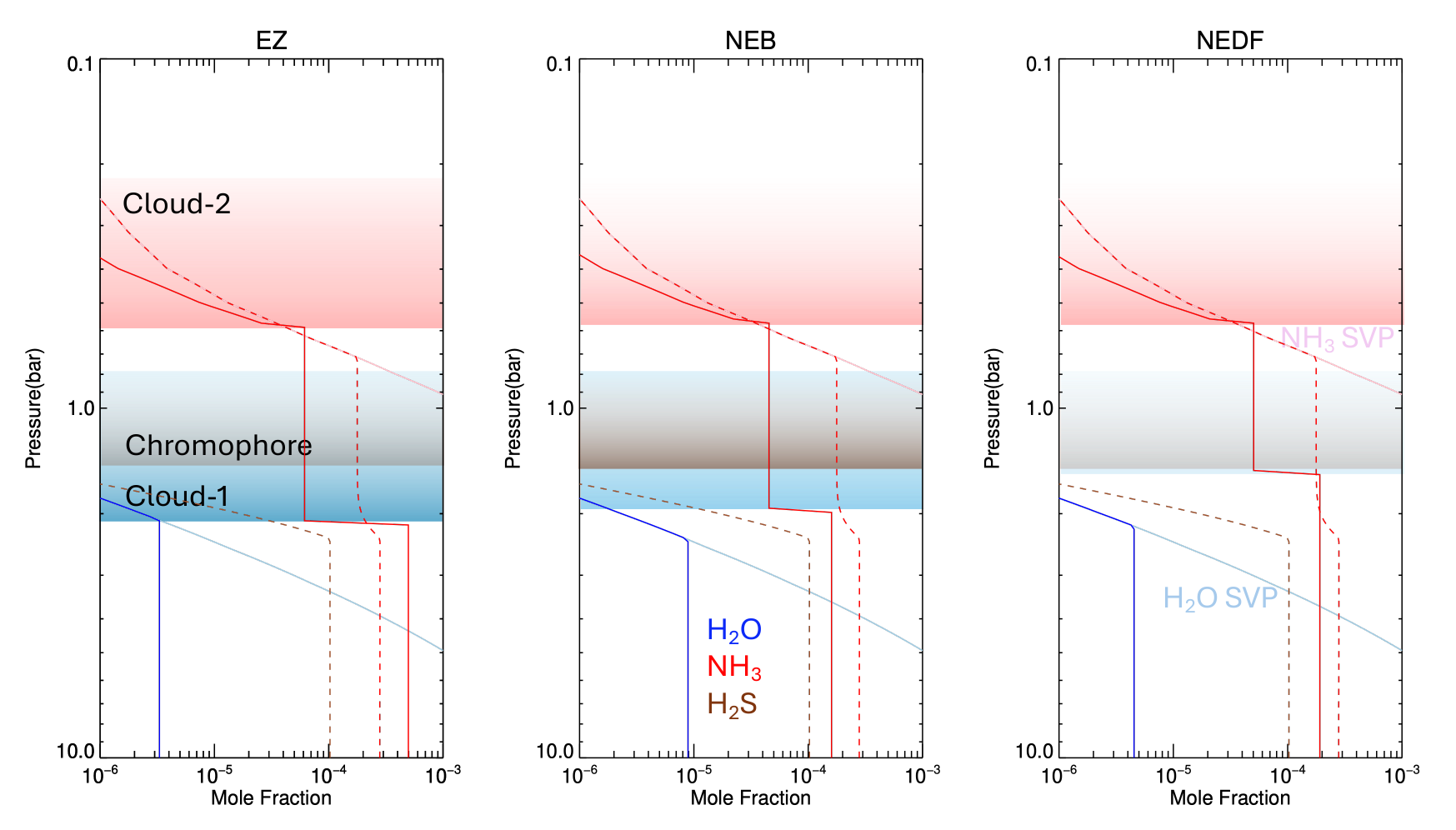}
    \caption{Gas mole fraction profiles (H$_2$O and NH$_3$) and cloud profile summary for the EZ, NEB and NEDF (gas abundances from $0^\circ -$zenith angle retrievals). The \textit{a priori} H$_2$S and NH$_3$ profiles are shown as coloured, dashed lines (the H$_2$S abundance was not fitted). The light-shaded blue and pink lines are the SVP curves for H$_2$O and NH$_3$ from our assumed $T(p)$ profile. We find a thick lower cloud (Cloud-1, shaded blue) based at the level where NH$_3$ first reduces in abundance at $p_1 = 1 - 2$ bar, and a second optically thinner upper cloud (Cloud-2, shaded pink) based  where NH$_3$ saturates near 0.55 bar. Above the saturation level, ammonia is reduced to a fitted relative humidity profile, which decays with height. Our chromophore layer (shaded brown) is based at 1.25 bar and is more blue-absorbing and less hidden by the lower cloud (Cloud-1) in the NEB. } 
    \label{fig:summary}
\end{figure*}

In this study we have determined possible solutions for the vertical profiles of ammonia and clouds in Jupiter's atmosphere that match the observed spectra of the EZ, NEB and NEDFs over a wide spectral range (0.48 -- 5.20 $\mu$m) and a range of viewing angles in our combined VLT/MUSE, Cassini/VIMS and Juno/JIRAM-SPE datasets. Our conclusions are summarised in Fig. \ref{fig:summary} and the detailed conclusions are listed below.

\begin{enumerate}

    \item We find that the vertical cloud structure across the planet can be explained by three main layers: 1) a lower cloud  (`Cloud-1') at 1 -- 2 bar; 2) an upper cloud (`Cloud-2') in the upper troposphere at $\sim$0.55 bar; and 3) a layer of chromophore within the Cloud-1 layer. Most of the opacity is in the lower cloud (Cloud-1) layer, which at 1.5 $\mu$m is over 10 times more opaque than the upper cloud (Cloud-2) layer and more than 100 times more opaque than the chromophore layer.

    \item We find that the ammonia profile is intimately linked with our cloud profile, with the lower cloud (Cloud-1) coinciding with an initial sharp drop in ammonia abundance, perhaps associated with the formation of a water-ammonia mixture of a composition similar to that proposed to lead to the theoretical `mushballs' of \citep{guillot20}, or combination with H$_2$S into solid NH$_4$SH. The upper cloud (Cloud-2) is based at the level where the remaining ammonia exceeds its saturation vapour pressure (SVP). Above this condensation level, the ammonia is modelled to decrease as a fraction of the base relative humidity (RH)  and is further reduced with height by photodissociation, modelled as an exponential reduction of RH with height.

    \item We find that the lower cloud (Cloud-1) particles have spectral properties common with water ice, being highly scattering at visible wavelengths, but rather absorbing at 5 $\mu$m, especially in the EZ. We also find that the Cloud-1 particles must be rather large ($r\sim10$ $\mu$m) in order that they are forward-scattering at visible wavelengths and thus allow light to scatter through and be Rayleigh-scattered back from deeper levels. This accounts for the higher than expected equivalent widths of the methane and ammonia bands at visible wavelengths, and thus unexpectedly deep cloud-top pressures noted from MUSE observations by \citet{irwin25}. In addition, such large particles are more effective absorbers at 5 $\mu$m.

    \item We find that the properties of the upper cloud (Cloud-2) particles have features in common with ammonia ice, being highly absorbing at 3 $\mu$m, but not at 2 $\mu$m.  This 3-$\mu$m absorption is necessary to match the low reflectivity in this wavelength region and was previously noted by \citet{brooke98}. 
    We also find that the particles in this layer are large ($r\sim10$ $\mu$m), which is needed to ensure that the reflectivity at methane-absorbing wavelengths does  not drop rapidly with wavelength, and also helps light penetrate to greater depths at visible wavelengths. Although this layer has spectral properties similar to large ammonia ice particles, as noted by \citet{brooke98}, such a condensate would give secondary features at other wavelengths, which are not seen. Hence, the upper cloud (Cloud-2) layer is likely composed of hybrid particles, probably containing ammonia ice, but also including other 3-$\mu$m-absorbing particles containing N--H absorption bonds \citep{biagiotti25}, contributing to the ``ubiquitious 3-$\mu$m  absorption" noted by \citet{sromovsky10a}. 

    \item The two-step ammonia profile assumed in our study lowers the ammonia abundance above the lower cloud (Cloud-1), necessary to reduce ammonia gas absorption at 1.5 and 2 $\mu$m, but leaves it high below Cloud-1, which is necessary to match the visible wavelength absorption of ammonia. Our retrieved deep ammonia abundances are more consistent with Juno/MWR \citep{guillot20li} and VLA \citep{moeckel23} determinations. Using a two-step ammonia profile, rather than a single-step one, also explains why the ammonia abundances determined using a single-step ammonia profile by \citet{braude19}, \citet{alexander24,Alexander2026}, and \citet{irwin25} are lower than those determined from Juno/MWR and VLA by a factor of $\sim$1.5.
 
    \item We find that the bulk of the chromophore particles lie within the main lower cloud (Cloud-1) layer, rather than above it, and that changes in red colouration between the EZ and NEB are caused partly by increased blue-absorption of these particles in the NEB combined with a lower opacity of the Cloud-1 particles. Hence, our preferred solution is more analogous to a `Chocolat Li\'egeois' distribution than the `Cr\`eme Br\^ul\'ee' model of \citet{baines19}. We find that the chromophore particles must be reasonably small (we assumed a fixed value of $r = 0.2$ $\mu$m in this study) which limits their effect to wavelengths shorter than than 600 nm. 

    \item We find that we do not need a separate, detached haze at $p < 0.5$ bar. Instead, we find that the reflectivity variations seen at methane-absorbing wavelengths, such as 890 nm for VLT/MUSE, are accounted for by variations in the opacity and vertical extent of our upper cloud (Cloud-2) into these pressure levels.

    \item Assuming that we are correct in believing the upper cloud (Cloud-2) to contain a significant component of ammonia ice, we conclude that ammonia ice clouds are in fact widely present in Jupiter's atmosphere, but have much lower opacity than the lower cloud (Cloud-1) layer at 1--2 bar and are mixed with other particulates. We suggest that this thin upper hybrid-ammonia-ice `cirrus' cloud has previously been mistaken for the detached haze layer.

    \item The main lower cloud deck at 1--2 bar, Cloud-1, has properties similar to, but not identical to water ice. It may be that this cloud is composed of a H$_2$O--NH$_3$ mixture consistent with that leading to the `mushball' precipitation scenario of \citep{guillot20}, but it may also contain condensates such as NH$_4$SH. Unfortunately, a definite identification of the composition of the Cloud-1 particles is beyond the scope of this study.

    \item Identifying the main clouds to be based at 1 -- 2 bar, rather than the assumed ammonia condensation level of 0.7 bar has implications for using cloud tracking to infer wind speeds, since the wind speeds derived would now appear to be at substantially deeper pressures than previously estimated. \citet{harkett24} found that wind speeds determined from HST were only consistent with JWST/NIRCAM winds \citep{hueso23} around the GRS at $\sim$240 mb if the HST winds refer to a pressure level of $\sim$1.2 bar, instead $\sim$0.7 bar. This conclusion appears to be fully consistent with our results.

\end{enumerate}

Although we have found a good solution to cloud and ammonia profiles in Jupiter's equatorial atmosphere, the computational challenge of modelling so many spectral points at differing observations geometries is very significant, and with our existing NEMESIS code we have only been able to explore a limited range of possible parameter space. Fortunately, a Python implementation of NEMESIS, called archNEMESIS \citep{alday25}, has recently been released. The Python structure of archNEMESIS allows it to use multiple cores for the same task (unlike NEMESIS) and its speed on a single core exceeds that of the Fortran NEMESIS through the use of `just in time' (JIT) compilers, and also by restricting the number of read/writes to the disk. As well as Optimal Estimation, archNEMESIS can also use `Nested Sampling' minimisation to fit the observations, which allows a wider exploration of the parameter space and also gives a clearer understanding of the various parameter degeneracies. Hence, in a future paper we plan to use archNEMESIS to widen the parameter search and also extend the analysis to vortex features such as the Great Red Spot.

Finally, this study was inspired by the `holistic' cloud model of Uranus and Neptune determined from observations over a wide spectral range by \citet{irwin22}. Having found a model that matches the observed spectra of Jupiter well, we suspect that a similar model setup may also be suitable for modelling the observed spectra of Jupiter's close cousin, Saturn. Hence, we intend to extend this approach to VLT/MUSE, Cassini/VIMS and Cassini/CIRS observations of Saturn in the near future.

\section*{Acknowledgments}
Joseph Penn and Michelle Colantoni were funded by the Doctoral Training Programme of the UK Science and Technology Facilities Council (grant numbers ST/Y509474/1 and UKRI1772, respectively). Santiago P\'{e}rez-Hoyos was supported by the Basque Government, Spain (Grupos de Investigación, IT1742-22), Elkartek, Spain KK-2025/00106 and by Grant
PID2023–149055NB–C31 funded by MICIU, Spain/AEI/10.13039/501 100011033 and by FEDER, UE. Asier Anguiano-Arteaga was supported by the Programa de Perfeccionamiento de Personal Investigador Doctor 2024-2027 of the Basque Government. Leigh Fletcher was supported by STFC Grant reference UKRI1205.  For the purpose of open access, the authors have applied a Creative Commons Attribution (CC BY) license to the Author Accepted Manuscript version arising from this submission.

\section*{Data Availability}
The data underlying this article will be shared on reasonable request to the corresponding author. The spectral fitting and retrievals were performed using the NEMESIS radiative transfer and retrieval algorithm \citet{irwin08} and can be downloaded from \citet{irwin22a} (or \url{https://github.com/nemesiscode/radtrancode}), with supporting website information at \citet{irwin22b} (or \url{https://github.com/nemesiscode/nemesiscode.github.io}) The calibrated Cassini/VIMS-VIS and VIMS-IR FITS products used in this study are available from Zenodo at \url{https://doi.org/10.5281/zenodo.19223781}.




\bibliographystyle{mnras}
\bibliography{bibliography} 




\appendix

\section{VIMS Wavelength Calibration}

\begin{figure}
	\includegraphics[width=0.7\columnwidth]{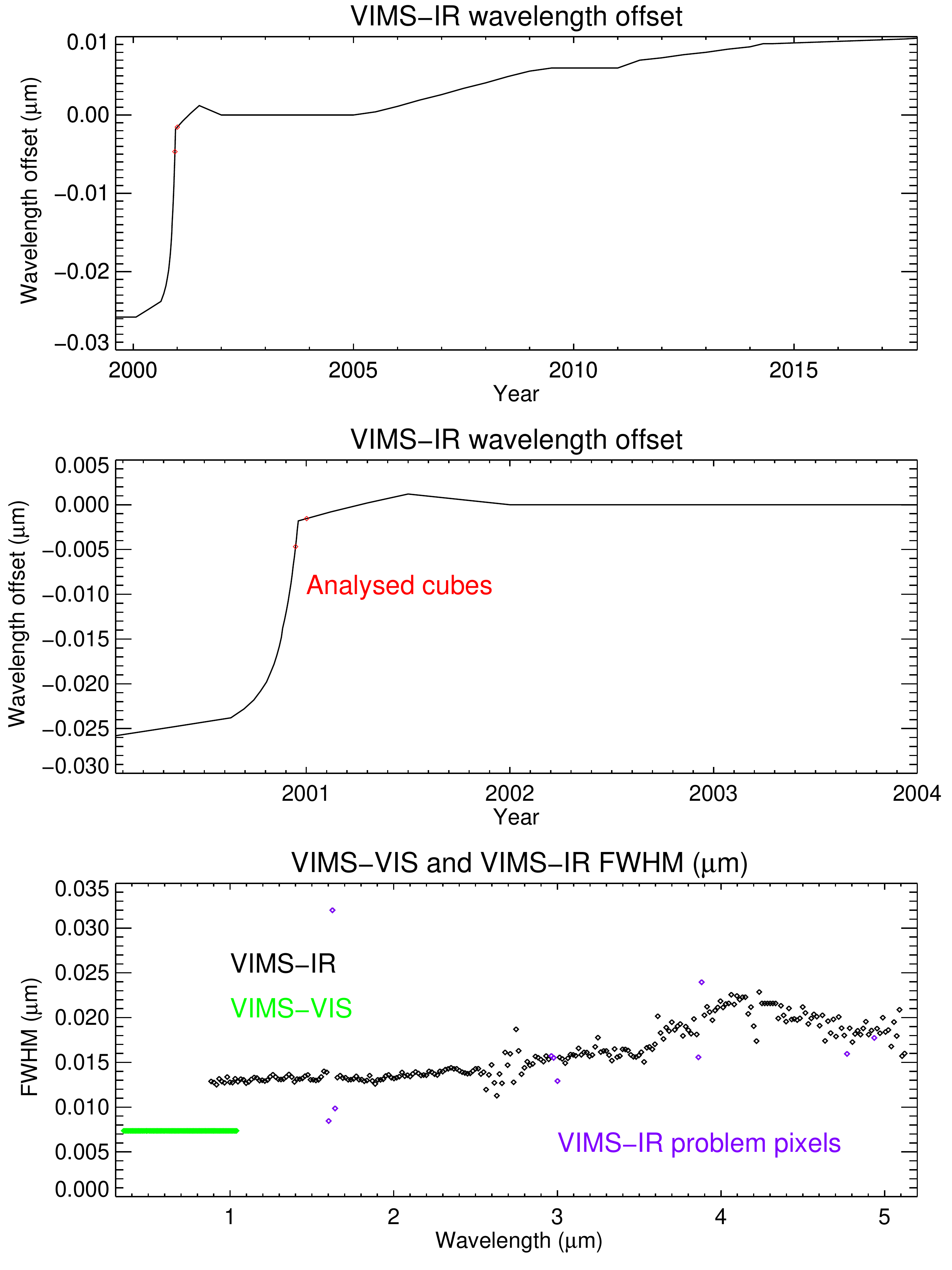}\centering
    \caption{VIMS Calibration variation. Top plot shows the variation in the wavelength offset of VIMS-IR with time as reported by \citet{clark18}. Middle plot shows a close-up of this time variation around the time of Cassini's flyby of Jupiter in December 2000. The red dots in these two plots shows the interpolated offsets for cubes V1355256529\_3 and V1356976257\_3. Bottom plot shows how the full-width-half maximum (FWHM) of the VIMS-IR Instrument Line Shape (ILS) varies with wavelength and is compared with the assumed FWHM of VIMS-VIS of  0.007368 $\mu$m. The FWHM values for pixels with known calibration problems are coloured purple. Both VIMS-VIS and VIMS-IR are assumed to have a Gaussian-shaped Instrument Line Shapes (ILS). }
    \label{fig:vims_calibration}
\end{figure}

The spectral calibration of VIMS-IR varies with wavelength and was also found to vary with time, as summarised in the final calibration report of VIMS \citep{clark18}. Figure \ref{fig:vims_calibration} shows how the wavelength offset of VIMS-IR (relative to its nominal calibration in 2004) was found to vary with time, and also how the full-width-half-maximum (FWHM) of the VIMS-IR Instrument Line Shape (ILS ) varies with wavelength. Highlighted in Fig. \ref{fig:vims_calibration} are the recommended interpolated offsets for our analysed cubes V1355256529\_3 and V1356976257\_3, which can be seen to be significantly offset from the reference 2004 calibration. 

\section{Combined MUSE/VIMS data in NEB}

\begin{figure}
	\includegraphics[width=0.9\columnwidth]{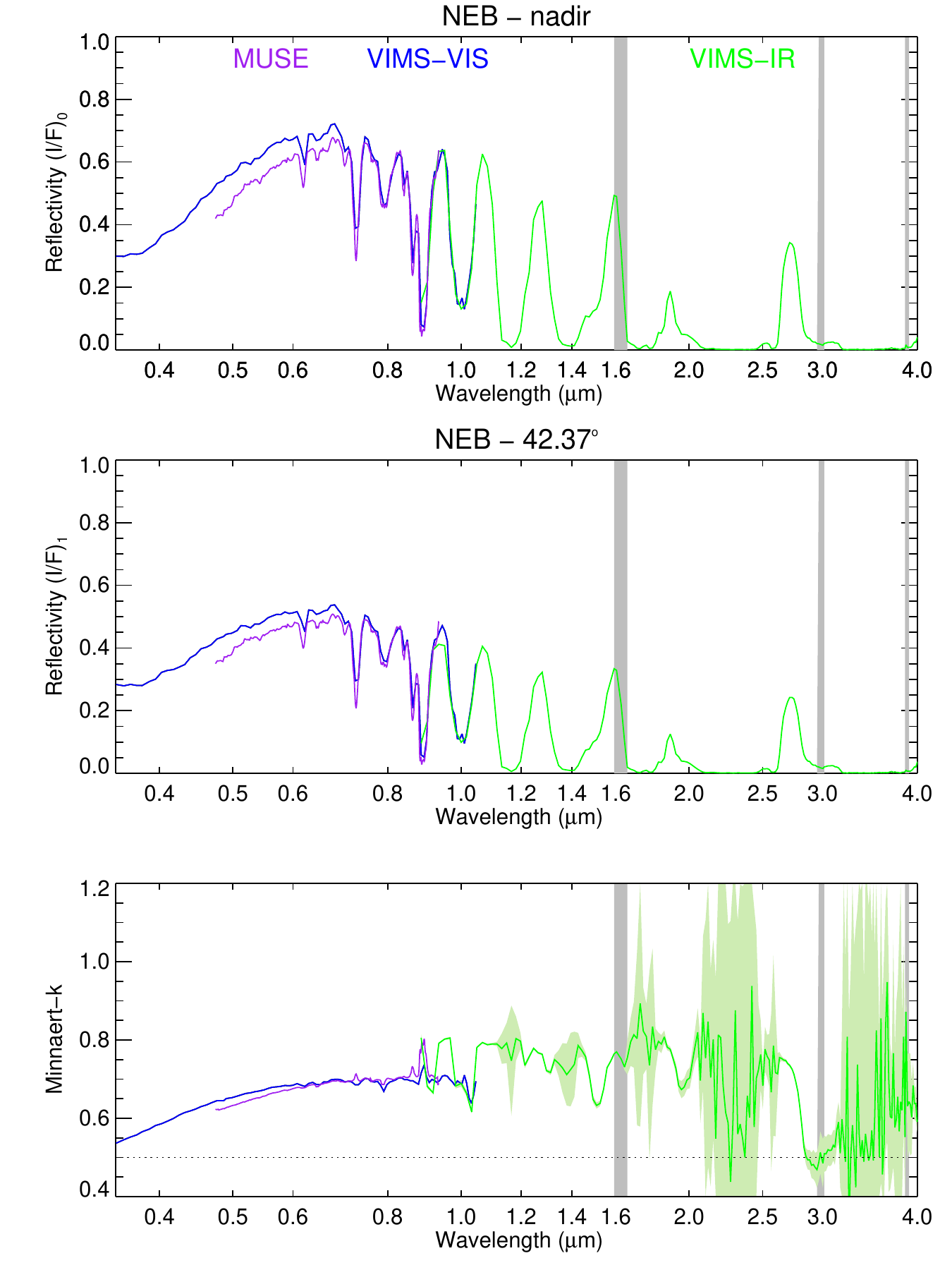}
    \caption{Comparison of MUSE, VIMS-VIS and VIMS-IR spectra from 0.3 to 4 $\mu$m in the NEB (12$^\circ$N). Top row shows extracted spectra at nadir, middle row shows extracted spectra at a zenith angle of 42.37$^\circ$, and bottom row shows the fitted Minnaert-$k$ spectra. A log scale has been used for wavelength to better compare all data. In the Minnaert-$k$ spectra, the estimated noise for VIMS-IR is shaded in light green. For the other instruments, the noise is too small to see. The greyed regions indicate known problems in the VIMS-IR spectra due to order-sorting filter boundaries, etc. }
    \label{fig:jupiter_spectra_comparison1}
\end{figure}

Figure  \ref{fig:jupiter_spectra_comparison1} compares the MUSE, VIMS-VIS and VIMS-IR spectra for the NEB (12$^\circ$N).

\section{Comparison of CH$_4$, NH$_3$ and H$_2$S absorption line strengths}\label{Section:absorption}

\begin{figure}
	\includegraphics[width=\columnwidth]{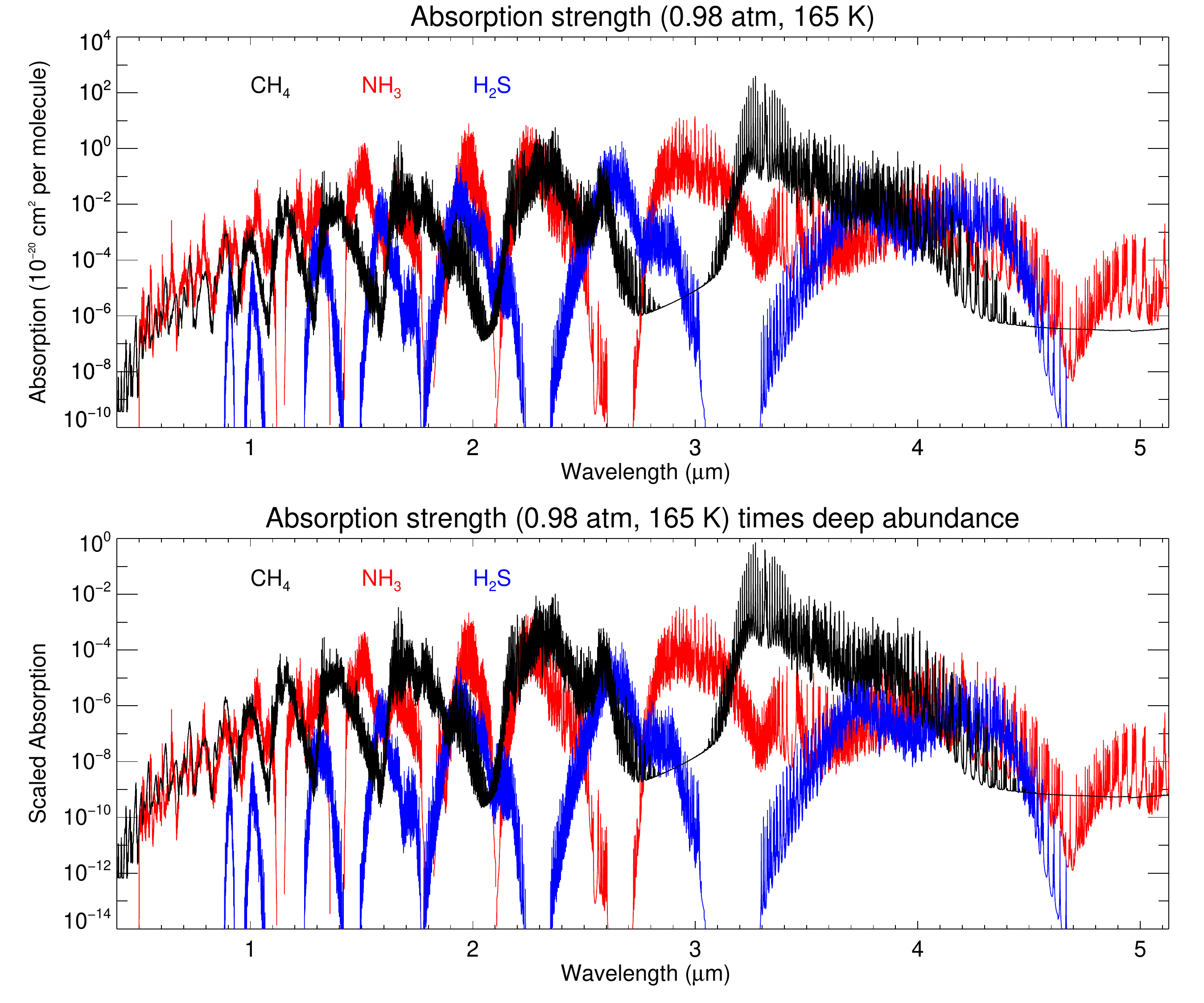}
    \caption{Absorption line strengths of methane (CH$_4$), ammonia (NH$_3$) and hydrogen sulphide (H$_2$S) in our Line-by-Line (LBL) look-up tables, compiled for $p = 0.98$ atm, and $T=175$ K. In the second plot, the strengths have been scaled by the \textit{a priori} assumed Jovian deep abundances of $1.82\times 10^{-3}$, $2.8\times 10^{-4}$
    and $1.0\times 10^{-4}$, respectively. }    \label{fig:absorption}
\end{figure}

Figure \ref{fig:absorption} compares the absorption line strengths of methane (CH$_4$), ammonia (NH$_3$) and hydrogen sulphide (H$_2$S)  at $p = 0.98$ atm and $T=175$ K. 

\section{VIMS-IR Comparison of North Equatorial Dark Feature with the EZ and NEB}\label{Section:NEDF}

\begin{figure}
	\includegraphics[width=\columnwidth]{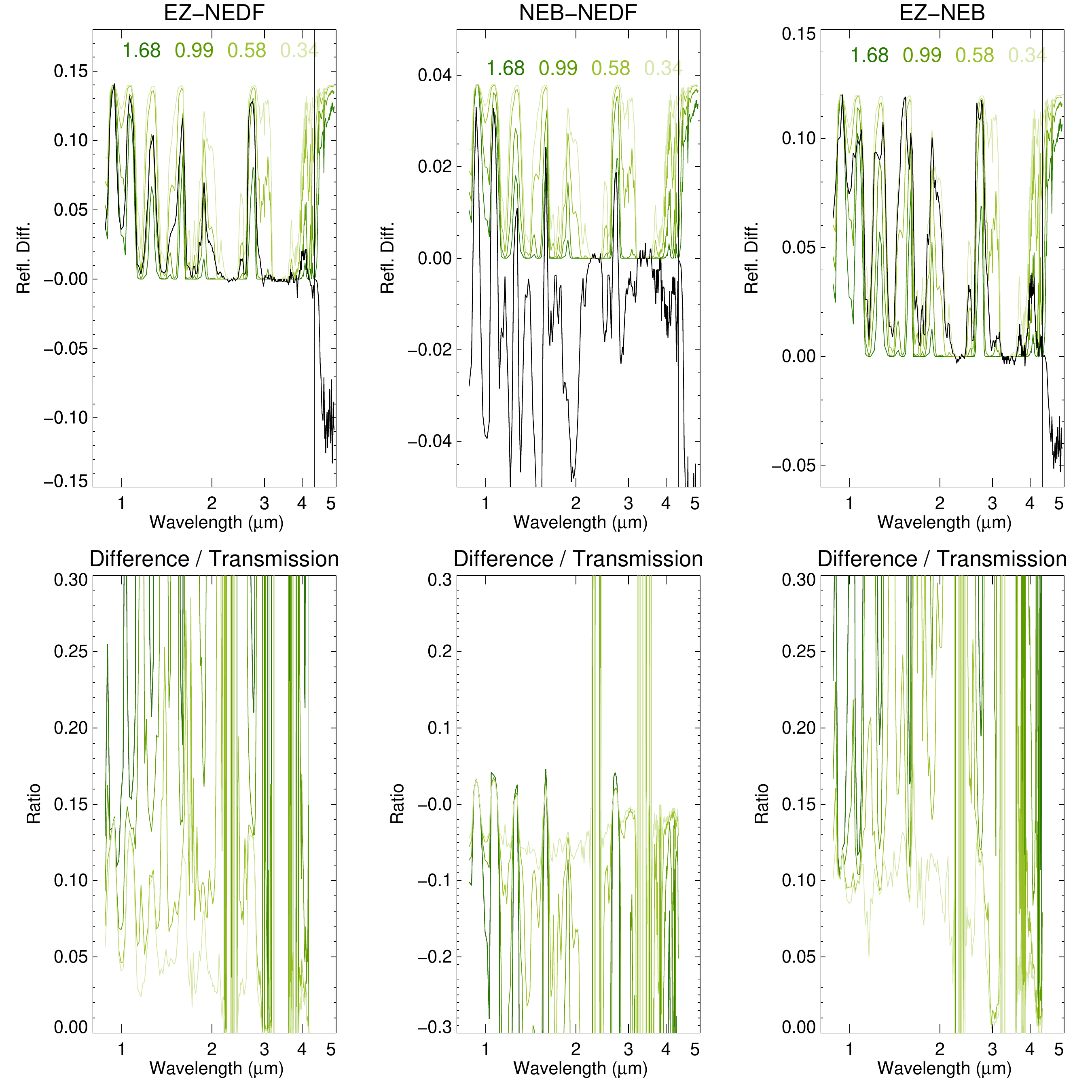}
    \caption{Analysis of Cassini/VIMS-IR spot spectra, comparing the difference between the EZ and NEDF spectra, the difference between the NEB and NEDF spectra, and the difference between the EZ and NEB spectra.   The top row compares the observed difference spectra with the two-way transmission spectra of our standard cloud-free atmosphere to various pressure levels, roughly scaled to be consistent with the peak differences below 1 $\mu$m. The transmissions are colour-coded, as defined in the panels. The bottom row shows the observed difference spectra divided by the calculated transmission spectra. The spectra have been shown on a log wavelength scale to make the shortwave peaks clearer. At wavelengths longer than 4.4 $\mu$m, the radiance difference (W m$^{-2}$ sr$^{-1}$ $\mu$m$^{-1}$) is scaled by 0.5. In the lower row we have omitted the 5-$\mu$m emission wavelength region for clarity. }
    \label{fig:spot_vims_analyse}
\end{figure}

Following on from our analysis of the MUSE difference spectra between the EZ, NEB and NEDF in the main paper, we performed a similar analysis of the VIMS-IR difference spectra, which we show here in Fig. \ref{fig:spot_vims_analyse}. In this figure we have  used a log scale for wavelength to make the shorter wave peaks easier to see. The top row compares the reflectivity difference spectra with the calculated two-way nadir transmission spectra to different pressures, roughly scaled to match the peak of the reflectivity difference spectra. In this spectral region we can see that the analysis is far less simple, which is due to several reasons. First of all, at longer wavelengths the absorption of ammonia becomes very strong and so in the VIMS-IR range the difference spectra depend both on the cloud differences \textbf{and} ammonia abundance differences. Secondly, at these wavelengths there are several strong absorption bands of methane in which we cannot see the deeper clouds at all, but only the aerosols in the upper layer. In contrast, at MUSE wavelengths, only the 890 nm region is solely sensitive to this upper layer and this is where there is most confusion in the ratio plots of Fig. 13 in the main paper. Hence, at VIMS-IR wavelengths we need to consider both changes in the opacity in the upper haze layer \textbf{and} and at deeper levels, and also ammonia abundance differences. The EZ--NEDF and EZ--NEB spectra are mostly sharply peaked at wavelengths of minimum gaseous absorption, which is consistent with reflectivity changes at deeper pressures and for the shortwave peaks we again find best correspondence at pressures of $\sim$1.5 bar. However, at longer wavelengths, where the gaseous absorption is generally increasing, we see better correspondence between the shape of the reflectivity difference peaks at the two-way transmission spectra to lower pressures. This tells us that the aerosol layers, peaking at 1.5 -- 1.7 bar, must be vertically extended, or we need two components - one near 1.5 bar and one  near 0.5 bar. In addition, we see complicated behaviour near the ammonia absorption bands, showing that the ammonia profile must also be changing between these three locations. Finally, for the EZ--NEB difference spectrum we find that the ratio of the difference with the transmission `flattens out' at a pressure of $\sim$0.35 bar, indicating that opacity of the upper level component changes significantly between these locations. This is to be expected, looking at slices of the MUSE and VIMS-IR cubes at wavelengths of strong methane absorption. In contrast the EZ--NEDF ratio spectra do not flatten out in the same way, indicating that the opacity of the overlying aerosols in the upper haze is similar, which is again to be expected from looking at the methane-absorbing slices of the MUSE and VIMS-IR cubes.

\section{Onion Peeling Method}\label{app:onion}

\begin{figure}
	\includegraphics[width=\columnwidth]{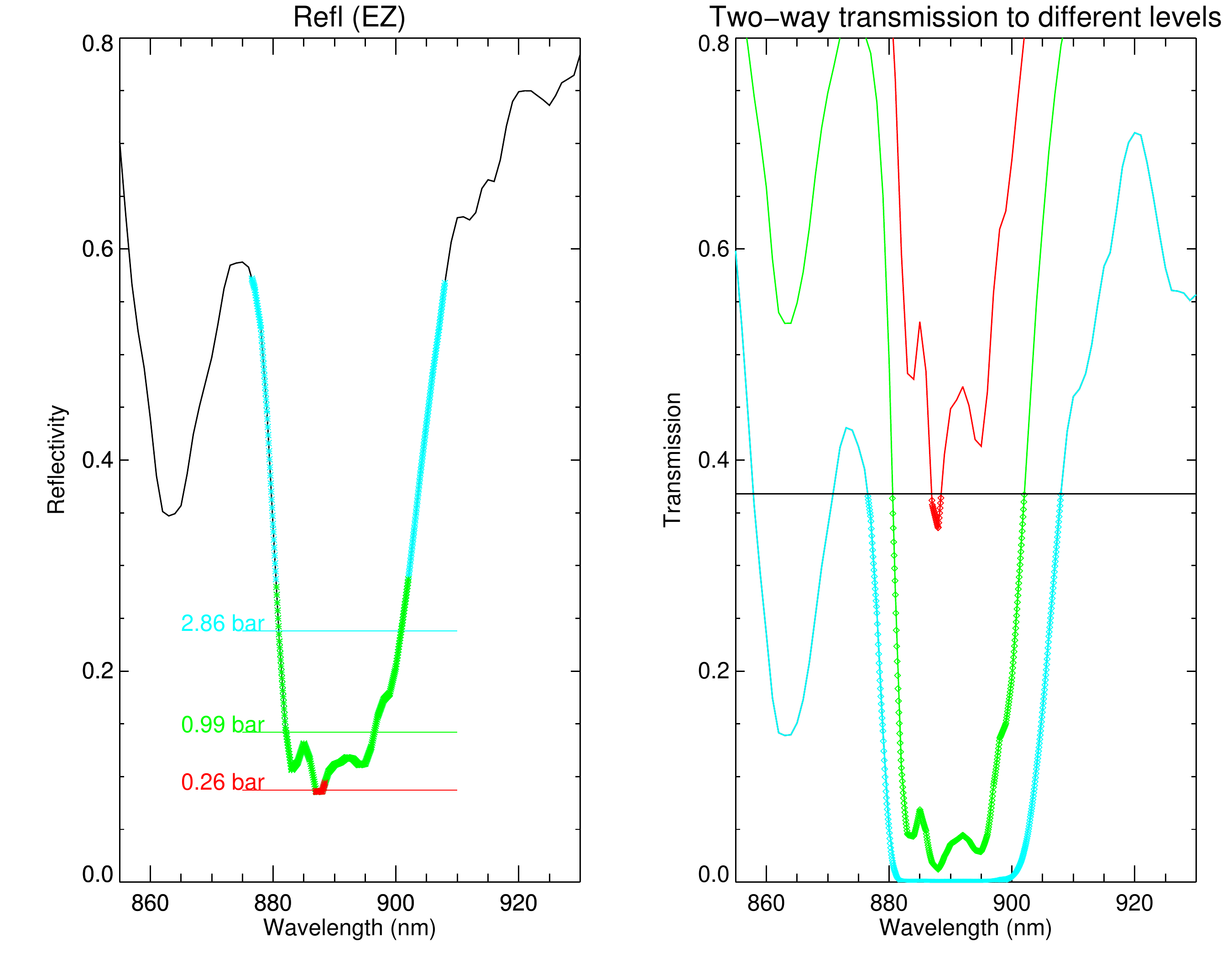}
    \caption{Example of `onion-peeling' for a typical MUSE spectrum measured in the EZ. The left hand panel shows the measured spectrum, while the right hand panel shows the calculated two-way nadir transmission spectra (cloud-free) to three representative pressure levels. Taking the transmission spectrum for each pressure level, we select wavelengths where the transmission is less then $e^{-1}$ (i.e., opacity $\tau > 1$) and then calculate the mean reflectivity of the measured spectrum across these wavelengths, which we show as the horizontal coloured lines in the left-hand panel.}
    \label{fig:oniontest1}
\end{figure}

\begin{figure}
	\includegraphics[width=\columnwidth]{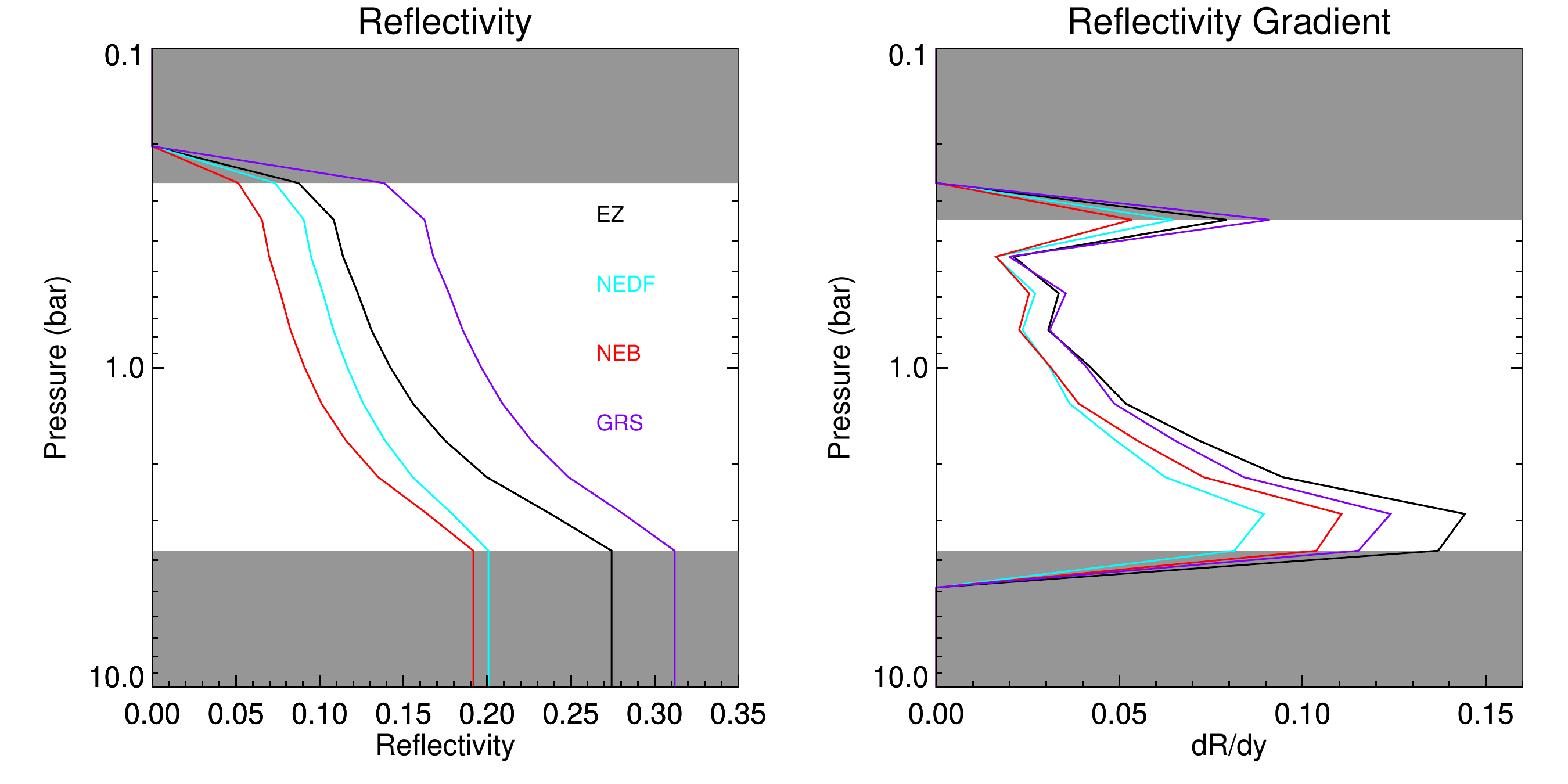}
    \caption{`Onion-peeling' calculations. Left hand panel shows the mean reflectance of the sample MUSE spectra in the 875 -- 910 nm band in four different regions, averaged over wavelengths for which the calculated two-way nadir opacity exceeds 1.0 (i.e., $\tau>1$) for each base pressure level, plotted as a function of base pressure. The right hand panel shows the rate of change of this mean reflectivity with respect to log(pressure), i.e., $y = -\log(p)$. The greyed area in these plots are where we have no information. Four regions plotted are: the Equatorial Zone (EZ), a North Equatorial Dark Feature (NEDF), the North Equatorial Belt (NEB), and the Great Red Spot (GRS). }
    \label{fig:oniontest2}
\end{figure}

Our previous band-depth analysis  of VLT/MUSE observations \citep{irwin25} has shown that much can be learned about Jupiter's atmospheres with relatively simple models, provided these can be interpreted reliably. One such method is `onion-peeling', where spectra are analysed by first estimating properties in the upper atmosphere, subtracting or `peeling away' the effects of these layers to leave just the effects of lower layers, and then working downwards iteratively. We found the region near 890 nm in the VLT/MUSE spectra was especially well suited for such an analysis since the region is not much affected by ammonia absorption and the absorption of gaseous methane varies enormously from the centre of the band to its edge. This can be seen in Fig. 10 of the main paper, where in the centre of the band the atmosphere `blacks out' (i.e., the two-way transmission to space drops to near zero) at pressures of about 0.3 bar, while at the edge of the band the transmission does not black out until pressures of several bars. From the computed transmission spectra to different pressure levels in the 875 -- 910 nm region, a selection of which are shown in Fig. \ref{fig:oniontest1}, we selected the MUSE wavelengths where the two-way opacity $\tau > 1$ (and thus transmission $< \exp(-1.0) = 0.3679$) and then calculated the average MUSE reflectivity of these wavelengths at all points on Jupiter's disc. As we go to deeper pressures, there is a wider range of wavelengths where the two-way opacity exceeds 1.0 and the mean reflectivity increases. Hence, if we plot the averaged reflectivity against the pressures of the calculated transmission spectra used to select the averaging wavelengths, we can estimate how the average reflectivity increases with pressure, which we show in Fig. \ref{fig:oniontest2} for a range of typical locations: 1) Equatorial Zone (EZ); 2) North Equatorial Dark Feature (NEDF); 3) North Equatorial Belt (NEB); and 4) the Great Red Spot (GRS). As can be seen, the reflectivity increases monotonically with pressure, but the rate of increase will be fastest when there is a cloud layer present. Figure \ref{fig:oniontest2} also shows the rate of increase of reflectivity with respect to $y=-\log(p)$, where $p$ is the pressure. Here it can be seen that there is evidence of an upper cloud/haze at $p < 0.5$ bar and also a deeper cloud at $p > 1$ bar. This can be seen from the raw MUSE cubes themselves. At the centre of the 890 nm band we see just the upper level component, which has a very different distribution from the main aerosol features, as can be seen in Fig. 2 of the main paper. As we move away from the centre of the 890 nm band, the atmosphere becomes less opaque and we start to see the lower, main cloud features. Hence, the cloud profile shape derived from this onion-peeling method is consistent with the raw observations. It is also remarkably consistent with the Galileo Probe Nephelometer results shown in Fig. 1 of the main paper, although the onion-peeling profile here peaks at higher pressures. However, the absolute pressure level where the reflectivity gradient peaks depends on our assumed value for the minimum two-way opacity, which we chose here to be $\tau = 1$. Choosing a lower value of $\tau$ moves the inferred clouds to lower pressures and thus the absolute pressure levels of the inferred clouds is not well defined.  

\begin{figure}
	\includegraphics[width=\columnwidth]{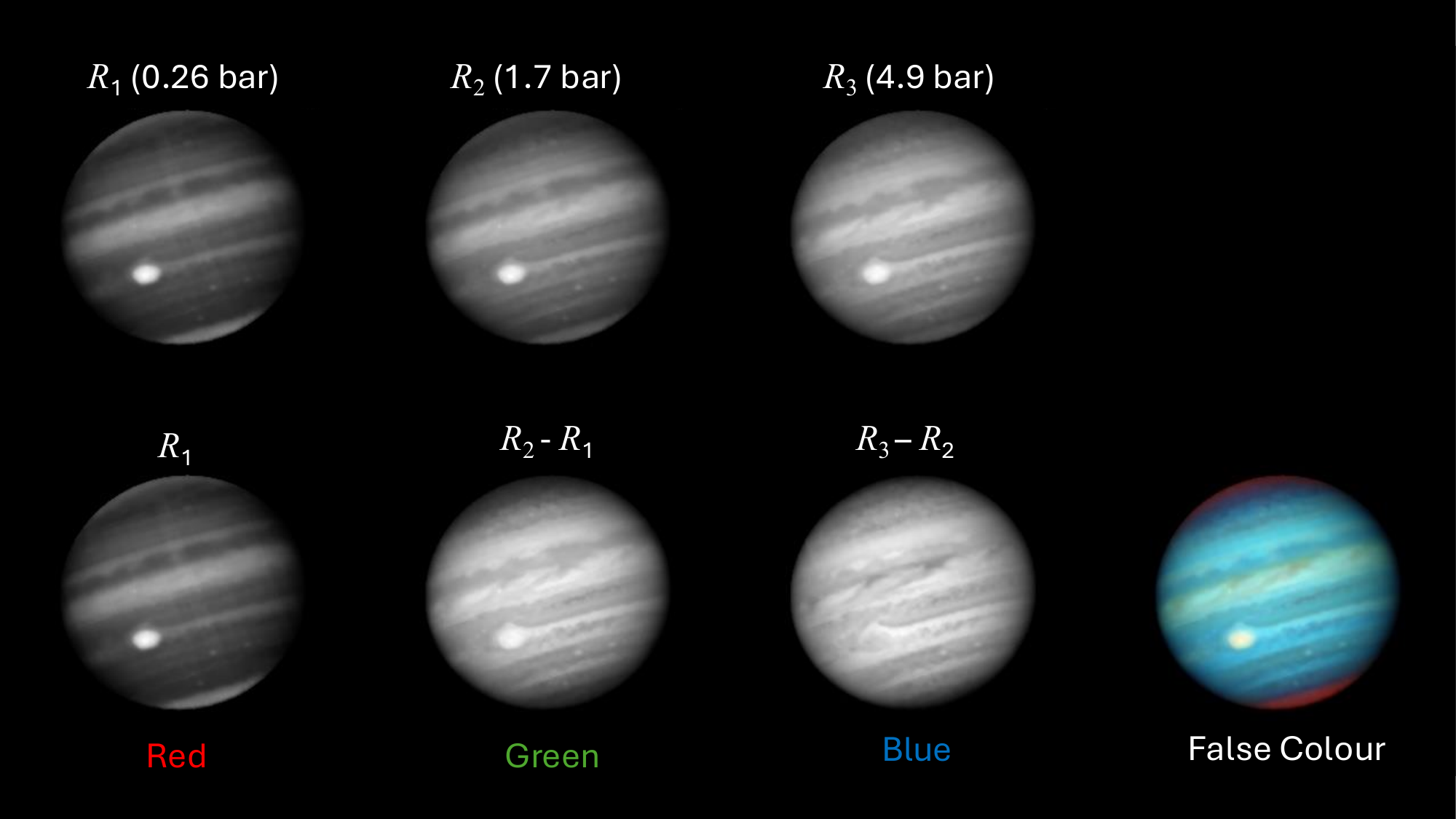}
    \caption{Top row shows the averaged reflectivities for wavelengths where the calculated two-way opacity spectra to pressure levels: 0.26 bar ($R_1$), 1.7 bar ($R_2$) and 4.9 bar ($R_3$) exceed 1.0, respectively, scaled to the same overall brightness. In the bottom row we show $R_1$ again, then $R_2 - R_1$ and $R_3-R_2$, again all scaled to the same overall brightness. The fourth image on the bottom row is false colour composite of the three previous  images.}
    \label{fig:onionfigure}
\end{figure}

To visualise the results of Fig. \ref{fig:oniontest2} more easily, the top row of Fig. \ref{fig:onionfigure} shows the averaged reflectivity distributions across Jupiter's disc for wavelengths where the two-way opacity spectra to the 0.26, 1.7 and 4.9 bar pressure levels (illustrative of upper, intermediate, and deeper-level contributions) are greater than 1.0, which we will label as $R_1$, $R_2$ and $R_3$, respectively. The 0.26-bar image shows just the upper component, while the following two images also show the deeper cloud structure, which is clearest in the third image where we average over most of the 875--910 nm range and thus the image is dominated by the reflection from regions of low methane absorption. Small differences between these images can be used to probe for variations of cloud reflectivity with depth. In the bottom row of Fig. \ref{fig:onionfigure} we show $R_1$ again, but then $R_2 - R_1$ and $R_3-R_2$. The fourth image on the bottom row is an RGB composite of the previous three images. This figure reveals the power of this onion-peeling method. The second image on the bottom row can be thought of as the contribution to the overall reflectivity from clouds between 1.7 and 0.26 bar, while the third image can be thought of as the contribution to the overall reflectivity from clouds between 4.9 and 1.7 bar. The middle image shows most of the usual cloud features, which points to their origin likely arising from clouds in this region. We can also see that that this image is distinctly different from the 3rd image, where the GRS is less obvious, but the NEDF features are enhanced relative to the NEB. Hence this points to the NEDFs being caused by differences in cloud reflectivity at greater pressures.

\section{JIRAM-SPE 5-micron window}\label{sec:JIRAM-5micron}

\begin{figure}
	\includegraphics[width=\columnwidth]{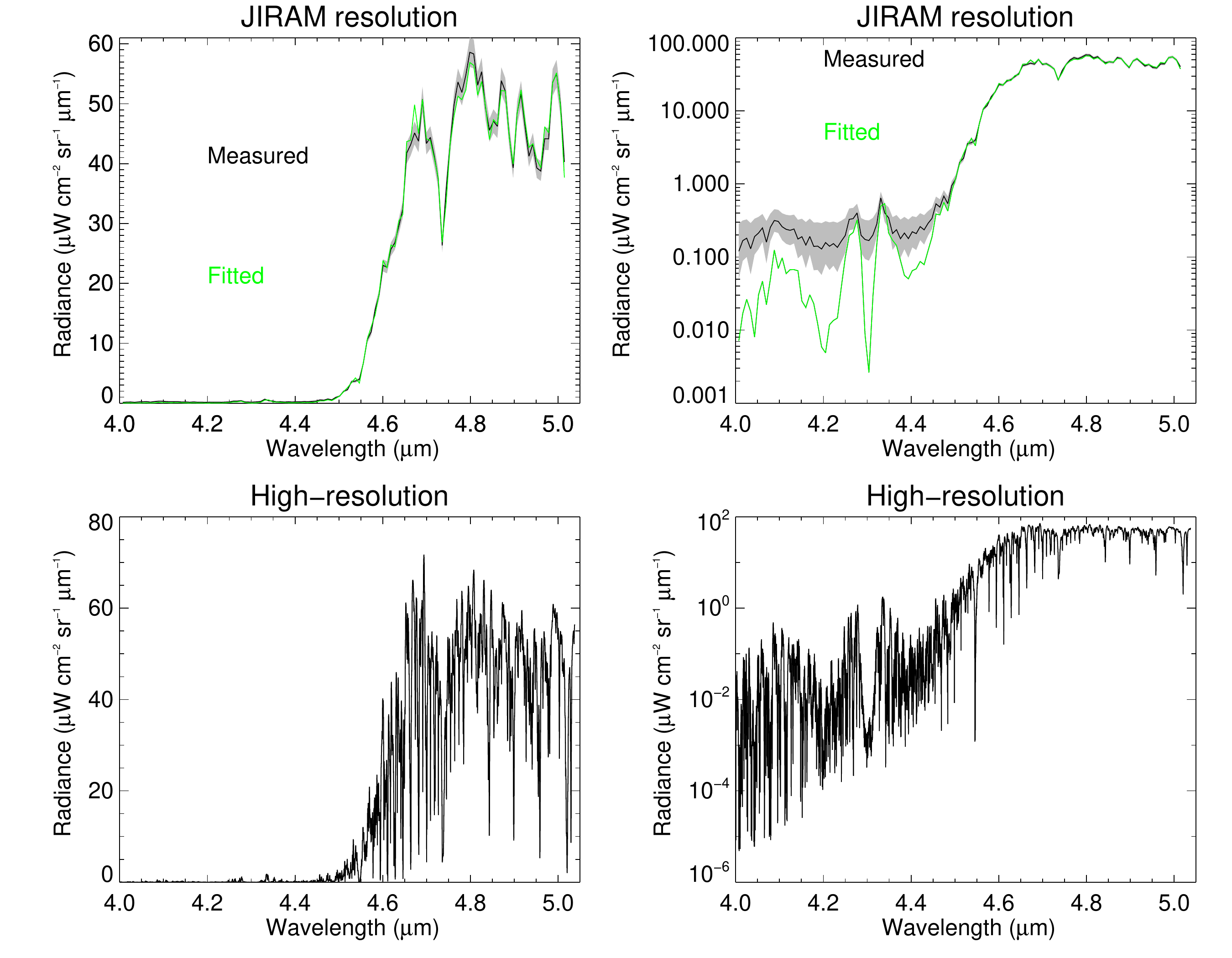}
    \caption{Example fit to a STB JIRAM-SPE 5-$\mu$m spectrum  (JIR\_SPE\_RDR\_2016240T164555, \#158) analysed by \citet{grassi20}, considering no scattered sunlight and showing a similar goodness of fit as achieved by \citet{grassi20}. Top row compares the measured and fitted spectra. The assumed error on the measured spectrum is indicated in grey. Bottom row shows the computed high-resolution LBL spectrum prior to smoothing down to JIRAM-SPE resolution. The left-hand column shows the spectra plotted on a linear radiance scale, while the right-hand column shows the same spectra plotted on a log radiance scale. The model under-fits the data at shorter wavelengths as reflected sunshine was not included in this retrieval.}
    \label{fig:JIRAM_comparison}
\end{figure}

\begin{figure}
	\includegraphics[width=0.95\columnwidth]{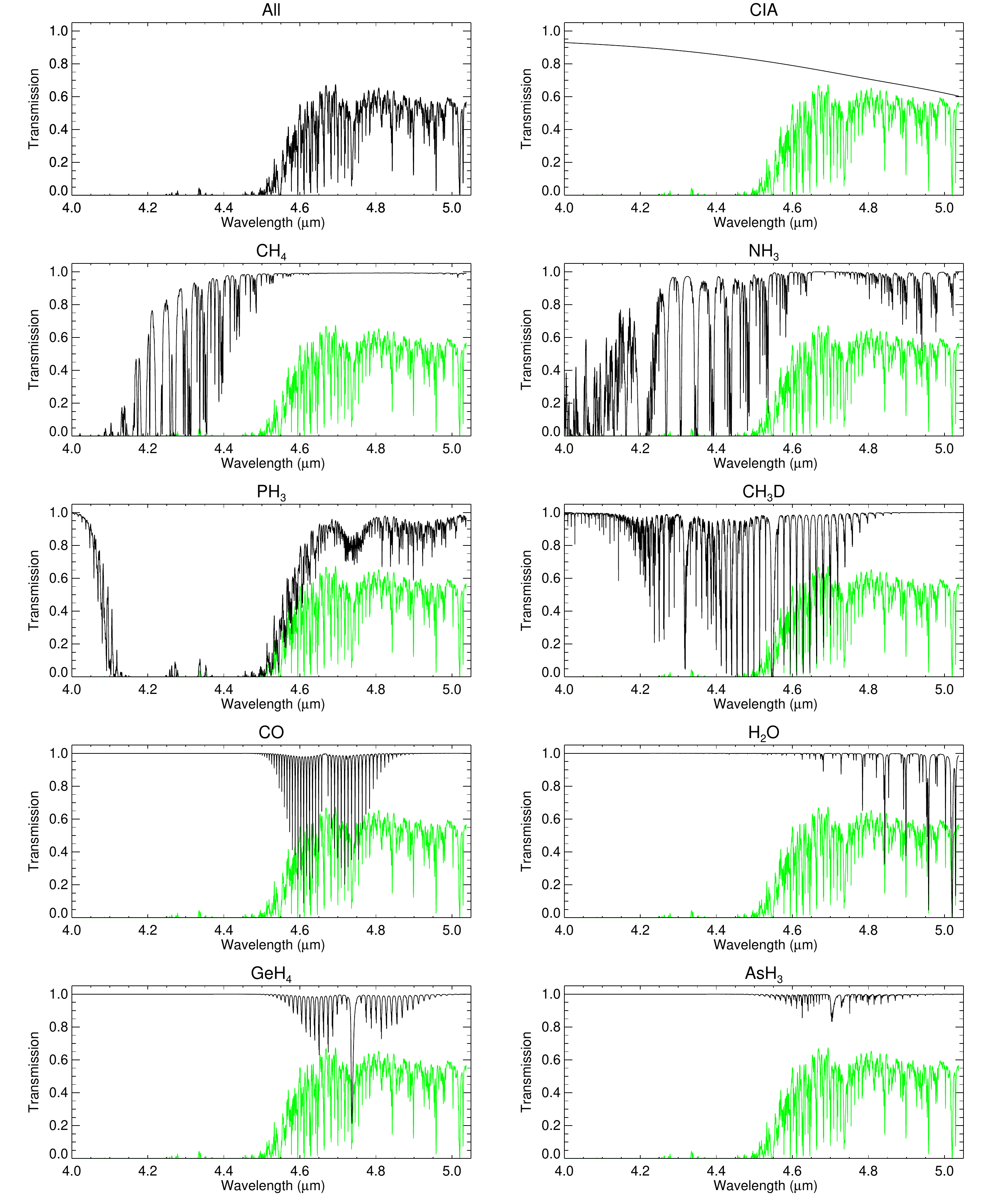}
    \caption{Line-by-line nadir transmission calculation down to 4.75-bar pressure level for the atmospheric model fitted to the sample STB JIRAM-SPE spectrum (Fig. \ref{fig:JIRAM_comparison}). The first panel shows the transmission of all gases, while in subsequent panels this calculation (in green) is compared to the transmission of the  individual components, including  collision-induced absorption (CIA), showing which gases are most responsible for which features.}
    \label{fig:5mic_trans_hires}
\end{figure}

\begin{figure}
	\includegraphics[width=0.95\columnwidth]{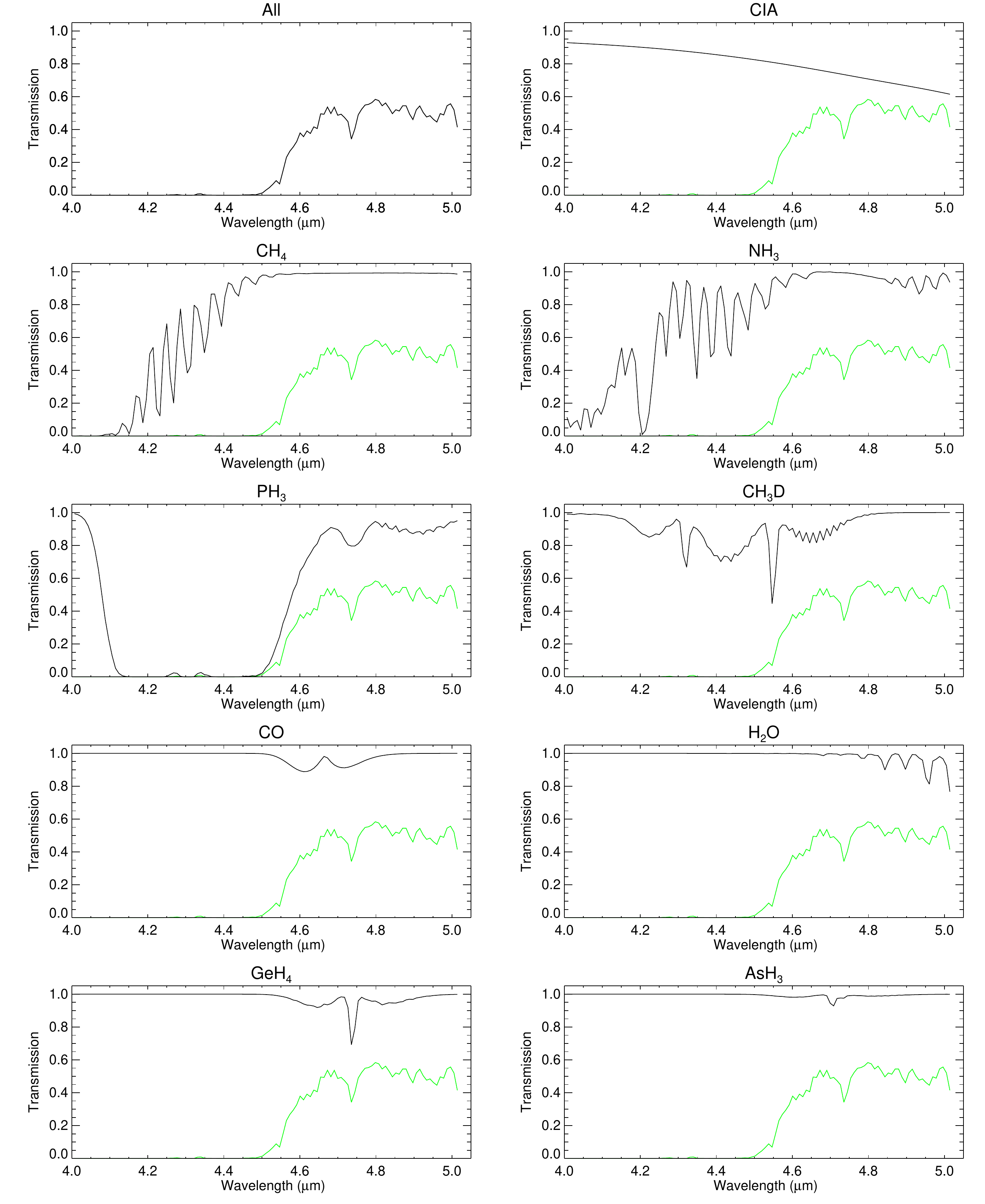}
    \caption{As Fig. \ref{fig:5mic_trans_hires}, but smoothed to JIRAM-SPE spectral resolution}
    \label{fig:5mic_trans_lores}
\end{figure}

\begin{figure}
	\includegraphics[width=\columnwidth]{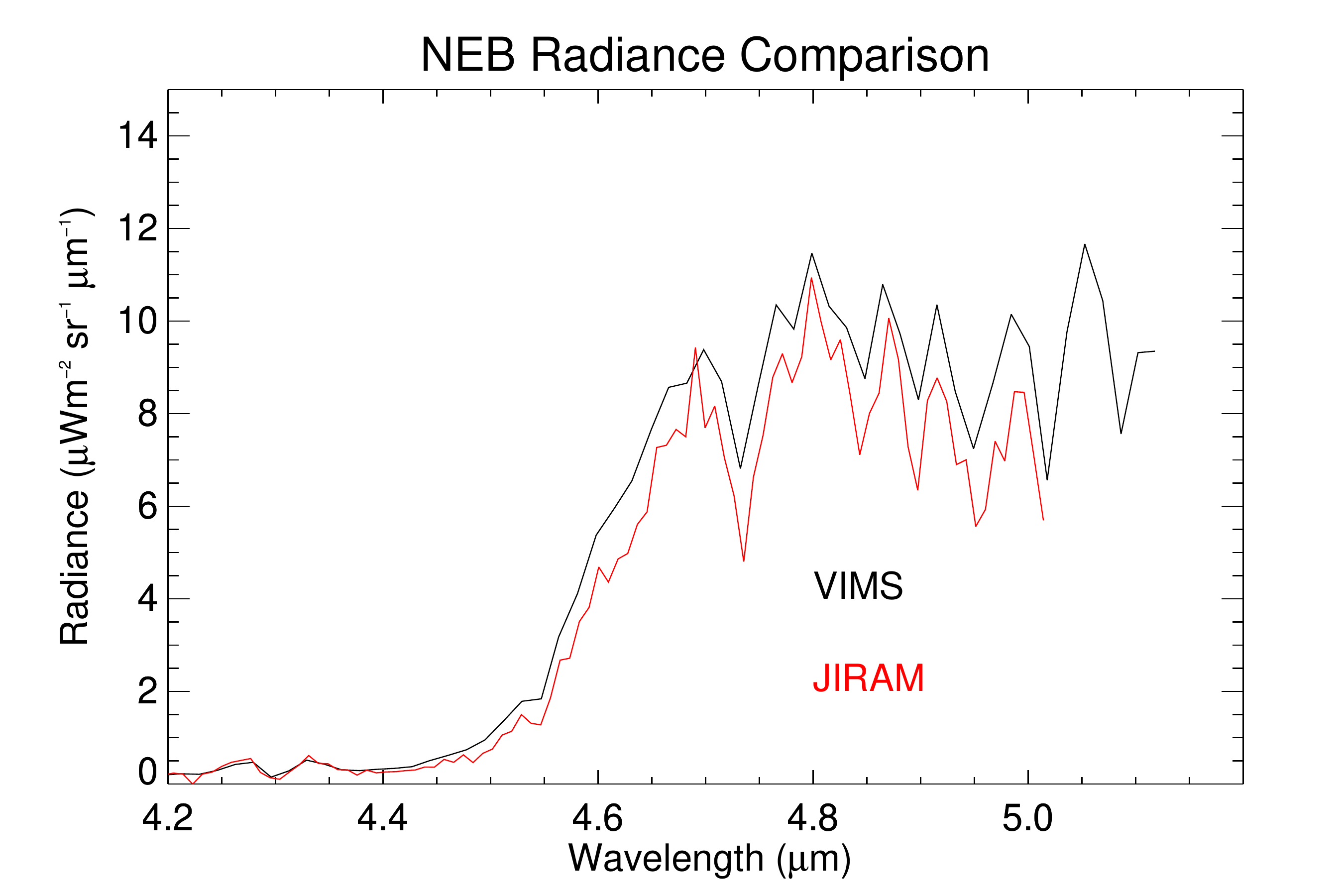}
    \caption{Comparison of typical VIMS and JIRAM-SPE spectra in the NEB (12$^\circ$N), showing that JIRAM-SPE has greatly superior spectral resolution, and also superior wavelength sampling. The VIMS standard 2004 wavelength calibration \citep{clark18} has been used here and appears to be satisfactory for this observation date.}
    \label{fig:jiram_vims_comparison}
\end{figure}

In this section we test whether our radiative transfer model is sufficiently accurate to model Jovian 5-$\mu$m spectra previously analysed by \citet{grassi20}. We analysed JIRAM-SPE spectra from the first Juno perijove passage (PJ1, 27th August 2016, 12:50 UTC), following \citet{grassi20}, concentrating on the planning period JM0003, measured after 26th August, 10:58 UTC. We found we are able to achieve similarly good fits to these data using our model, using the nominal spectral resolution of 13.6 nm at 5 $\mu$m \citep{adriani17}. Figure \ref{fig:JIRAM_comparison} shows our fit to the South Temperate Belt spectrum \\
JIR\_SPE\_RDR\_2016240T164555, \#158, also analysed by \citet{grassi20}. This spectrum is from near the terminator with a solar zenith angle of 79.6$^\circ$ and viewing zenith angle of 21.42$^\circ$. In this fit, and following \citet{grassi20}, we assumed the solar reflected component was negligible and considered only thermally emitted radiation.  We used the cloud model of \citet{grassi20}, consisting of a single thin cloud at 1 bar, of variable opacity with a flat extinction spectrum, a single scattering albedo of 0.9 and an asymmetry parameter  of $g=0.7$ \citep{giles15}. Similarly, we had variable scaling factors for the abundances of H$_2$O, PH$_3$, NH$_3$, AsH$_3$, GeH$_4$, CO and CH$_3$D. The top panel of  Fig. \ref{fig:JIRAM_comparison} shows the fit at JIRAM-SPE resolution, which was calculated using our line-by-line (LBL) look-up tables, which we found to give superior fitting accuracy compared with correlated-$k$ models that have previously been used to model such spectra. 
The bottom panel of Fig. \ref{fig:JIRAM_comparison} shows the underlying high-resolution LBL calculation before smoothing with the JIRAM-SPE Gaussian instrument function. This high-resolution spectrum reveals the complexity of gaseous line absorptions that make up the 5-$\mu$m spectrum. The contribution of individual gases to the calculated spectrum is shown in Figs. \ref{fig:5mic_trans_hires} and \ref{fig:5mic_trans_lores}. Finally, in Fig. \ref{fig:jiram_vims_comparison} we compare the extracted VIMS approach nadir spectrum in the NEB at 12$^\circ$N with a single JIRAM-SPE spectrum at the same latitude and similar 5-$\mu$m brightness, where we see good overall correspondence and the superior spectral resolution of JIRAM-SPE. Here, the standard 2004 calibration of the VIMS spectra appears adequate, although the VIMS-IR spectrum is quite undersampled in the rapidly changing 5-$\mu$m range making the accurate fitting of any wavelength correction difficult.

\bsp	
\label{lastpage}
\end{document}